\documentclass[11pt]{article}

\usepackage[numbers]{natbib}
\usepackage{aastex_hack}

\usepackage{fancyhdr}
\usepackage{mdframed} % E Kara added

\newcommand{\reqfilled}[1]{}

\newcommand{\Msun}      {\mbox{$\,M_{\mathord\odot}$}}
\newcommand{\strobex}{\textit{STROBE-X}}
\newcommand{\xmm}{\textit{XMM-Newton}}
\newcommand{\nustar}{\textit{NuSTAR}}
\newcommand{\nicer}{\textit{NICER}}
\newcommand{\swift}{\textit{Swift}}
\newcommand{\rxte}{\textit{RXTE}}
\newcommand{\chandra}{\textit{Chandra}}
\newcommand{\fermi}{\textit{Fermi}}
\newcommand{\rosat}{\textit{ROSAT}}
\newcommand{\eROSITA}{\textit{eROSITA}}
\newcommand{\athena}{\textit{Athena}}

\usepackage[small,bf]{caption}
\usepackage[small,compact,center]{titlesec}
\usepackage{paralist}
\usepackage{rotating}
\usepackage{float}
\usepackage{eurosym}
\usepackage{hyperref}
\usepackage{wrapfig}
\usepackage{pdfpages}
\usepackage{amsmath}
\usepackage{amssymb}
\usepackage{colortbl}

\newcommand{\fixme}[1]{}

\usepackage{graphicx}

\begin{document} 

%%%%%%%%%%%%%%%%%%%%%%%%%%%%%%%%%%%%%%%%%%%%%%%%%%%%%%%%%%%%%%%%%%%%%%%%%%%%%%

\pagenumbering{roman}

%\vspace*{1ex}
\begin{center}
\includegraphics[width=4.0in]{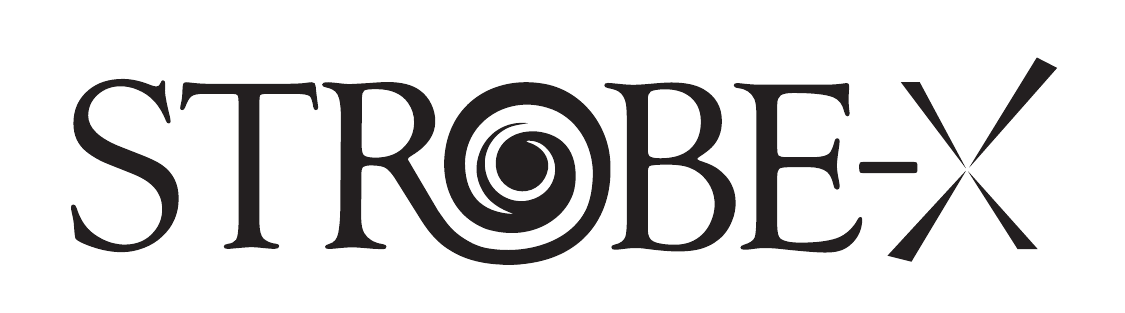}\\

  {\LARGE\bf \textit{STROBE-X}: X-ray Timing and Spectroscopy on Dynamical Timescales from Microseconds to Years}\\[0.2in]
  
  {\large White Paper Submitted to Astro 2020 Decadal Survey}\\[0.2in]
  
  Paul S. Ray\\
  U.S. Naval Research Laboratory\\
  \href{mailto:paul.ray@nrl.navy.mil}{\tt paul.ray@nrl.navy.mil}\\

\end{center}
\paragraph{\strobex{} Steering Committee:} Zaven Arzoumanian, David Ballantyne, Enrico Bozzo, Soren Brandt, Laura Brenneman,
Deepto Chakrabarty, Marc Christophersen, Alessandra DeRosa,
Marco Feroci, Keith Gendreau, Adam Goldstein, Dieter Hartmann, Margarita Hernanz, Peter Jenke,
Erin Kara, Tom Maccarone, Michael McDonald, 
Michael Nowak, Bernard Phlips, Ron Remillard, Abigail Stevens, John Tomsick, Anna Watts, Colleen Wilson-Hodge, Kent Wood, Silvia Zane

\paragraph{\strobex{} Science Working Group:}
%\fixme{Check for missing people and confirm that everyone is OK being listed.}
Marco Ajello,
Will Alston,
Diego Altamirano,
Vallia Antoniou,
Kavitha Arur,  
Dominic Ashton,
Katie Auchettl,
Tom Ayres,
Matteo Bachetti,
Mislav Balokovi{\' c},
Matthew Baring,
Altan Baykal,
Mitch Begelman,
Narayana Bhat,
Slavko Bogdanov,
Michael Briggs,
Esra Bulbul,
Petrus Bult,
Eric Burns,
Ed Cackett,
Riccardo Campana,
Amir Caspi,
Yuri Cavecchi,
Jerome Chenevez,
Mike Cherry,
Robin Corbet,
Michael Corcoran,
Alessandra Corsi,
Nathalie Degenaar,
% Imma Donnarumma, % Can't sign on without ASI permission
Jeremy Drake,
Steve Eikenberry, 
Teruaki Enoto,
Chris Fragile,
Felix Fuerst,
Poshak Gandhi,
Javier Garcia,
Adam Goldstein,
Anthony Gonzalez,
Brian Grefenstette,
Victoria Grinberg,
Bruce Grossan,
Sebastien Guillot,
Tolga Guver,
Daryl Haggard,
Craig Heinke,
Sebastian Heinz, 
Paul Hemphill,
Jeroen Homan,
Michelle Hui,
Daniela Huppenkothen,
Adam Ingram,
Jimmy Irwin,
Gaurava Jaisawal,
Amruta Jaodand,
Emrah Kalemci,
David Kaplan,
Laurens Keek,
Jamie Kennea,
Matthew Kerr,
Michiel van der Klis,
Daniel Kocevski,
Mike Koss,
Adam Kowalski,
Dong Lai,
Fred Lamb,
Silas Laycock,
Joseph Lazio,
Davide Lazzati,
Dana Longcope,
Michael Loewenstein,
Dipankair Maitra,
Walid Majid,
W. Peter Maksym,
Christian Malacaria,
Raffaella Margutti,
Adrian Martindale,
Ian McHardy,
Manuel Meyer,
Matt Middleton,
Jon Miller,
Cole Miller,
Sara Motta,
Joey Neilsen,
Tommy Nelson,
Scott Noble,
Paul O'Brien,
Julian Osborne,
Rachel Osten,
Feryal Ozel,
Nipuni Palliyaguru,
Dheeraj Pasham,
Alessandro Patruno,
Vero Pelassa,
Maria Petropoulou,
Maura Pilia,
Martin Pohl,
David Pooley,
Chanda Prescod-Weinstein,
Dimitrios Psaltis,
Geert Raaijmakers,
Chris Reynolds,
Thomas E. Riley,
Greg Salvesen,
Andrea Santangelo,
Simone Scaringi,
Stephane Schanne,
Jeremy Schnittman,
David Smith,
Krista Lynne Smith,
Bradford Snios,
Andrew Steiner,
Jack Steiner,
Luigi Stella,
Tod Strohmayer,
Ming Sun,
Thomas Tauris,
Corbin Taylor,
Aaron Tohuvavohu,
Andrea Vacchi,
Georgios Vasilopoulos,
Alexandra Veledina,
Jonelle Walsh,
Nevin Weinberg,
Dan Wilkins,
Richard Willingale,
Joern Wilms,
Lisa Winter,
Michael Wolff,
Jean in 't Zand,
Andreas Zezas,
Bing Zhang,
Abdu Zoghbi
%######################

\clearpage

%%%%%%%%%%%%%%%%%%%%%%%%%%%%%%%%%%%%%%%%%%%%%%%%%%%%%%%%%%%%%%%%%%%%%%%%%%%%%%
\renewcommand{\contentsname}{}
\setcounter{tocdepth}{3}

\newcommand{\newpart}[1]{\addtocontents{toc}{\protect\vspace{-2ex}}\part{#1}}

{\footnotesize \tableofcontents}

\clearpage

%%%%%%%%%%%%%%%%%%%%%%%%%%%%%%%%%%%%%%%%%%%%%%%%%%%%%%%%%%%%%%%%%%%%%%%%%%%%%%

\pagenumbering{arabic}

\section{Executive Summary}

\begin{wrapfigure}{r}{0.51\textwidth}
%\begin{figure}
\includegraphics[width=0.51\textwidth]{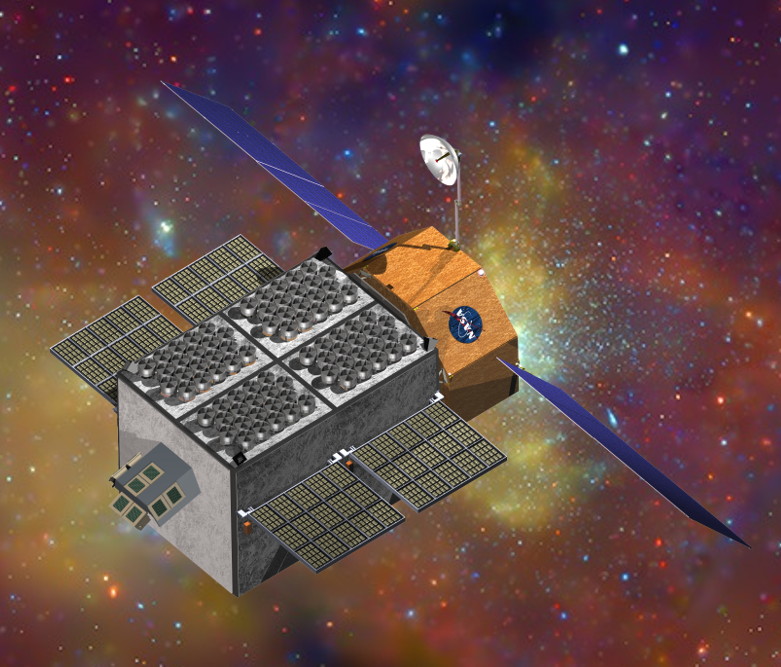}
%\caption{Rendering of \strobex}
%\end{figure}
\end{wrapfigure}

We present the \textit{Spectroscopic Time-Resolving Observatory for Broadband Energy X-rays} (\strobex{}), a probe-class
mission concept selected for study by NASA. It combines huge collecting area, high throughput, broad energy coverage, and excellent spectral and temporal resolution in a single facility. \strobex{} offers an enormous increase in sensitivity for X-ray spectral timing, extending these techniques to extragalactic targets for the first time. It is also an agile mission capable of rapid response to transient events, making it an essential X-ray partner facility in the era of time-domain, multi-wavelength, and multi-messenger astronomy.  Optimized for study of the most extreme conditions found in the Universe, its key science objectives include:
%\vspace{0.01in}
\begin{itemize}
\itemsep 0in
\topsep 0in
\parskip 0in
\partopsep 0in
    \item Robustly measuring mass and spin and mapping inner accretion flows across the black hole mass spectrum, from compact stars to intermediate-mass objects to active galactic nuclei.
    \item Mapping out the full mass-radius relation of neutron stars using an ensemble of nearly two dozen rotation-powered pulsars and accreting neutron stars, and hence measuring the equation of state for ultradense matter over a much wider range of densities than explored by \nicer{}. 
    \item Identifying and studying X-ray counterparts (in the post-\textit{Swift} era) for multiwavelength and multi-messenger transients in the dynamic sky through cross-correlation with gravitational wave interferometers, neutrino observatories, and high-cadence time-domain surveys in other electromagnetic bands. 
    \item Continuously surveying the dynamic X-ray sky with a large duty cycle and high time resolution to characterize the behavior of X-ray sources over an unprecedentedly vast range of time scales.
\end{itemize}
\begin{figure}[h]
\includegraphics[width=6.5in]{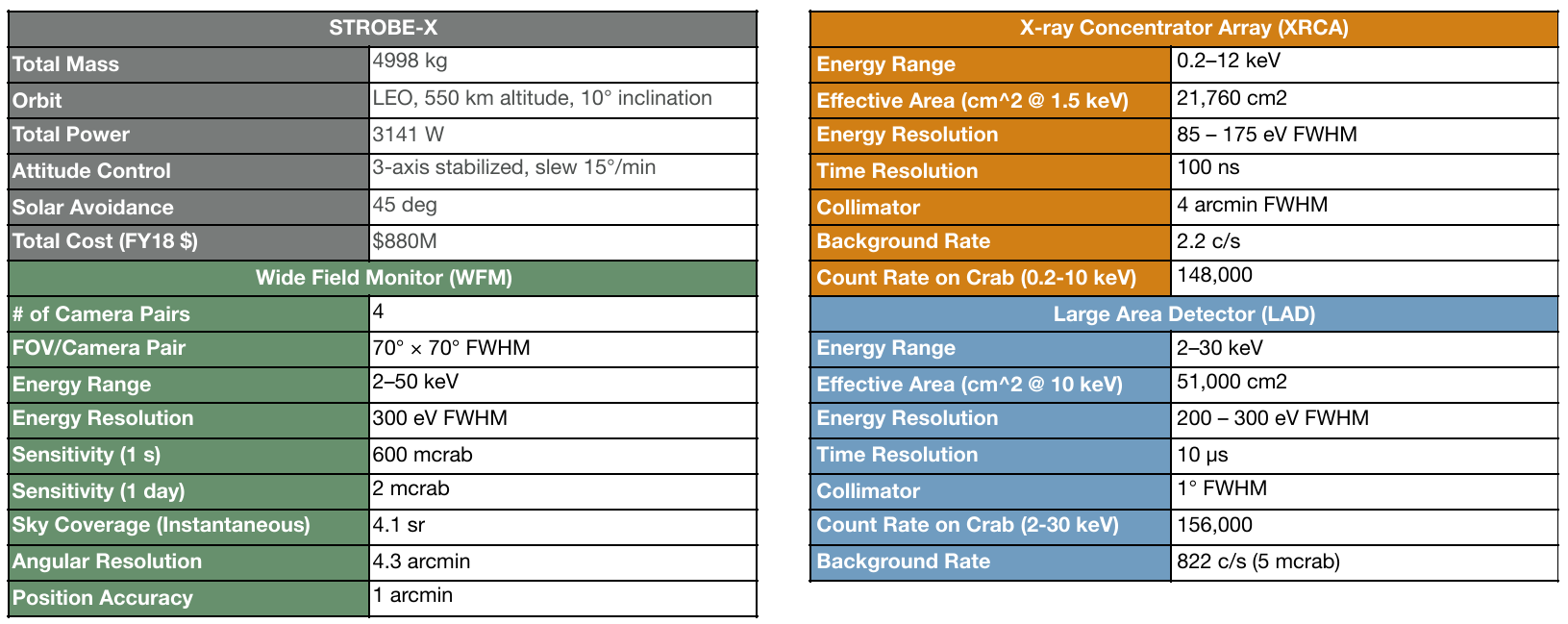}
\end{figure}

\clearpage
\strobex{}'s formidable capabilities will also enable a broad portfolio of additional science including accretion physics, stellar evolution, stellar flares, gamma-ray bursts, tidal disruption events, active galactic nuclei, clusters of galaxies, and axion searches.

\medskip
\strobex{} carries three instruments: 
\begin{itemize}
\itemsep 0in
\topsep 0in
\parskip 0in
\partopsep 0in
\item The {\bf X-ray Concentrator Array (XRCA)} covers the soft or low-energy band (0.2--12 keV) with an array of lightweight optics (3~m focal length) that concentrate incident photons onto small solid-state detectors with CCD-level (85--175 eV) energy resolution, 100~ns time resolution, and low background rates. This technology has been fully developed for \nicer{} and will be scaled up to take advantage of the longer focal length of XRCA, which provides an order-of-magnitude improvement in effective area over \nicer{} with over 2.1 m$^2$. 
\item The {\bf Large Area Detector (LAD)} covers the harder or higher-energy band (2--30 keV or beyond), with modules of Si drift detectors and micropore collimators originally developed for the European \textit{LOFT} mission concept. LAD provides an order-of-magnitude improvement in both effective area (5.1 m$^2$) and spectral resolution
(200--300 eV) over \rxte/PCA.
\item The {\bf Wide-Field Monitor (WFM)} will act as a trigger for pointed observations of X-ray transients and will also provide high duty-cycle, high time-resolution, and high spectral-resolution monitoring of the dynamic X-ray sky over the 2--50~keV band. WFM will have 15 times the sensitivity of the \textit{RXTE} All-Sky Monitor, enabling multi-wavelength and multi-messenger investigations with a large instantaneous field of view, down to a new, order-of-magnitude lower flux regime.
\end{itemize}

The \strobex{} mission does not require any new technologies to be developed. The XRCA is a small modification
of the flight-proven optics and detectors from \nicer, while the LAD and WFM are based on large-area silicon 
drift detectors already used in experiments at the Large Hadron Collider as well as microchannel plate collimators that have multiple commercial vendors available. In addition, the spacecraft relies only on high-Technology Readiness Level (TRL) components.

\medskip
During our study, we produced detailed instrument and mission designs working with the Integrated Design Center (IDC) at NASA/GSFC. We constructed master equipment lists (MELs) down to the component level for both the instruments and the mission, and we 
validated our parts acquisition and screening strategy and optimized our design to facilitate manufacture, assembly, integration and test flow. Based on these efforts, we produced a realistic development schedule assuming a Phase A start of October 1, 2023 that yields a launch date of January 1, 2031. The final result of our study is a mission cost estimate. The instrument and spacecraft costs are parametric cost estimates from PRICE-H and SEER, driven by the detailed MELs and using a common set of assumptions for a Class B mission. To that hardware cost,
we applied standard percentage multipliers for the other work breakdown structure (WBS) elements and 25\% reserves,
a \$150M fixed charge for launch services giving a total mission lifecycle cost estimate of \$880M (FY2018 dollars). This fits within the maximum probe-class budget of \$1000M with an additional 13\% margin beyond the reserves, giving us high confidence that this mission is executable as a probe.

\medskip
\strobex{} is a highly executable probe-class mission that is ready for construction in the 2020s. This mission is poised to deliver high-impact science in the 2030s that will address some of the highest priority science questions about the formation, evolution, and accretion processes of black holes, the nature of dense matter and gravity, and a wide range
of cosmic explosions.

\clearpage

\section{Introduction}
\label{sec:intro}

%\fixme{NASA Suggested content:\\
%- State of the Art in the Field\\
%- Compelling Outstanding Questions\\
%- Needed Capabilities for Progress
%}

\begin{wrapfigure}{r}{0.50\textwidth}
\centering\includegraphics[width=4.0in]{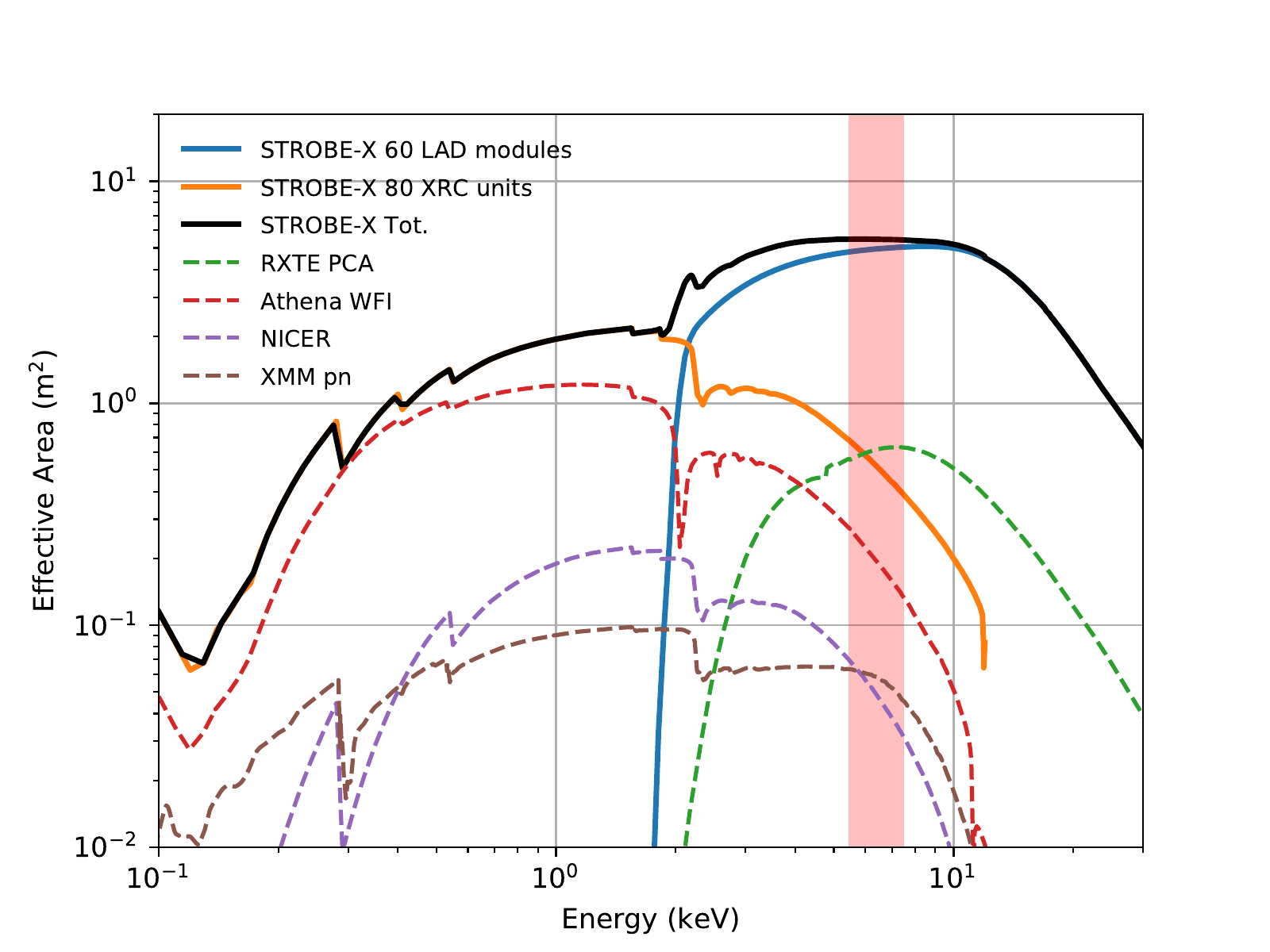}
\caption{Effective area of the \strobex{} pointed instruments (solid curves), compared to some previous and planned missions (dashed curves). \strobex{} has the largest area over its entire bandpass. The Fe-K line region near 6.4 keV is denoted by the pink band. 	\label{fig:effarea}}
\end{wrapfigure}
X-ray astronomers have made great strides in achieving high spatial resolution and extremely high spectral resolution. \chandra{}  led the way with subarcsecond imaging and grating spectroscopy, and \textit{Hitomi} demonstrated the power of microcalorimetry. Over the next decade, these achievements will be built upon by \textit{XRISM} and \athena.  These missions cover important observational phase space, but their high cost and demanding technical requirements conspire to limit their collecting area. Furthermore, their long focal lengths and lack of all-sky monitoring capability leave them unable to detect and respond to transient events and source state changes on very short time scales.
There is thus a great need for the complementary capability of high-throughput spectroscopy and timing on an agile platform with rapid re-pointing and onboard sky monitoring.

\rxte{} demonstrated the power of high-throughput X-ray timing during its highly successful 16~year mission (1995--2012). However, its antiquated detector technology (gas proportional counters) provided poor spectral resolution and was unsuitably heavy and bulky for scaling up to considerably larger collecting area. 
This has changed with the advent of solid-state X-ray detectors, particularly Si drift detectors (SDDs). These detectors make large areas achievable with vastly reduced mass and volume compared to gas detectors, and are intrinsically higher in spectral resolution, enabling a massive increase in mission capability at a modest cost.  Because of the larger number of detectors, the SDD approach also can handle much higher count rates before detector deadtime hampers sensitivity.

High-throughput, flexibly scheduled missions are crucial for understanding bright, variable sources.  For spectroscopy, this combination allows {\it many} high quality spectra to be made for strongly variable sources, so that the spectral evolution of the sources can be used to understand them.  Even more crucially, for understanding source variability, we have:
\begin{equation}
(S/N) = \frac{1}{2} I r^2 (T/\Delta f)^{1/2}
\end{equation}
where $S/N$ is the signal to noise of a variability feature, $I$ is the count rate, $r$ is the fractional rms variability amplitude, $T$ is the exposure time, and $\Delta f$ is the frequency width in the Fourier spectrum for the feature %\cite{1989ASIC..262...27V,2004AIPC..714..371V}.  
\cite{1988SSRv...46..273L}.
The exposure time needed thus scales with the count rate (and hence effective area) to the $-2$ power for variability features, unlike for source detections for faint sources.  Thus, even small changes in effective area are important, and large changes, as proposed for \strobex (see Fig.~\ref{fig:effarea}), are revolutionary.  Furthermore, for many sources in the classes like those probed by \strobex, important variability features can be short-lived transient phenomena like the highest frequency oscillations seen during the nuclear burning phases on the surfaces of neutron stars, or the highest frequency oscillations from accretion disks around black holes and neutron stars.  These can evolve or even disappear on timescales from minutes to days, so that it can often be impossible to offset lower collecting area with greater exposure times; this is fundamentally different from the areas of astrophysics aimed at detection of faint, but steady sources, where longer exposure times can compensate for less powerful facilities.

\nicer{} has demonstrated the efficacy of this approach by flying an instrument with twice the collecting area of \xmm{}/EPIC-pn at a small fraction of the cost.  Its highly-modular design is straightforward to scale up by a large factor, while its non-imaging X-ray concentrators with short focal lengths allow rapid slewing without
placing expensive demands on the spacecraft systems.

The time is right for a new mission with an order of magnitude more collecting area than \nicer{} in the soft band, and than \rxte{} in the hard band, and with a wide-field monitor with 10 times the field of view of \textit{BeppoSAX}, the last X-ray all-sky monitor with a large instantaneous field of view. Time-resolved X-ray spectra
with CCD-class spectral resolution and high time resolution can reveal detailed information
about the geometry, composition, ionization state, and velocities of accretion flows,
as well as emission from neutron star surfaces, AGN jets, stellar coronae, white dwarf accretion columns, and diffuse gas. As \rxte{}/ASM, \textit{Swift}/BAT, \textit{Fermi}/GBM, and \text{MAXI} have shown, a wide-field monitoring capability
is essential for discovery and characterization of transient sources, monitoring source state changes, and many other studies. Again, SDDs enable a major improvement in capability
with sensitive continuous monitoring of a large fraction of the sky, all with full timing
and spectral information on timescales from microseconds to years, in contrast to scanning
monitors (like \rxte/ASM) that miss a large range of timescales, and in contrast with monitors like \textit{Fermi}/GBM that cover those timescales, but with much worse localization.

The powerful ensemble of new capabilities offered by \strobex will revolutionize our understanding of many outstanding questions in astrophysics, including:
\begin{itemize}
\itemsep 0in
\itemindent 0in
\topsep 0in
\parskip 0in
\partopsep 0in
    \item What are the spin and mass distributions of accreting stellar mass black holes?
    \item How do supermassive black holes form and grow, and what fraction is obscured?
    \item What determines the masses, radii, and spins of neutron stars? Is there new physics in the super-dense interior of neutron stars?
    \item What are the properties of the precursors and electromagnetic counterparts to gravitational wave sources and neutrino sources?
    \item What powers relativistic jets and disk winds? How are accretion disks and jets coupled in AGN and stellar mass black holes?
    \item How are the most powerful stellar flares generated and what are their implications for development of life?
    \item How do metals grow in abundance in the Universe?
    \item Could axions be the primary source of dark matter?
\end{itemize}

%\fixme{Here,  maybe give a list of `needed capabilities' to answer the above questions, that \strobex{} will bring to the table. e.g. High-sensitivity X-ray timing and spectral observations of known neutron stars and black holes; high-throughput X-ray spectroscopy; simultaneous measurements of thermal and non-thermal components and their relationship; ... }

In this report,  we present the 
\textit{Spectroscopic Time-Resolving Observatory for Broadband Energy X-rays} (\strobex{}), which brings these two technologies 
together into a uniquely powerful observatory with two pointed instruments that, combined, have over an order of magnitude more effective area than {\em NICER} in the soft band and
{\em RXTE} in the hard band. This combination allows simultaneous spectral and variability measurements of thermal and non-thermal emission processes and precise
characterization of the relationship between the two, and it also enables the effects of absorption to be clearly separated from 
continuum emission. A third instrument monitors the X-ray sky over a range of timescales that will both trigger
observations by the pointed instruments and be a powerful instrument for discovering and characterizing rare transients on its own. In the following sections, we describe the science drivers, the designs of each of the three instruments, and the mission design resulting from our studies in the NASA/GSFC Integrated Design Center (IDC).

\section{Science Drivers}
\label{sec:sci}

%\fixme{NASA Suggested content for this section:\\
%- Science Goals and Objectives\\
%- Perceived Scientific Impact\\
%- Observations, Measurements, with Science Yield Estimates
%}

The \strobex{} mission includes a versatile set of instruments optimized for fast timing and broadband, time-resolved spectroscopy of compact objects.  The key science drivers are threefold: (1) measuring the spin distribution of accreting black holes, (2) understanding the equation of state of dense matter in neutron stars, and (3) observing the properties of the precursors and electromagnetic counterparts of gravitational wave sources.  A wide range of additional science is also enabled by \strobex{}'s unique combination of instruments, from studies of the inner solar system to the high redshift Universe.  Herein we outline the key science cases, with particular attention given to probing the counterparts of gravitational wave sources, as this area of study has expanded rapidly in the past two years.

%\fixme{In the following sections, should we bold some sentences, or add some slick-looking text boxes that make the most important points, to help those readers that are skimming and just need the elevator version?}

\subsection{Key goals}
\subsubsection{Black Hole Spins}
\label{subsub:spins}
%{\it\bf John Tomsick for binaries}
%{\it\bf Erin Kara for AGN}

%Status of the BH spins section as of 10/29/18 (JAT/EK).  This is a fairly complete first draft.  We removed the lag sensitivity plot with the curves and currently, this is a 2-panel figure with simulations in both panels.  If the lag sensitivity plot is in the introductory material, we may want to refer to it.  Dan Wilkins is working on simulations to show the time lag pattern that is predicted when the emission from the plunging region is included.  Chris Reynolds suggested that it might be worthwhile to check on whether STROBE-X could detect the loops that are visible in Fig. 1d of Reynolds et al. (1999, ApJ, 514, 164).

%We are assuming that the fact that STROBE-X probes regions of strong gravity is covered before this section.  The strong gravity discussion should have a curvature/potential figure (Figure 1a in the original STROBE-X proposal)

%Introductory paragraph
%\begin{mdframed}
%STROBE-X, with its broad energy coverage, large effective area and exquisite time resolution will provide the most robust measurements of black hole spin in stellar mass black holes, and will allow us to measure spins in supermassive black holes that are not accessible with current instruments.
\begin{compactitem}
\item \strobex{} will measure the spins of stellar mass black holes using three independent techniques with different systematic uncertainties.
\item \strobex{} will map the inner regions of stellar mass black holes and high-mass AGN through X-ray reverberation mapping. This technique is currently limited mostly to low-mass AGN.
\item X-ray reverberation lags will provide important constraints on the geometry of the X-ray corona, currently, one of the biggest sources of systematic uncertainty in spin measurements.
\end{compactitem}
%{\em STROBE-X} will make a groundbreaking contribution to these science questions by measuring the spins of stellar mass black holes in three independent ways with different systematics.  The huge increase in effective area will allow {\em STROBE-X} to measure X-ray lags on short enough time scales to measure the source geometry, removing a key uncertainty in the most widely used black hole spin measurement technique, which is currently the only measurement technique for supermassive black holes.  Robust spin measurements are also important for understanding the implications of the LIGO spin constraints.
%\end{mdframed}

%Currently, a major on-going effort in black hole (BH) studies is to measure their spins. 
%\begin{figure}
\begin{wrapfigure}{r}{0.50\textwidth}
\centering\includegraphics[width=3.5in]{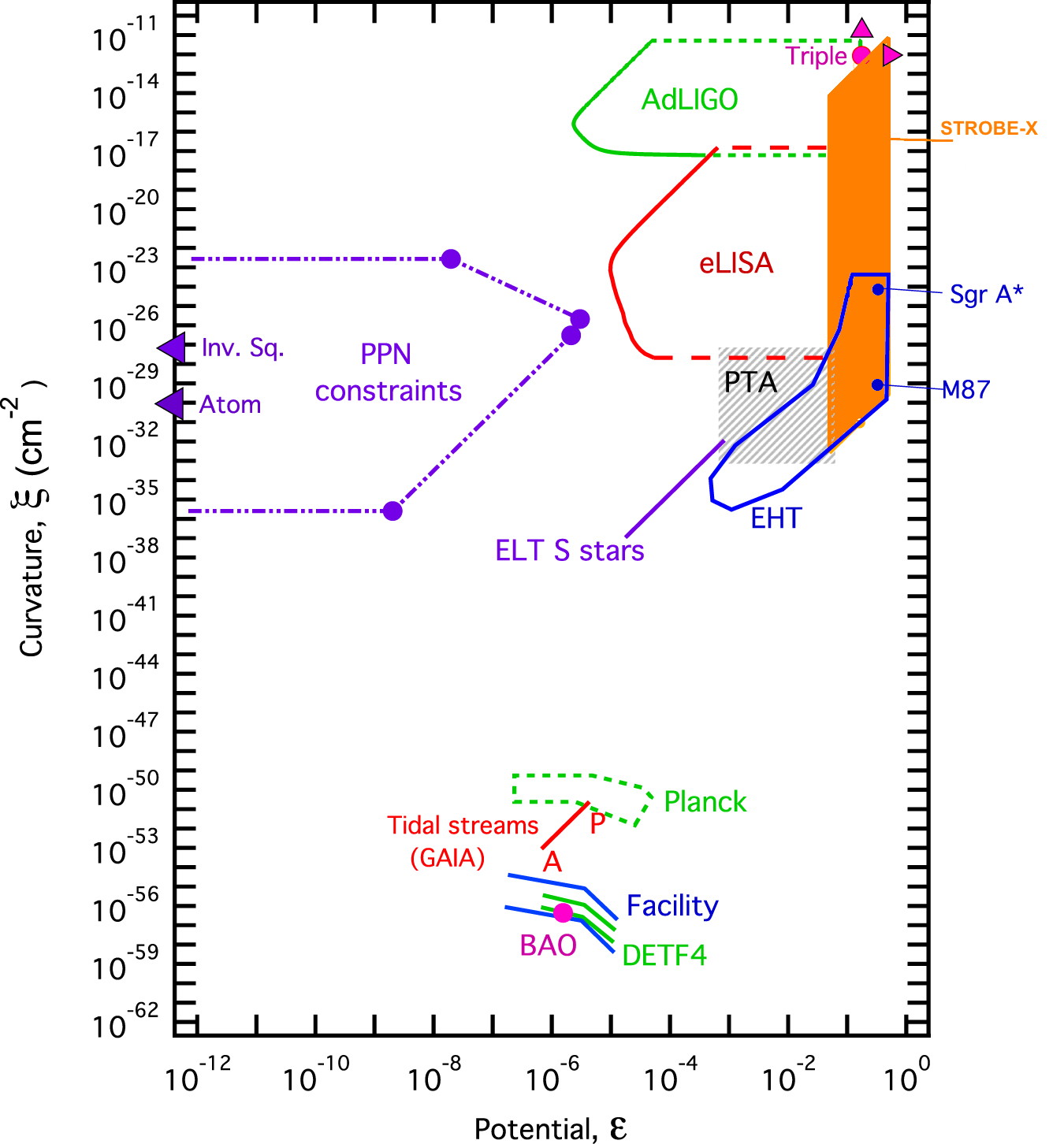}
\caption{Range of strong-gravity parameter space probed by astrophysical measurements \cite{Baker_etal_2014}. \strobex{} probes a broad region complementary to gravitational wave interferometers.\label{fig:curvature}}
%\end{figure}
\end{wrapfigure}
\paragraph{Background and Motivation:} 
Strong-field gravity results in gross deviations from Newtonian physics and qualitatively new behavior for motion near compact objects, including the existence of event horizons and an innermost stable circular orbit (ISCO). As shown in Figure~\ref{fig:curvature}, \strobex{} will probe strong gravitational fields of black holes (BHs) in a way that is complementary to gravitational wave interferometers like LIGO/Virgo \cite{Baker_etal_2014}. Accretion flows and the X-ray photons they emit are ``test particles'' that probe the stationary spacetimes of compact objects, while gravitational waves carry information about the dynamical evolution of these spacetimes. As a result, \strobex{} will enable mapping of the stationary spacetimes of BHs and test the no-hair theorem \cite{Johannsen_Psaltis_2010, Johannsen_Psaltis_2011, Johannsen_Psaltis_2013, Bambi_Barausse_2011, Bambi_2013}.

Strong-field gravity is parametrized by BH mass and spin, which are encoded in X-ray emission from the inner accretion flow. There are powerful astrophysical motivations for measuring these quantities and the flow geometry. A slowly-rotating accreting BH can only reach the maximal spin allowed in general relativity (GR) by doubling its mass \cite{Bardeen_1970}. Understanding the spin distribution of stellar-mass BHs (where such mass growth is unlikely) thus yields essential information on the BH formation process, and hence on the evolution of massive stars and the supernova mechanism \cite{Miller_etal_2011, Fragos_McClintock_2014}, while
 measurement of spins in active galactic nuclei (AGN) will probe cosmic BH spin evolution.
 
X-ray observations of the inner accretion flows around BHs are currently our most precise means of measuring a BH's spin. 
% changed to singular black hole, so LIGO people don't fuss...
%EK removed for space: A non-zero BH spin changes the space-time around the BH, requiring the Kerr rather than the Schwarzschild metric to describe its geometry.  
Spin measurements are possible for accreting BHs in X-ray binaries and in AGN because the Kerr metric predicts that the location of the inner edge of a non-truncated disk (i.e., the ISCO) depends monotonically on the spin \cite{bardeen72}.  BH spin is commonly expressed in terms of the dimensionless quantity $a_{*} = cJ/GM^{2}$ with $|a_{*}| < 1$, where $M$ and $J$ are the mass and angular momentum of the BH, and $G$ and $c$ are the gravitational constant and the speed of light, respectively.  For a BH with maximal prograde\footnote{Prograde means that accretion disk and BH angular momentum vectors are in the same direction, and retrograde means that they are opposite.} spin, $a_{*} = 1$ and the ISCO is near 1 gravitational radius, $R_{g} = GM/c^{2}$.  For a non-rotating BH ($a_{*} = 0$), the ISCO has a much larger value of 6\,$R_{g}$, and for a BH with maximal retrograde spin ($a_{*}$ = --1), the ISCO is at 9\,$R_{g}$.  Especially at lower luminosities, there is the possibility that the disk is truncated at a radius larger than the ISCO.  In this more general case, the measurement of the location of the inner edge of the disk provides a lower limit on $a_{*}$.

\begin{figure}
\vspace{-2cm}
\hspace{-0.4cm}
\includegraphics[width=6.8in]{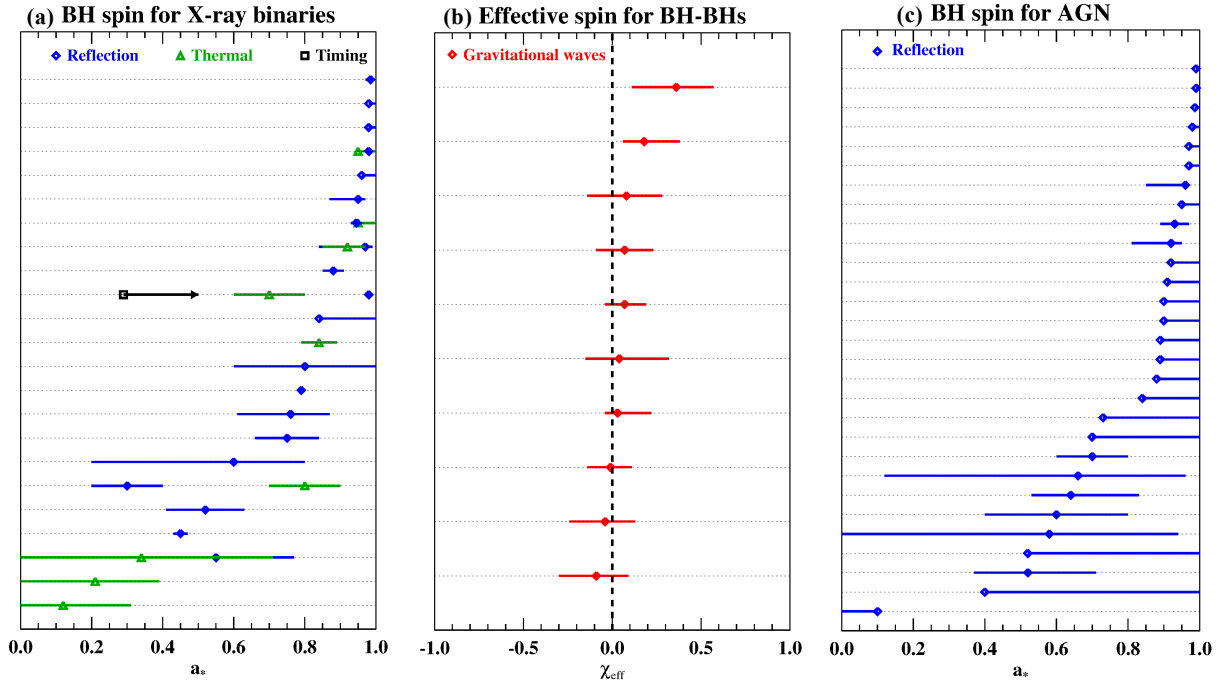}
\vspace{-2.8cm}
\caption{Current black hole spin measurements for X-ray binaries ({\em a}) and AGN ({\em c}) and effective spin ($\chi_{\rm eff}$) measurements for gravitational wave mergers ({\em b}).  For the X-ray binaries, measurements or constraints have been possible using three techniques, and \strobex{} will be capable of using all three.  It is important to note that the errors shown only include statistical uncertainties.  For AGN, only the reflection technique has been used.  Many of the X-ray binary measurements come from Ref. \citenum{middleton16}.  Other measurements were taken from the available literature.  The $\chi_{\rm eff}$ measurements are surprisingly low if the BH-BH mergers evolve from high-mass X-ray binaries. \label{fig:spin_measurements}}
\end{figure}

In addition to its effect on the space-time geometry around the BH, spin is thought to play a critical role in determining the evolution of all BHs and their host galaxies, as we describe below:
\begin{compactitem}
\item \textbf{What is the growth history of supermassive black holes?} The growth history of a supermassive BH is imprinted on its final spin, and thus, measuring a large sample of BH spins throughout the universe can constrain the dominant mechanism for growing supermassive BHs. 
%For instance, if BHs predominantly grow through prolonged prograde accretion (via an accretion disc), we expect the distribution of BH spins to be weighted more heavily to higher spin values (Ref.~\citenum{Berti_Volonteri_2008}). Whereas, if growth is driven predominantly by mergers, then there should be no preferential spin. 
It is vital to probe black hole spins in different environments in order to disentangle the importance of different spin-evolution mechanisms. 
%Currently, most BH spins measured in AGN are high (see Figure~\ref{fig:spin_measurements}c). 
However, currently, supermassive black hole spin is measured almost exclusively in local Seyfert galaxies (see Figure~\ref{fig:spin_measurements}c). 
\strobex{} will allow us to measure black hole spins in different environments, where it is currently not possible. 
\strobex{} measure spin via both spectroscopy and X-ray reverberation in AGN with masses $M_{\mathrm{BH}}>10^8M_{\odot}$, and will constrain BH spin in normally quiescent BHs through the detection of tidal disruption events (TDEs) with the Wide Field Monitor. 
Furthermore, \strobex{} spin measurements of single black holes will be highly complementary to the gravitational wave constraints from \textit{LISA}, which will measure the combined spin of SMBH mergers in a specific phase of evolution of the AGN, post-merger, when the environments are often very disturbed (e.g., [\citenum{blecha18}]).
% Because major mergers can trigger strong AGN activity, so that black hole binaries are often in disturbed environments (e.g. \citenum{blecha18}), the spins measured by LISA will be strongly biased toward the spins that result from this process.  
%LISA measurements on the combined spin of SMBH mergers will come only in one specific phase of evolution of the AGN, post-merger.
%Together, STROBE-X and LISA can probe black hole spins in different environments.
%First, LISA will have very poor sensitivity to mergers of black holes above $10^7 M_\odot$, precisely the range of black hole masses that STROBE-X will begin to probe that cannot currently be probed electromagnetically.  
Finally, in the cases where the post-merger black hole is accreting rapidly, \strobex{} and \textit{LISA} may have the opportunity to calibrate electromagnetic spin estimation techniques against the effective spin inferred from the gravitational radiation waveform.  
%Third, the LISA measurements of black hole spins will come only in one specific phase of evolution of the AGN, post-merger.  Because major mergers can trigger strong AGN activity, so that black hole binaries are often in disturbed environments (e.g. \citenum{blecha18}), the spins measured by LISA will be strongly biased toward the spins that result from this process.  The collection of a large sample of black hole spins in a different set of environments can thus be used to disentangle the importance of different spin-evolution measurements.  Similarly, in the context of 30-m telescopes, additional information about the stellar populations of the host bulges can provide still more information about the merger history of the host galaxy.
\item \textbf{How are relativistic jets powered?} 
%The question of how relativistic jets are powered and launched is relevant for both supermassive and stellar-mass BHs. 
Most theoretical models for jet production (e.g., [\citenum{bz77}]) rely on the BH spin.  While some observational studies of stellar-mass BHs have found evidence for correlations between BH spin and jet power \cite{nm12}, others find no significant relation \cite{rgf13}.  However, a large part of the disagreement is related to the BH spin determinations; thus, reliable BH spins are critical to clarify this issue. \strobex{} will make important contributions to the spin-jet connection in Galactic BHs by finding more BH transients with the WFM, and by measuring their spin using independent techniques with the XRCA and LAD (see below). %LAURA SUGGESTS LABELING SECTIONS.  JAT: I agree, but I think that the decision about subsubsubsections needs to be made for the document as a whole.  It would be good for these cases like this "see below"
In AGN, \strobex{} will measure reverberation lags in high-mass (more "quasar-like") systems for the first time (details below). %SECTION LABEL 
Our current measurements of reverberation in AGN are nearly exclusively in low-mass, radio-quiet Seyferts (e.g., \cite{Kara_etal_2016}).
\item \textbf{How do stellar mass black holes form and how do black hole binaries evolve?}  Measuring the distribution of stellar mass BH spins has become increasingly important for interpreting the LIGO BH inspiral results that constrain the effective spin, $\chi_{\rm eff}$, which is a mass-weighted combination of the aligned components of BH spins \cite{abbott16b,abbott17,lv2018}, Abbott et al., 2018, in prep.).  After the O2 LIGO-Virgo run, ten BH-BH mergers have been detected, and eight of them have values of $\chi_{\rm eff}$ that are consistent with zero, ranging from --$0.09^{+0.18}_{-0.21}$ to $0.08^{+0.20}_{-0.22}$.  The other two cases have $\chi_{\rm eff}$ values of $0.18^{+0.20}_{-0.12}$ (for GW151226) and $0.36^{+0.21}_{-0.25}$ (for GW170729), which do not necessarily indicate extremely large spin values either (see Figure~\ref{fig:spin_measurements}b).  However, as $\chi_{\rm eff}$ depends on the alignment between the BH spin axes and the orbital angular momentum axis, low values do not necessarily indicate low BH spins, but they do indicate that either the component spins are low or that there are significant spin misalignments \cite{farr17}.  Partial explanations for the low $\chi_{\rm eff}$ values include the possibility that the spin axis of the BH becomes tilted during the collapse of the progenitor star \cite{tauris17}, an evolutionary channel where a Wolf-Rayet star collapses to a BH before reaching tidal synchronization \cite{hp17a,hp17b}, or the possibility that very massive stars leave slowly rotating BHs \cite{belczynski17}.  Thus, the critical point is that if X-ray binary spin measurements (Figure~\ref{fig:spin_measurements}a) are indicating that most BHs are born with high spin (even including the high mass X-ray binary Cygnus~X--1), then it is unclear why BHs would have low spins (or strongly misaligned spins) at the times of their mergers.  While much theoretical work has already gone into trying to understand the apparent discrepancy, we must continue to increase the number of robust spin measurements for X-ray binaries.  In addition, it is important to probe BH spins in binaries with companion stars of different masses and evolutionary stages to infer whether measured BH spins are indeed intrinsic from birth or whether, in the case of long-lived systems with low-mass companions, they might be spun up by long-term accretion.
%\item How does matter and radiation behave in strongly curved space time?
\end{compactitem}

% EK: I moved the next few sentences to the boxed region above.
%{\em STROBE-X} will make a groundbreaking contribution to the effort of measuring BH spins by being uniquely capable of routinely using all three of the current techniques (described below) to determine the location of the ISCO.  The huge increase in effective area will also allow {\em STROBE-X} to measure X-ray lags on short enough time scales (see Figure~\ref{fig:zoghbi_cackett_fig}) to measure the source geometry, removing a key uncertainty in the most widely used BH spin measurement technique.

%Motivation for both supermassive and stellar-mass BHs

%Reliable BH spin measurements are needed to make progress on several scientific fronts %relevant to supermassive and stellar-mass BHs.
%AGN motivation in a couple paragraphs here...
%In the case of supermassive BHs in active galactic nuclei (AGN), the distribution of spins %encodes many epochs of growth via gas accretion and mergers with other BHs, thereby offering %an intriguing probe of galaxy formation and evolution (see, e.g., %Ref.~\citenum{Berti_Volonteri_2008}).

%Consider whether we need to work the following information in
%A slowly-rotating accreting BH can only reach the maximal spin allowed in general relativity by doubling its mass\cite{Bardeen_1970}.  Understanding the spin distribution of stellar-mass BHs (where such growth is unlikely) thus yields essential information on the BH formation process, and hence on the evolution of massive stars and the supernova mechanism\cite{Miller_etal_2011,Fragos_McClintock_2015}

%The three techniques for constraining or measuring the location of the ISCO

%...................................................................................
\begin{figure}
\includegraphics[width=0.45\textwidth,trim={0cm 4cm -1cm 0cm}]{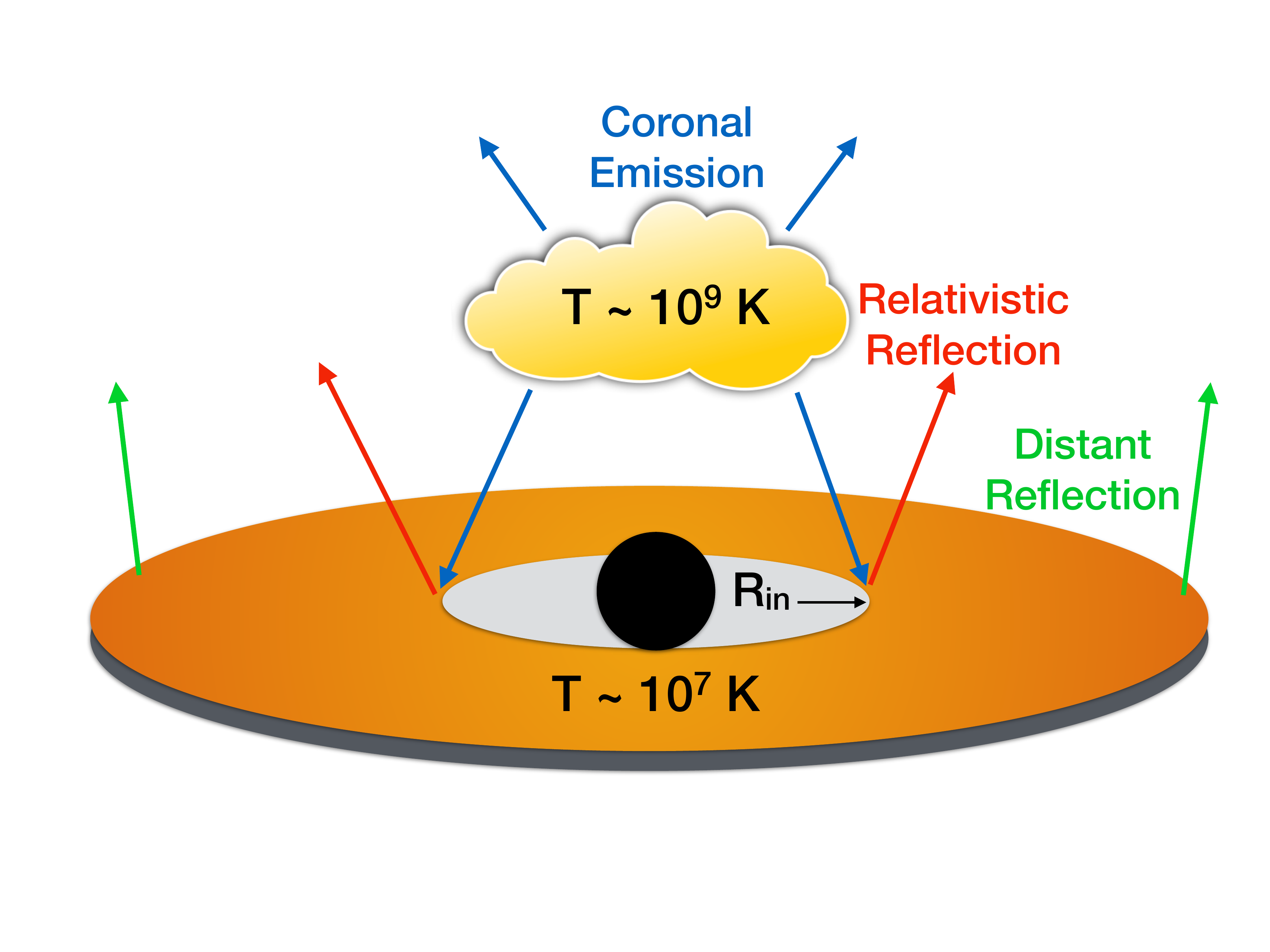}
\includegraphics[width=0.52\textwidth,trim={-1cm 1cm 0cm 1cm}]{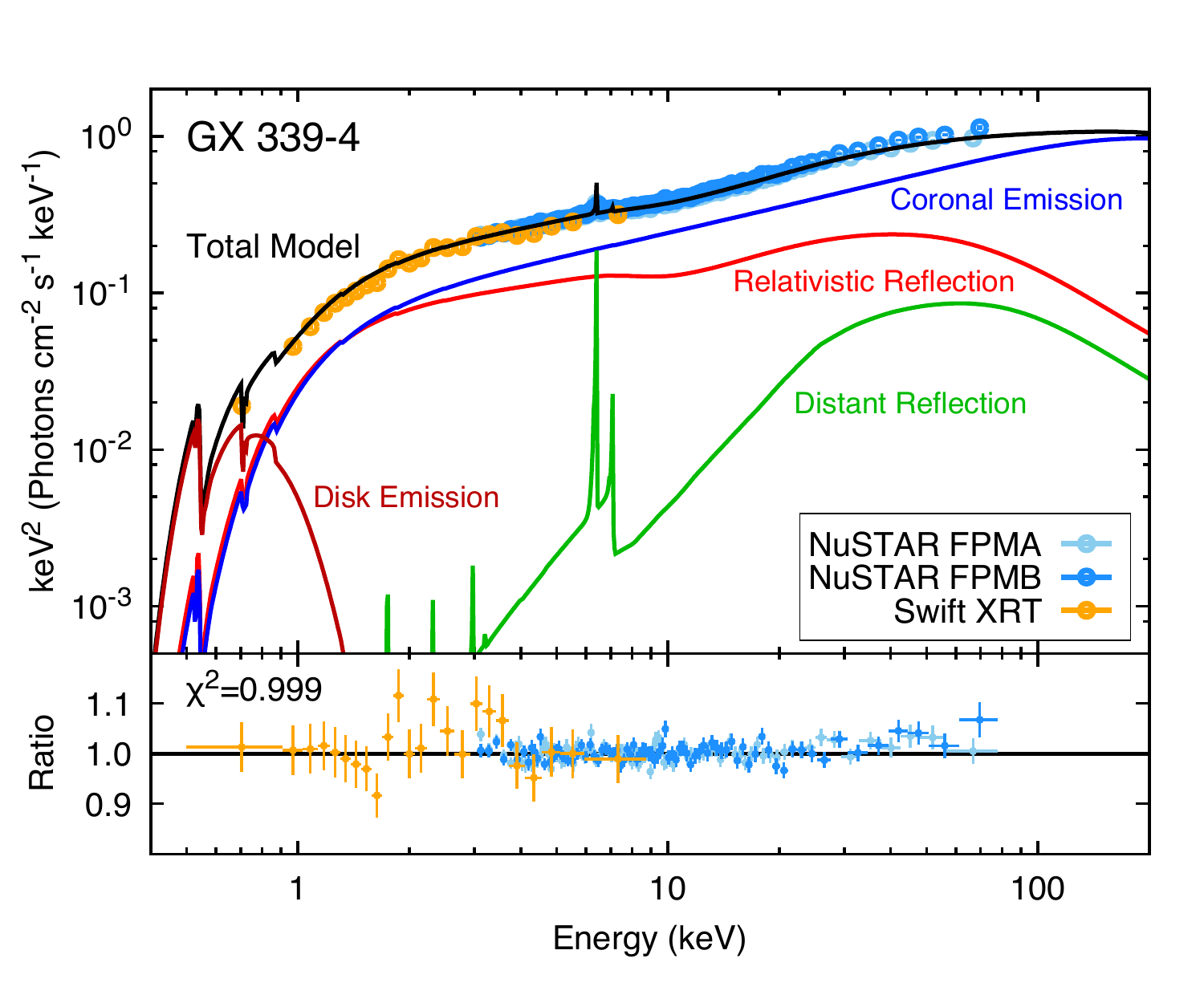}
\caption{{\it (left)} Cartoon of the X-ray source surrounding an accreting black hole.  Unless the BH is maximally rotating, there will be a gap between the inner edge of the accretion disk (orange) at $R_\mathrm{in}$ and the black hole's event horizon (black). Above the disk is a hot corona (yellow), which is the primary source of X-rays. The disk's thermal emission is Compton scattered into a power-law component (blue arrows) by electrons in the corona. About half of this component illuminates the disk thereby generating the relativistic reflection component (red arrows), as well as a distant reflection component (green). ({\it right}) The {\em Swift-XRT} and \nustar~spectrum of the BH X-ray binary GX 339-4 during its 2017 outburst, fitted with a Compton continuum (blue); relativistic reflection (red); and distant reflection (green).  The lower panel shows the fit residuals.  From Garcia et al., in prep.  }\label{fig:garcia_cartoon}
%\vspace{-25pt}
\end{figure}
%-=
%...................................................................................

\paragraph{Observational techniques and state of the field:}
There are three main ways to measure BH spin in the X-ray band:
%\begin{compactitem}

{\bf (1) Thermal Continuum Fitting} When stellar mass BHs emit strong thermal emission, the accretion disk is $\sim 10^7$~K and exhibits a multi-temperature blackbody component in the X-ray band (see schematic in Fig.~\ref{fig:garcia_cartoon}). The temperature profile of the disk is well understood, and thus provides an estimate of the inner edge of the disk, which likely extends to the ISCO in the high-soft state \cite{shafee06,mcclintock14,Schnittman2016}. While the implementation of this technique is straightforward, a challenge is that the determination of the spin depends on the distance to the source, the mass of the BH, and the inclination of the inner disk.  However, distance measurements are improving for Galactic sources with {\em Gaia} parallax measurements \cite{gandhi18}, and upcoming new or improved radio facilities (e.g., the Square Kilometer Array and the Next Generation Very Large Array) %\fixme{Does this imply a requirement on absolute effective area calibration that should be in the requirements matrix?  Is that dangerous, given that the dominant source of uncertainty is likely to be the overall flux scale, rather than what we do?}
%JAT: not sure if we should define these acronyms or remove them
will allow parallax distance measurements for many more sources.  Also, we can expect significant advances in measuring BH masses with new techniques, such as using the information from the double-peaked H$\alpha$ emission line in the optical to determine the binary mass ratio \cite{casares16}, and potentially by using sensitive near-IR spectroscopy with the {\em James Webb Space Telescope} to make measurements for sources in regions of the Galactic plane with higher extinction.

% Finally, modeling the reflection component with {\em STROBE-X} provides a means for constraining the inclination of the inner disk.
{\bf (2) Reflection and X-ray Reverberation} In both AGN and X-ray binaries, the X-ray corona can irradiate the inner accretion disk, producing broadened fluorescence lines (Refs.~\citenum{Fabian_etal_1989,miller02,Miller_2007,Kinch2016,Kinch2018}; and Fig.~\ref{fig:garcia_cartoon}). The red wing of the line is due to gravitational redshift, which is enhanced if the BH spin is high, i.e., $R_{\mathrm{ISCO}} \sim 1\,R_{g}$. While early spectral modeling could not distinguish between broadened emission and a partial-covering absorber, in recent years, measurements of short timescale reverberation time delays have confirmed that the iron~K features originate close to the central black hole \cite{Zoghbi_etal_2012,Kara_etal_2016}.
The biggest remaining systematic is the uncertainty of the geometry of the corona, which \strobex{} will probe with unprecedented sensitivity (see next section).  The reflection method has the advantage of being independent of the distance to the source and the mass of the BH, and the use of reverberation in addition to pure time-integrated spectral fitting can break the degeneracies in coronal geometry models.

\begin{figure}
    \centering
    \includegraphics[width=3.5in]{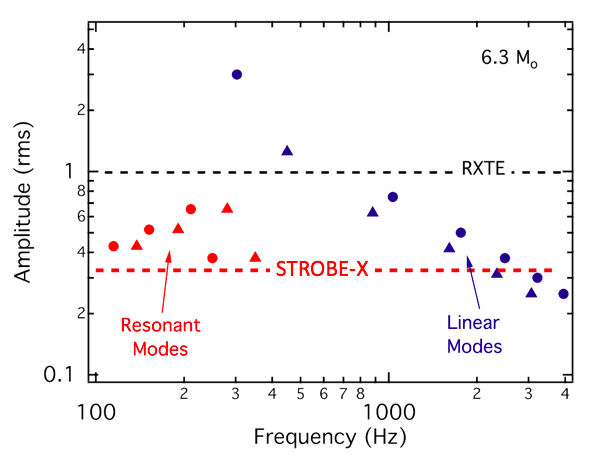}
    \caption{The two known HFQPO modes (above top dashed line) and the expected higher-order modes in the 6.3 $M_\odot$ stellar-mass BH GRO J1655$-$40, for the two model classes (linear and nonlinear/resonant modes) consistent with current observations \cite{Abramowicz_Kluzniak_2001, Wagoner_2012}. The horizontal dashed lines show the minimum detectable QPO amplitude for \rxte{} and \strobex. \strobex{} will reveal a large number of additional modes.}
    \label{fig:hfqpo}
\end{figure}

{\bf (3) High-frequency Quasi-periodic Oscillations (HFQPOs)} At frequencies of 41--450\,Hz, HFQPOs probe time scales close to or at the ISCO.  Although their origin is not fully understood, these HFQPOs have been interpreted as hot spots or disk warps near the inner edge of the disk.  In the eight systems \cite{rm06,altamirano12} where they have been seen, they appear at specific stable frequencies that are fundamental to the system (e.g., accretion disk modes or Keplerian orbital frequencies), and often with ratios of frequencies that are small integers.  We anticipate that \strobex{} will find weak signals due to modes even above 450\,Hz (see Figure~\ref{fig:hfqpo}), but it is likely that the HFQPOs strong enough for detection by {\em RXTE} were no higher than the Keplerian frequency at the ISCO.  This is the basis for the lower limit on the spin of the BH in GRO~J1655--40 \cite{strohmayer01} shown in Figure~\ref{fig:spin_measurements}a.  More model dependent approaches can already convert pairs of HFQPOs into black hole masses and spins, and identification of higher modes can break the model-dependent degeneracies both through strengths of the different modes and phase correlations between oscillations at different frequencies \cite{Schnittman2005,bispectrum}.

%As the origin or exact location of these HFQPOs is not understood, these constraints put a lower limit on the location of the ISCO.

%The maximum dynamical time scale in an accretion disk is the orbital frequency at the inner edge of the disk, and thus detections of HFQPOs, which have been seen in a handful of Galactic BH transients, provide a lower-limit on the BH spin. 
%\end{compactitem}

Figure~\ref{fig:spin_measurements} is a compilation of the BH spins currently measured with these three techniques for stellar mass BHs (left) and with the reflection technique for AGN (right). Most spin measurements are near maximal, and we need to understand if this is an observational bias, a modelling deficiency, or if this BH spin distribution is telling us something fundamental about the nature of the BH formation and growth processes \cite{vasudevan16}. For stellar mass BHs, \strobex{} will measure spin independently in these three ways in a greater number of systems and will be vastly more sensitive to HFQPOs, given that the exposure time needed to detect a QPO scales as the inverse of the square of the effective area.  These multiple spin measurements are key for identifying the important systematic uncertainties.  In AGN, \strobex{} will open up X-ray reverberation studies to a new class of AGN that are more like typical quasars, helping us understand possible selection biases.  As described in the following, for both stellar mass BHs and AGN, \strobex{} will provide very large improvements in the reflection method.

\paragraph{\strobex{} and stellar-mass black hole spin:}

%\begin{figure}
%\includegraphics[width=2.75in]{lag_sensitivity.pdf}\hfill%\includegraphics[width=3.5in]{lags_strobex_1ks.pdf}
%\caption{Sensitivity of {\em STROBE-X} for reverberation mapping.  Left: {\em STROBE-X} (solid curves) compared with other satellites (dotted curves) for an AGN black hole, assuming a 2 mCrab source, 100 ks exposure, and a $10^6 M_\odot$ black hole. {\em STROBE-X} is more sensitive over the entire bandpass. The Fe-K line region near 6.4 keV is denoted by the grey band. Right:  A simulated 1 ks {\em STROBE-X} observation of the XRB black hole GX~339-4 showing 1--10 Hz reverberation lags (XRCA shown as black circles, LAD as red triangles), with parameters following those of Ref.~\citenum{Uttley_etal_2011}.  The LAD lag precision at the Fe-K line is about 20~$\mu$s per spectral bin in a 1~ks observation, significantly better than 1~$R_g/c$ (a gravitational radius crossing time) for a 10\,$M_\odot$ black hole. \label{fig:zoghbi_cackett_fig}}
%\end{figure}

\begin{figure}
\includegraphics[width=\textwidth]{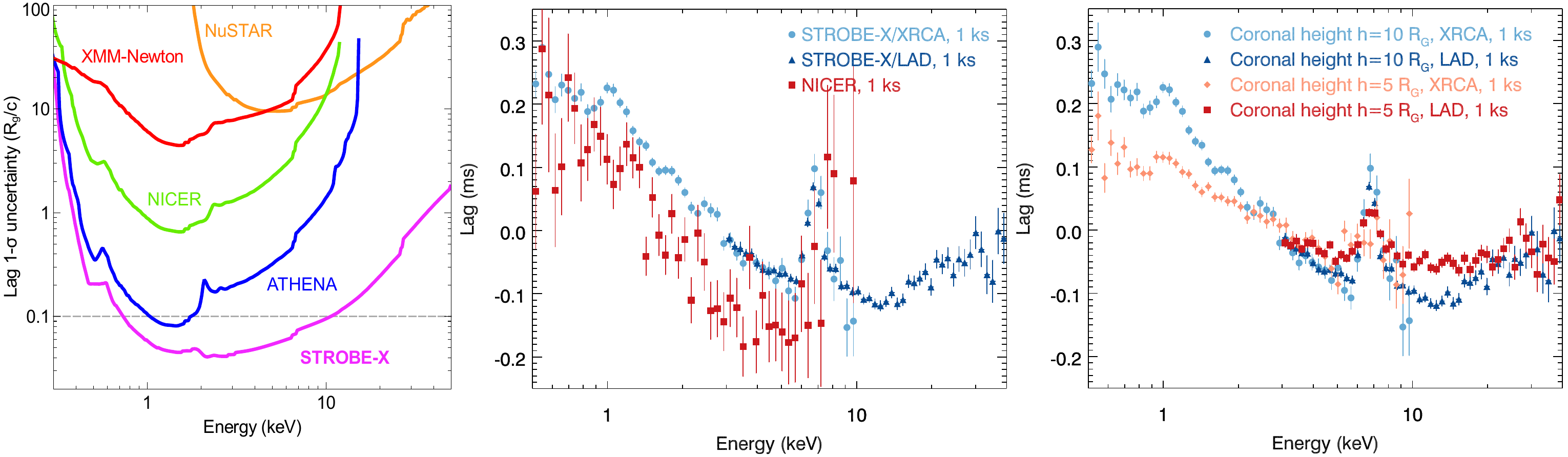}\hfill
\caption{\textbf{Advances with {\em STROBE-X} for X-ray reverberation mapping in black hole X-ray binaries}.  {\em Left:} Sensitivity curve for lags in X-ray binaries, comparing \strobex{} to other current and approved missions. {\em Middle:} The {\em STROBE-X} lag-energy spectrum (blue) compared to a simulation for \nicer{} (red). The reflected emission lags behind the continuum by $\sim0.5$~ms due to the light travel time between the corona and the disk. {\em Right:} A simulated 1 ks {\em STROBE-X} observation showing 1--8 Hz reverberation lags for a source as bright as the BH X-ray binary GX~339$-$4 \cite{Uttley_etal_2011}.  With {\em STROBE-X}, we will determine the geometry of the X-ray emitting corona, which is one of the largest sources of systematic uncertainty in measurements of black hole spin.
%The LAD lag precision at the Fe-K line is about 20~$\mu$s per spectral bin in a 1~ks observation, significantly better than 1~$R_g/c$ (a gravitational radius crossing time) for a 10\,$M_\odot$ BH. 
%The blue points are for a source height of 10~$R_g$, while the orange/red points are for a source height of 5~$R_g$.
\label{fig:lag_bhb}}
\end{figure}

BH spin measurements from X-ray binaries will be possible in seconds, rather than hours of {\em STROBE-X} data, making it possible to determine whether model parameters are stable as sources vary.  Such {\em STROBE-X} measurements would verify that the BH spin is constant, but it is physically reasonable for other parameters to change on short time scales.  For example, for a few optimal sources, spectra have been seen to change over oscillation phase for low-frequency QPOs \cite{ingram16}.

%Could consider a paragraph on evidence for misalignments between the BH spin and the binary angular momentum axis (e.g., Maccarone+jets, Tomsick+CygX1, Walton+V404Cyg, Miller-Jones+V404Cyg) and how well {\em STROBE-X} will measure inner disk inclination with reflection measurements.

With {\em STROBE-X}, X-ray reverberation mapping via the measurement of X-ray time lags can be used for both stellar mass and supermassive BHs, and even accreting neutron stars \cite{Uttley_etal_2014}, and there will be very large improvements for the stellar mass case.  While they are $\sim$1000 times brighter than more distant AGN, their characteristic variability timescales are $>10^5\times$ faster, and they require more cycles to obtain the same signal-to-noise level as the AGN. However, while X-ray binaries are more challenging to detect with our current telescopes (due to pile-up and telemetry drop-outs), the signal-to-noise of the lags scales linearly with the count rate and thus improvements with {\em STROBE-X} will be dramatic. With {\em STROBE-X}, we will measure X-ray reverberation in tens of X-ray binaries, and will map them during state transition. Furthermore, {\em STROBE-X} will measure lags associated with the relativistically broadened iron line and Compton hump, which cannot be precisely probed with current instruments. 

X-ray lags provide critical information for the reflection technique because the time delay provides a measurement of the physical distance between the locations of different emission components (e.g., the direct emission from the corona and the reflected emission from the disk).  As it is likely for the direct source to be close to the disk (either distributed above the disk or near the BH), it is necessary to be able to study lags as short as the light travel time across $10 R_{g} = 10 GM/c^{2}$, which is 0.5\,ms for a 10\Msun BH.  Figure~\ref{fig:lag_bhb} illustrates the capabilities of {\em STROBE-X} for measuring X-ray reverberation lags, showing that it will be able to achieve the necessary levels and will be sensitive to small differences in source geometry. 
Such short lag time scales have only been probed very recently, and this was only possible with the combination of the high time resolution of \nicer{} along with an extremely bright BH transient, MAXI~J1820+070 \cite{KaraNature}.  The detection of these short lags shows that the direct source is close to the disk, and strong constraints were obtained on the source geometry.  While \nicer{} was able to make this important measurement in one unusually bright system, such measurements would be routine for {\em STROBE-X}.  Figure~\ref{fig:lag_bhb} compares the {\em STROBE-X} and {\em NICER} capabilities.

HFQPOs will provide a  complementary approach to estimating BH spins.  The observed HFQPOs are often found to exist in 3:2 frequency ratios.  Under a given assumption about the mechanism for producing the QPOs, they can be used to place tight constraints on the BH spins (e.g., [\citenum{Motta_etal_2014}]).  In systems in which HFQPOs have already been detected with $\sim$1\% rms amplitudes by {\em RXTE}, deeper observations with {\em STROBE-X} will allow detection of the 5--10 additional oscillation modes predicted by theory \cite{Abramowicz_Kluzniak_2001,Wagoner_2012} (see Fig.~\ref{fig:hfqpo}), 
which can discriminate between competing models and yield masses and spins.  For the strongest oscillations, amplitude measurements across the 0.2--30\,keV band would strongly constrain the HFQPO emissivity mechanism. 

At the present time, the results from HFQPO modeling, thermal continuum fitting and the reflection fitting method are not in universal agreement with one another, so it is vital to collect new and better data to understand the systematics in the methods.  It is also important to note that measurements of supermassive BH spins have so far only been achieved with the reflection fitting method, whereas thermal continuum fitting, HFQPO modeling and reflection fitting have all yielded spin constraints in stellar-mass BHs.  Cross-calibrating these techniques on XRBs (where, additionally, mass constraints are usually much tighter) will therefore prove broadly beneficial, yielding greater confidence that the measured supermassive BH spins are accurate.

\begin{figure}
%\vspace{-2cm}
\begin{center}
\includegraphics[width=0.9\textwidth]{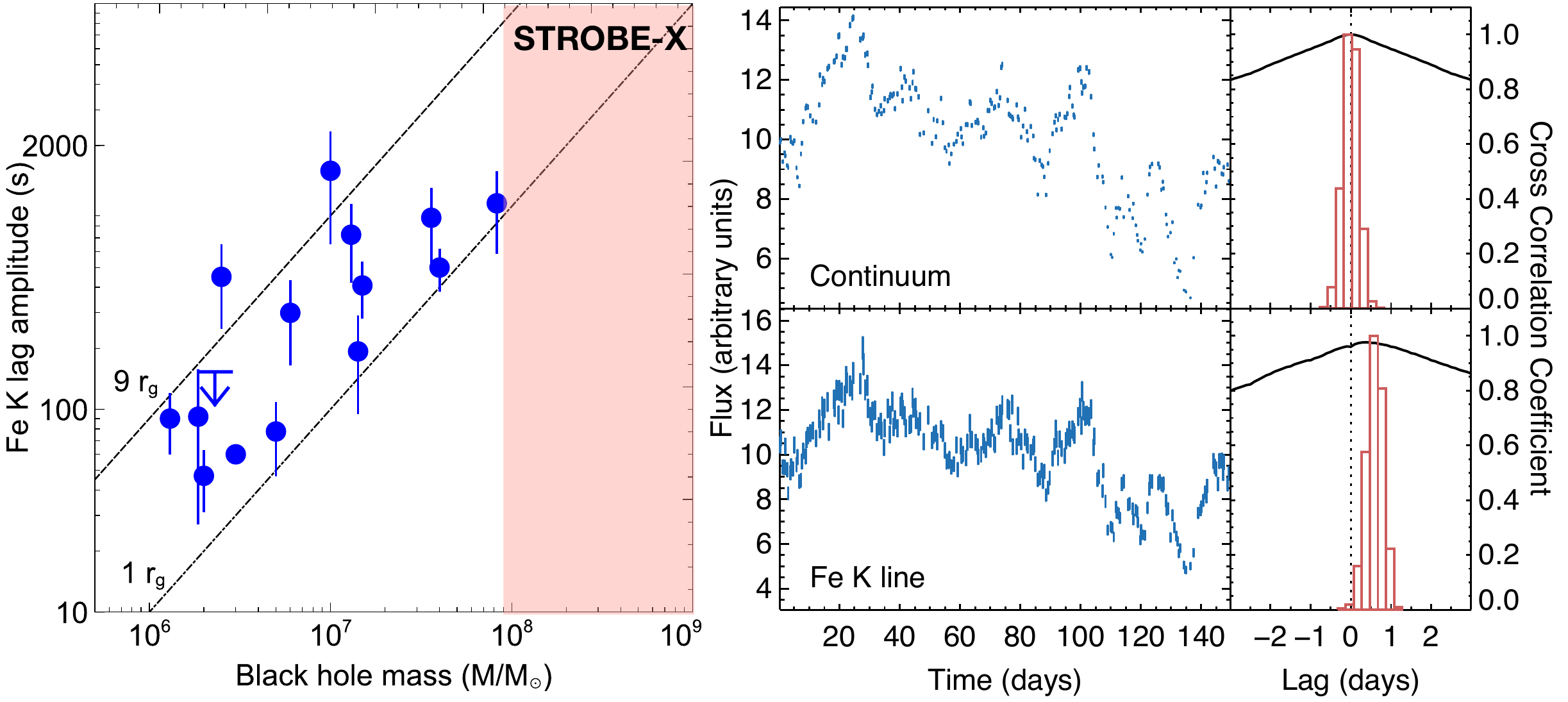}
\end{center}
\vspace{-0.5cm}
\caption{\textbf{Advances with {\em STROBE-X} for X-ray reverberation mapping in AGN.} {\em Left:} The current landscape for iron~K reverberation in AGN. We are limited to low-mass Seyferts, but with {\em STROBE-X} we can measure iron~K lags in typical mass black holes ($>10^{8} M_{\odot}$). {\em Right:} A simulated {\em STROBE-X} X-ray reverberation mapping campaign for a high mass AGN ($2\times 10^{8} M_{\odot}$). The iron~K line lags behind the continuum by 0.5~days. Understanding the inner accretion flow and BH spins in high-mass, radio-loud AGN (which are closer analogs to $z=2$ quasars) is important for using AGN spin distributions as a probe of the growth history of SMBHs.  \label{fig:highmass_reverb}}
\end{figure}

\paragraph{{\em STROBE-X} and supermassive black hole spin} ~

{\em X-ray Reverberation in more typical AGN:} 
%Spectral studies of the broad iron line produced in the inner accretion flow around AGN greatly benefit from the inclusion of the arrival time of the photons. With this information, we can put better constraints on the geometry of the corona, which is an important element in making a reliable measure of the BH spin. 
X-ray reverberation measurements in AGN have been made exclusively with {\em XMM-Newton} and \nustar{} because of their large effective areas in the Fe~K band. Unfortunately, both of these satellites have relatively severe Sun-angle pointing constraints, meaning that most sources are unobservable for large fractions of the year, limiting the range of cadence possible for monitoring campaigns. This will also be an issue for {\em Athena}. Without monitoring capabilities, X-ray reverberation will continue to be limited to low-mass, local Seyfert galaxies, which may not necessarily be perfect analogs to the typical quasar (typically $10^{8-9} M_{\odot}$). {\em STROBE-X}, with its large effective area {\em and} fast slewing/acquisition capabilities will monitor highly variable and high-mass AGN to measure the iron~K reverberation lags (Fig.~\ref{fig:highmass_reverb}-left). 

As a proof of concept, in Fig.~\ref{fig:highmass_reverb}-right, we simulate a {\em STROBE-X} XRCA and LAD monitoring campaign to measure iron~K reverberation in an AGN with $M_{BH}=2\times10^{8} M_{\odot}$ and an X-ray flux of $10^{-11}$~erg~cm$^{-2}$~s$^{-1}$. We assume a lamppost corona at 10~$R_g/c$ irradiating an accretion disk inclined at 45$^{\circ}$, with an ISCO that extends to 1~$R_g/c$ for a maximally spinning BH. The anticipated reverberation time delay between corona and disk for this configuration is $\sim 0.5$~days (where typically, in Seyfert galaxies measured with {\em XMM-Newton}, we measure time delays of tens of seconds). {\em STROBE-X} will measure this relatively long lag at $3\sigma$ confidence level through a 150~day monitoring campaign, at a sampling rate of two 2~ks visits per day, for a total 600~ks campaign. {\bf X-ray reverberation mapping of high-mass AGN is science that only {\em STROBE-X} can do because it has the ability to monitor frequently along with the sensitivity to get high fidelity continuum and line flux measurements in short exposures.} This will be important for making robust measurements of BH spin in quasars. 

{\em Iron line and Compton hump -- Synergies with Athena:}
\strobex\ will launch around the same time as ESA's {\em Athena} observatory, and together these missions will revolutionize our understanding of relativistic reflection in AGN. {\em STROBE-X} provides the broadband coverage from 0.2--50 keV that is necessary for robustly determining the continuum and has an effective area at the iron line that is 25x that of {\em Athena}. {\em Athena}, on the other hand, will provide unprecedented spectral resolution at the iron line band, which will help to constrain narrow lines from distant reprocessing. 
\strobex\ will use its unprecedented effective area and broadband coverage to measure black hole spin via spectroscopy in 20 AGN with fluxes $>10^{-11}$~erg~cm$^{-2}$~s$^{-1}$ to an accuracy of $<10\%$.

\subsubsection{Neutron Star Equation of State}
\label{subsub:eos}
%{\it\bf Anna Watts}

Densities in the cores of neutron stars can reach up to ten times that of normal nuclear matter. In addition to nucleonic matter in conditions of extreme neutron richness, neutron stars may contain stable states of strange matter, either bound in the form of hyperons \citep{Ambartsumyan60} or in the form of deconfined quarks \citep{Collins75}.  Neutron stars are unique laboratories for the study of strong and weak force physics in cold, ultradense matter (see \cite{Oertel17,Baym18} for recent reviews).

The extremes of the QCD phase diagram cannot presently be explored using first principles calculations, due to the numerical challenges involved;  we rely instead on phenomenological models of particle interactions and phase transitions. These are tested by experiment and observation. Heavy ion collision experiments explore the high temperature and lower density parts of the phase diagram while neutron stars provide exclusive access to the low temperature high density region.    

\begin{figure}[tb]
\centering
\includegraphics[width=5in]{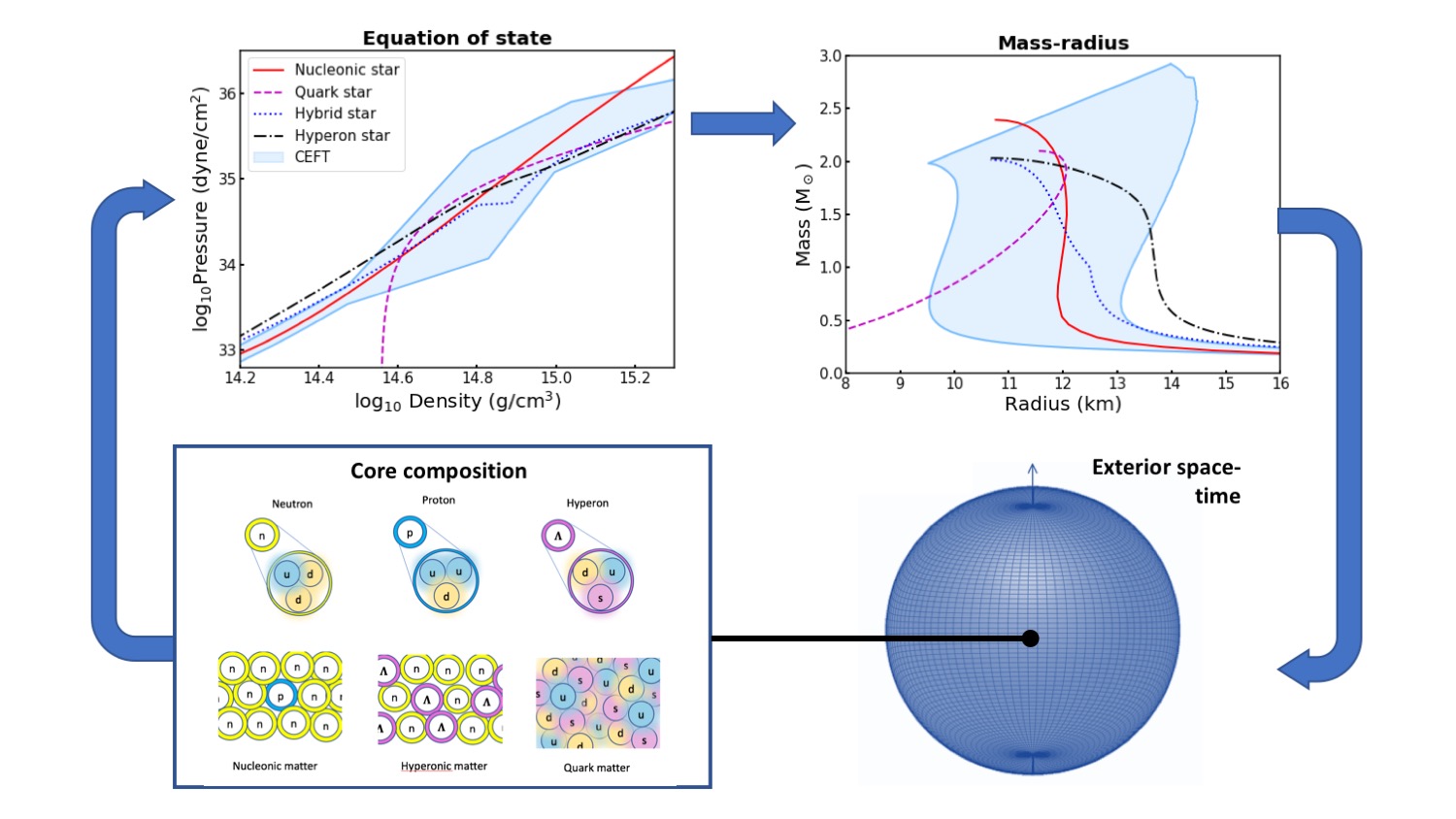}
\caption{The nature of the matter in the high density, low temperature environment of the neutron star core is highly uncertain: both the composition of the matter and the interactions between the particles are poorly constrained by theory.  Our uncertainty about these microphysical aspects (lower left:  uds = up down strange quarks) is encoded in the EOS (top left).  Here we show several currently viable EOS models: Red - nucleonic star \cite{Akmal97}; Magenta - quark star, composed entirely of quark matter \cite{Li16}; Blue - hybrid star with nucleonic outer core and quark matter inner core \cite{Zdunik13}; Black - hyperon star with nucleonic outer core and hyperonic matter inner core \cite{Bednarek_etal_2012}. The pale blue band shows the rough range covered by the full set of currently viable models, as computed from Chiral Effective Field Theory \citep[CEFT,][]{Hebeler_etal_2013}.  The different EOS then affect macroscopic properties of the star such as mass, radius and oblateness for a given rotation rate, via their influence on stellar structure (top right). These determine the exterior space-time properties of the star, exerting measurable influences on radiation propagating from the stellar surface that we exploit to infer information about the EOS and the associated microphysics.}	\label{fig:eosphysics}
\end{figure}

Our uncertainties about the microphysics of particle interactions in the conditions that prevail inside neutron stars are codified in the Equation of State (EOS), the relation between pressure and (energy) density.  The EOS forms part of the relativistic stellar structure equations that enable us %(given a central density and a spin rate) 
to compute model neutron stars.  The EOS parameters are mapped by these equations to parameters such as mass and radius that determine the exterior space-time of the star.   Using observational quantities that are affected by these properties of the space-time, we can in principle follow this chain in reverse to infer the properties of the EOS - and hence the microphysics \citep{Lattimer16}.  These connections are summarized in Figure \ref{fig:eosphysics}.   

\begin{figure}[tb]
\centering
\includegraphics[width=5in]{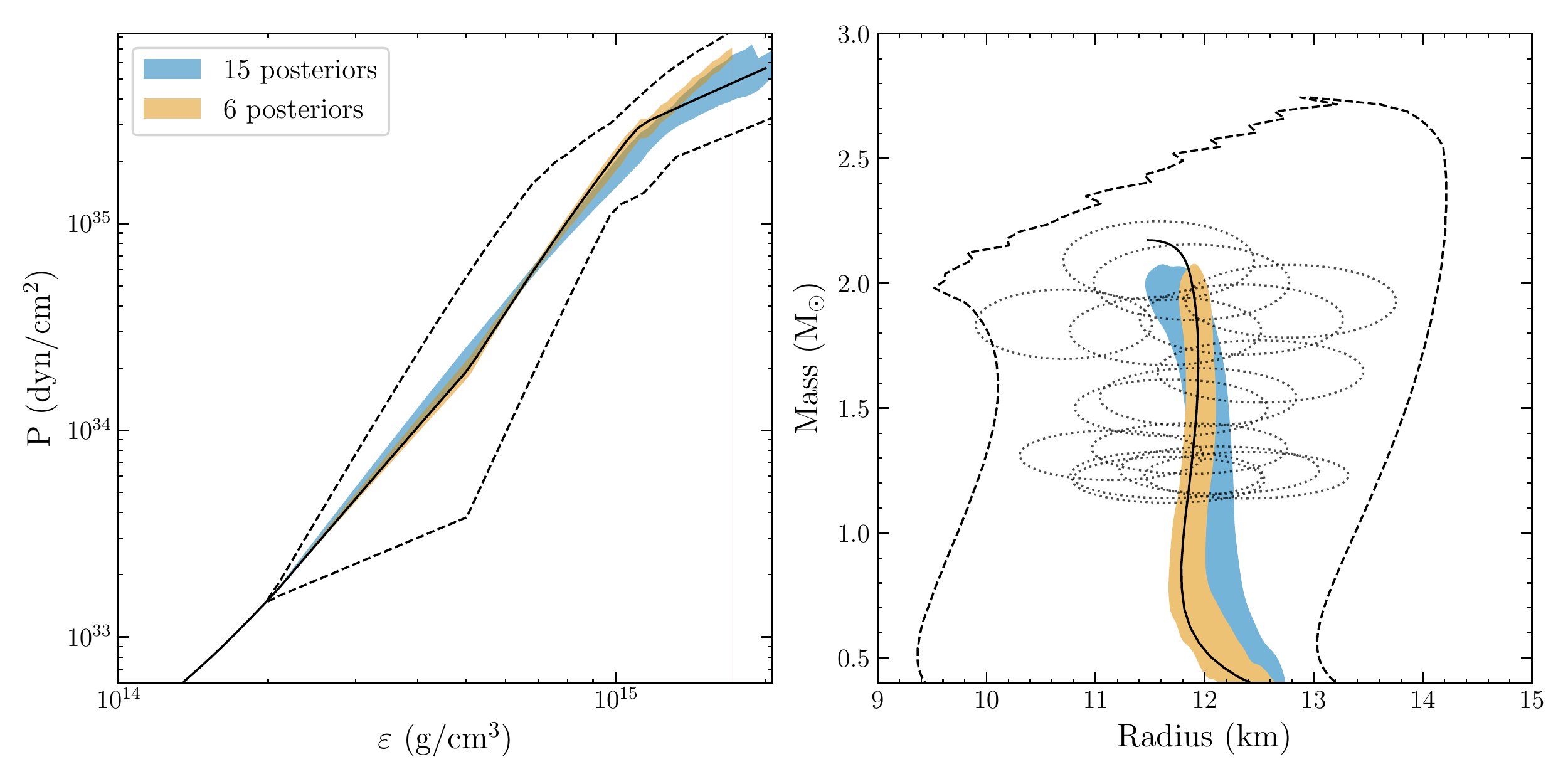}
\caption{Constraints on the Equation of State and associated Mass-Radius relation representative of those that we might expect from \strobex{} measurements.  The waveform modelling technique delivers mass-radius posteriors: here we use simple simulated posteriors (modelled as bivariate Gaussians - real posteriors will be more complex and degenerate in mass-radius, but incorporating this is not expected to change the outcome significantly, see for example \cite{Raaijmakers18}). We assume an initial set of 15 posteriors at $\pm 5$\% accuracy, shown as dashed ellipses on the right-hand panel scattered around an underlying Mass-Radius relation that is shown in black. The associated EOS is shown on the left hand panel, with the dashed lines indicating the range permitted by current models.  We assume a piecewise-polytropic parameterization for the EOS with fixed transition densities.  The 1 $\sigma$ constraints resulting from inference using these 15 measurements is shown by the blue band: we follow the procedure outlined in \cite{Greif18}. The orange band shows the results if we select 6 of these stars, based on ensuring an even spread over the mass range $1.2 - 2.0$ M$_{\odot}$, for deeper observations that result in $\pm 2$\% accuracy posteriors (not shown).}	\label{fig:eosconstraints}
\end{figure}

{\em STROBE-X} will deploy several techniques to study the equation of state (EOS) of ultradense matter in neutron stars \citep[see][for a review]{Watts16}.  The pulse profile modelling that can be done for a few pulsars with \nicer{} (sufficient to provide a proof of concept, \cite{Bogdanov13}) will be feasible for around 20 pulsars with {\em STROBE-X} (see Figure~\ref{fig:eosconstraints}).   By sampling more neutron stars, we can study the EOS across a wider range of central densities, mapping the EOS more fully and probing with finer resolution any potential phase transitions. 

%\begin{figure}
%\centering
%\includegraphics[width=3.5in]{EOS_marco.png}
%\caption{NEW FIGURE BEING PREPARED TO REPLACE THIS ONE, showing simulated dense matter EOS and neutron star mass-radius constraints obtained with {\em STROBE-X}.  Considering two panels - one showing 15-20 measurements at $\pm 5$\%, ostensibly acquired from different source classes with the goal of checking systematics - and a second showing tighter constraints acquired using 5-8 NS with much tighter constraints, 2-3\%, of the type we might make with a deeper pass of the most promising sources. Show both the EOS posteriors (the P-$\rho$ band, compared to the full range of viable possibilities) and the associated inferred mass-radius band.}	\label{fig:eosconstraints}
%\end{figure}

\begin{figure}[tb]
\centering
\includegraphics[width=5in]{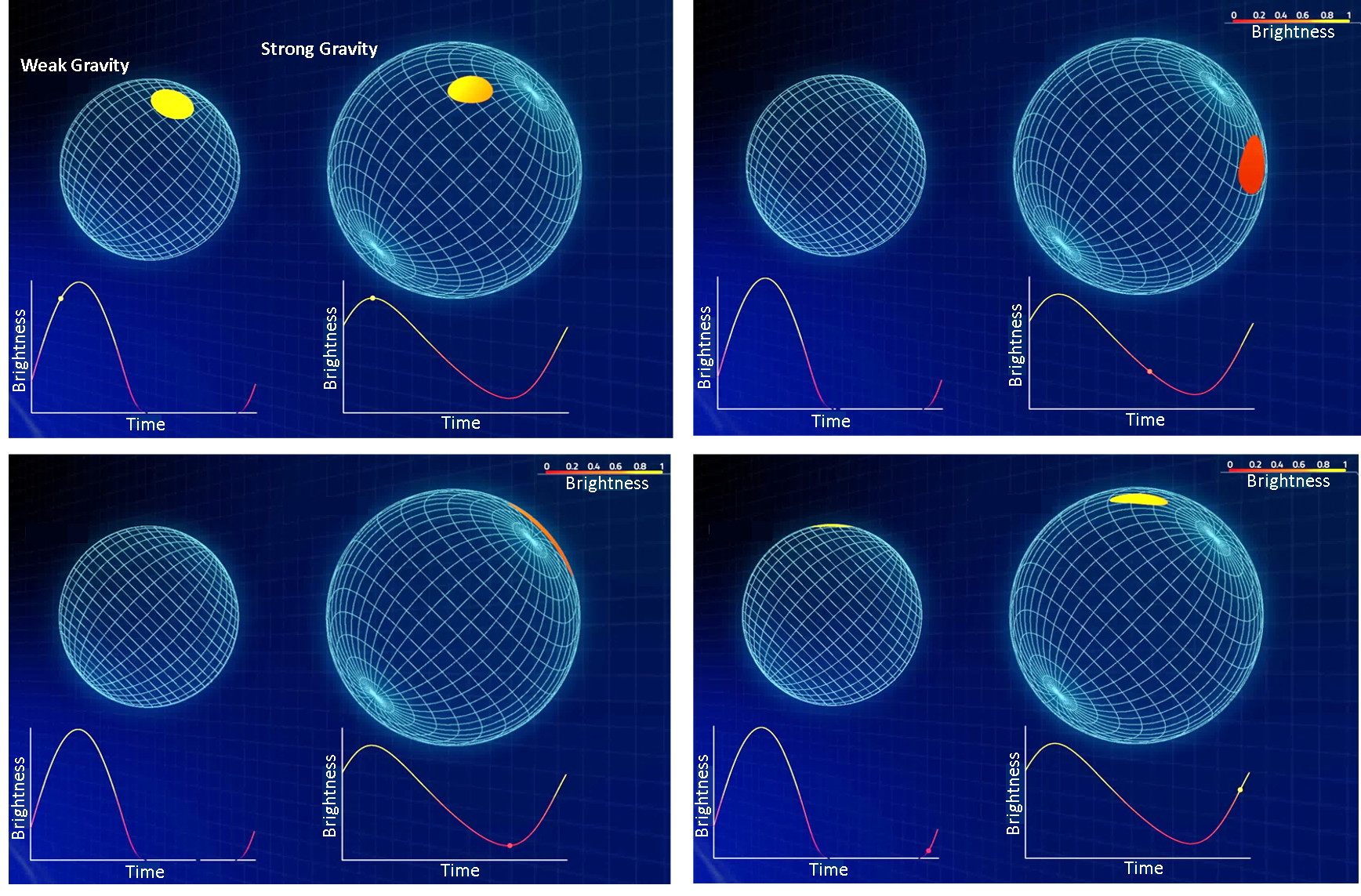}
\caption{A schematic illustration of a rapidly rotating $R=12$ km, $M=1.4$ M$_{\odot}$ neutron star with an X-ray emitting ``hot spot'' of angular radius 0.5 radians at a colatitude $\pi/4$. The four panels show snapshots at four different rotational phases.  In Newtonian (Weak) Gravity, the hotspot is occulted.  In the Strong Gravity case, where relativistic effects are included, the bending of photon trajectories due to the immense gravitational field causes most of the star to be visible to an observer at any given time. This is why the Strong Gravity star appears larger \cite{Nollert89}, and ensures that for this particular configuration the hot spot is visible at all times. Note that Doppler boosting and aberration due to rapid stellar rotation cause the peak intensity to occur before the closest approach of the spot to the observer. Since the severity of the relativistic light bending and Doppler effects depend on the neutron star mass and radius, these effects encode information about M and R in the observed energy-resolved pulse profile, which can be extracted via careful modeling. Image credit: Morsink/Moir/Arzoumanian/NASA}	\label{fig:raytracing}
\end{figure}

The pulse profile modelling technique (also known as waveform or lightcurve modelling) exploits the effects of General and Special Relativity on rotationally-modulated emission from surface hot spots (Figure \ref{fig:raytracing}).  A body of work extending over the last few decades has established how to model the relevant aspects - which include gravitational light-bending, Doppler boosting, aberration, time delays and  the effects of rotationally-induced stellar oblateness - with a very high degree of accuracy \citep{Pechenick83,Miller98,Poutanen03,Poutanen06,Cadeau07,Morsink07,Baubock13,Algendy14,Psaltis14,Nattila18}.  Given a model for the surface emission (surface temperature pattern, atmospheric beaming function, geometry) we can therefore predict the observed waveform.   By coupling relativistic ray-tracing models incorporating these effects to high-performance inference routines we are able to infer either exterior space-time parameters (like mass and radius) or EOS parameters directly from pulse profile data \citep{Lo13,MillerLamb15,Riley18,Raaijmakers18}.  High performance computing facilities are required to render such inference calculations feasible on reasonable ($\sim$1 week) timescales since the individual likelihood calculations are intensive and the models have many parameters, but work for {\em NICER} has spurred the development of highly efficient codes tailored specifically for this application.

%Successful application of the pulse profile modelling technique requires high quality energy and phase-resolved waveforms; in effect, a minimum number of photons.  
Application of the pulse profile modelling requires a minimum of about $10^6$ pulsed photons to produce sufficiently high-quality energy and phase-resolved waveforms to yield the level of precision necessary to distinguish EOS models.
%The precise number needed to deliver constraints at the level of precision necessary to distinguish EOS models depends on the precise geometry of a given source, but is roughly of the order of a million pulsed photons \cite{Lo13,Psaltis14}.   
The large collecting area of \strobex{} enables us to collect the requisite photons in reasonable observing times.   \strobex{} will target three different types of neutron stars with surface hotspots:  rotationally-powered pulsars (RPPs), accretion-powered pulsars (APPs), and thermonuclear burst oscillation (TBO) sources.   The RPPs are the sources targeted by \nicer; the hard-band  performance of  \strobex{}  opens up the other two classes for study.

RPPs have the advantage of an extremely stable spin and pulse profile, which makes it straightforward to build up high quality data sets.  The biggest source of uncertainty when doing inference is associated with the emission model (surface pattern and beaming): \nicer{} is addressing this by considering a range of models informed by pulsar theory, and performing formal Bayesian model comparison. \nicer{} is expected to deliver constraints at the 5--10\% percent level for four relatively bright sources (PSR J0437$-$4715, PSR J0030+0451, PSR J1231$-$1411 and PSR J2124$-$3358).  Of these, only PSR J0437$-$4715 has an independently-known mass; a tighter prior on the mass results in a tighter posterior on the radius.   \strobex, being $\approx$10 times larger in effective area in the waveband of most interest (0.2-2 keV), can expect to make (2-3 times) tighter measurements for these sources and can also target fainter sources, including several for which the mass is known very precisely and independently from radio timing: PSR J1614$-$2230, PSR J2222$-$0137, PSR J0751$+$1807 and PSR J1909$-$3744. Since the \strobex{} background will be lower than that of \nicer{} we can expect to deliver constraints at the few percent level on radius after observations of duration $\sim$ 1 Ms for these pulsars.  Due to the exceptional rotational stability of these MSPs and the superb absolute timing capabilities of  \strobex, these exposures can be accumulated over years without any adverse effect on the desired measurements.
% Are we concerned about reponse matrix changes over that timespan?  I think it doesn't really matter, but it might raise a red flag to the wrong person.

%\fixme{Actually STROBE-X will have considerably LOWER background than NICER, for three reasons: (1) there is more collecting area per unit detector area, (2) the orbit spends much more time in benign background regions (away from the polar horns), (3) and \textit{possibly} the XRCA will have a smaller FOV than NICER, lowering the X-ray diffuse background contribution} \fixme{Response from Anna - modified from 1-2Ms assuming NICER background to $\sim 1$ assuming better, Slavko has confirmed that this is OK.}

%  \fixme{So, for the nominal science program, how much time do we need to dedicate to this? 15 sources at 1 Ms each is a lot!}. \fixme{Answer from Anna - for the RPPs you need 1Ms per piece (8 sources). For the APPs and TBO sources a bit less (a few hundred ks) per source (for APPs this allows sampling several times as the accretion flow evolves; for TBOs you need this just to build up a data set) so to get to 20 sources total you are at about 6 Ms. So 15 Ms total is about right. You could downgrade, only sample an APP once (100 ks) if you wanted? Or aim for 15 sources total? But then less reliable and less chance of disposing of unfavourable geometry sources. }

In APPs, where accreting material is channeled towards the magnetic poles of the star, the pulsed emission has two main components: one from hotspots at the polar caps where the accreting material impacts the star, and one from the accretion shock in the accretion funnel \citep[see for example][]{Poutanen06}.  A third pulsed component may arise due to reflection from the accretion disk \citep{Wilkinson11}.  These additional parameters must be incorporated into the surface emission model; thereafter, the ray-tracing and inference proceed as described above \citep[see for example][]{Salmi18}.  Scaling from the results of this study, we anticipate that \strobex{} should be able to deliver constraints at the $\pm 5$ \% level on mass and radius after observations of $\sim$ 100 ks in duration.   Since the known AMXPs are transient accretors, with pulse profile variations as the accretion flow changes (on timescales of order days) we have the option of using several independent observations at different parts of the outburst to verify that the inferred mass/radius or EOS parameters remain unchanged as the accretion flow and hotspot geometry vary.  This is one of the reasons why the large effective area is so important: to keep observations short enough to ensure minimal variation during an individual snapshot.  There are sixteen known AMXPs, several of which have regular outbursts and which would hence be good targets:  SAX J1808.4$-$3658, XTE J1751$-$305 and IGR J00291+5934.  New AMXP discoveries during the mission lifetime may also be anticipated, given the very long recurrence times for some of the known AMXPs (meaning that many have not had an outburst while a wide-field monitor was on) and the improved sensitivity of the WFM relative to past wide field instruments.

Thermonuclear burst oscillations (TBOs) are hot spots that form \citep[via a mechanism that remains unclear, for reviews see][]{Galloway08,Watts12} during thermonuclear bursts (Type I X-ray bursts) on accreting neutron stars.  There are 19 known TBO sources.  They form a particularly interesting class for pulse profile modelling sources, since they have a well understood beaming function due to their thermal origin \citep[see for example][]{Suleimanov11}.  New challenges arise because the bursts themselves are relatively short, meaning that data from several bursts must be combined, and because the signal frequency is not always stable, meaning that we have to account for motion of the hotspot in the emission model.  These issues are however in principle surmountable \citep[see the discussion in][]{Lo13}.  The observing time necessary to accumulate sufficient TBO photons for few percent contraints on mass and radius can be estimated  from the burst properties observed by \rxte{} over its lifetime (burst recurrence times, burst fluxes, the percentage of bursts that show TBOs, and typical TBO amplitudes).  Estimated observation times are of a few hundred ks.  

The ability to target multiple source classes is very important, since it allows us to cross-check techniques, compare the source populations, and hence identify and combat model systematics.  A number of neutron stars have both accretion-powered pulsations and TBOs, enabling us to perform pulse profile modelling for the same source using two different types of hotspot.  For the bursters we also have the option of applying burst spectral modelling techniques \citep[see for example][]{Nattila17}.   This is a major advance over \nicer, which targets a single source class (RPPs) with a single type of pulsation.   Given the known source populations, making $\sim 20$ measurements is achievable in a straightforward manner.  This would place  unprecedented constraints on the properties of dense matter (Figure \ref{fig:eosconstraints}).

The large area of \strobex{} will also deliver unprecedented sensitivity for measuring neutron star spin via the detection of accretion-powered pulsations and burst oscillations.  EOS models are associated with a certain maximum spin rate (break-up, see \cite{Haensel09}), and the discovery of a neutron star with a spin rate of $\sim$ 1 kHz or above would provide a very clean constraint on the EOS (the only model dependency being the assumption that GR is correct).  There are also prospects for finding more rapidly spinning NSs in future radio surveys \citep{Watts_etal_2015}, however since the standard formation route for the millisecond radio pulsars is via accretion-induced spin-up \citep{Alpar82,Radhakrishnan82,Bhattacharya91}, it is clear that we should look in the X-ray as well as the radio. And interestingly, the drop-off in spin distribution at high spin rates seen in the millisecond radio pulsar sample is not seen in the current (albeit small) sample of accreting NSs whose spin has been measured \citep{Watts16}. \strobex{} would target transient (short-lived) and weak pulsations.   Although weak pulsations have proved elusive in current data \citep{Dib05,Messenger15,Patruno18}, transient accretion-powered pulsations and burst oscillations have been found all the way down to the sensitivity limit of current instruments, implying that detection prospects for a larger-area X-ray timing instrument are good \citep{Altamirano_etal_2008,Casella_etal_2008,Ootes17,Bilous18}.

\paragraph{X-ray constraints in the context of the gravitational wave era:}
The gravitational wave telescopes Advanced LIGO and Advanced VIRGO have now made the first direct detection of a binary NS merger \citep{Abbott17_BNS}. Gravitational waves from the late inspirals of binary NSs are sensitive to the EOS, with departures from the point particle waveform due to tidal deformation encoding information about the EOS \citep{Read09}. The statistical constraints from the first detection are comparable to and in agreement with those obtained from X-ray spectral fitting: some calculations take into account only the tidal deformability, whilst others incorporate inferences about the NS maximum mass from the accompanying kilonova \citep{Margalit17,Shibata17,Bauswein17,Coughlin18,Radice18,Rezzolla18,Abbott18_BNS,Annala18,Most18,Tews18,Lim18}. GW170817 appears to have been an unusual event in terms of its counterpart, so it remains to be seen how the GW situation evolves and how rapidly additional detections accumulate: estimates prior to GW170817 indicated that mapping the EOS at the desired (few percent) level would likely require a few tens of detections \citep{DelPozzo13,Agathos15,Lackey15,Chatziioannou15}. There are also modelling dependencies due to approximations made or higher-order terms neglected in the templates that may introduce systematic errors of comparable size \citep{Favata14,Lackey15}, and the GW results may still be in the regime where priors dominate inference of EOS parameters \citep{Greif18}. The coalescence can also excite post-ringdown oscillations in the supermassive NS remnant that may exist very briefly before collapse to a black hole. These oscillations are sensitive to the finite temperature EOS \citep{Bauswein12,Bauswein14,Takami14}, but detection will be difficult because there are no complete waveform models for the pre- and post-merger signal \citep{Clark16}.  The eventual detection of NS-black hole binary mergers may also yield EOS constraints \citep{Lackey14}.

\subsubsection{Multi-Messenger Astrophysics: Gravitational-Wave Source Progenitors and Counterparts}
% Adam Goldstein responsible
 \strobex{} will provide a vital electromagnetic (EM) complement to all classes of gravitational wave studies.  In the high-frequency (LIGO/VIRGO third generation) bands, the WFM on \strobex{} would instantly provide positions accurate enough for ground-based spectroscopy with integral field units (IFUs) for the subset of sources which are either beamed toward Earth or within about 40 Mpc, or for single-field imaging followed by spectroscopy on telescopes without IFUs.  The ability to do nearly-immediate spectroscopy, without the wide-field imaging searches, will open up new parameter space both through brighter, higher S/N spectra, and earlier phenomenology, and this capability will apply to all transient phenomena discovered by the STROBE-X WFM.
 
\begin{figure}
\centering
\includegraphics[width=3.5in]{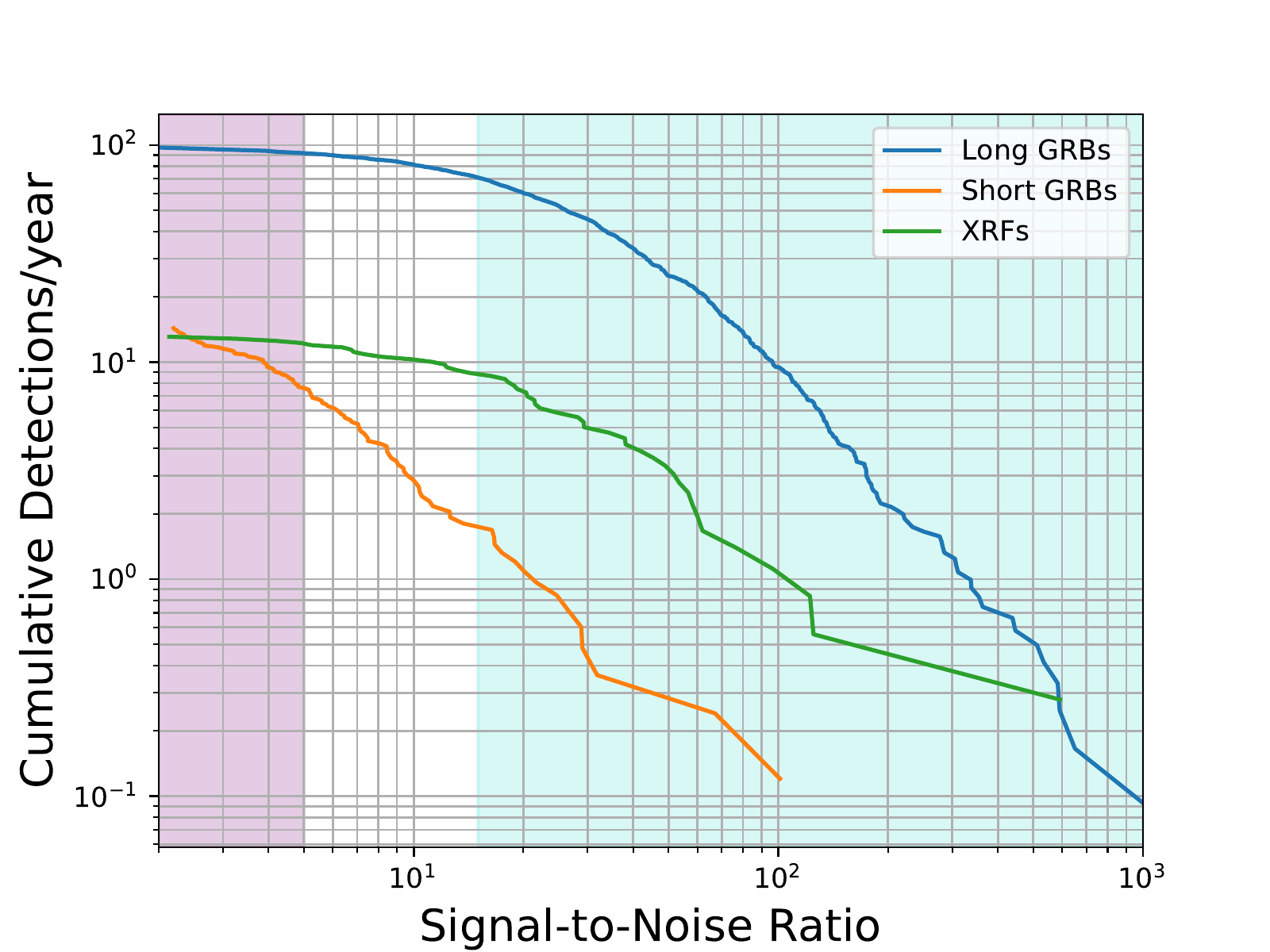}
\caption{The expected cumulative onboard detection rate of canonical gamma-ray bursts (GRBs) and X-ray flashes (XRFs) by the {\em STROBE-X}/WFM as estimated by folding observed GRB and XRF spectra through the WFM responses and accounting for the effective field-of-view of the WFM. It will detect $\sim 100$ long duration GRBs, $\sim 7$ short duration GRBs, and $\sim 12$ XRFs onboard per year. The onboard detection rate of long GRBs exceeds that of the {\em Swift}/BAT, while the short GRB detection rate is comparable.  A unique capability is the downlink of event data to the ground for the WFM, enabling sub-threshold searches to double the number of short GRB detections (purple shading).  The detection rate of XRFs exceeds that of previous instruments and is a particular science focus for the WFM.  The blue shading shows the region of signal-to-noise where high-fidelity spectroscopy can be performed in the prompt X-ray for these sources.\label{fig:grb_detections}}
\end{figure}

 With the ground-based gravitational-wave community discussing trade-offs on the localization capability to deepen the detection horizon for third-generation instruments (e.g., Einstein Telescope, Cosmic Explorer), the prompt and precise localization of X-ray transient counterparts to compact binary mergers will be crucial.  Current estimates indicate that the binary neutron star detection rate by LIGO/Virgo will likely be $\lesssim$ 10/year for the next few years (although more optimistic scenarios, with $\sim30$/yr remain viable \citep{kruckow18}), and because some fraction of those may not have prompt EM signals, uncertainties related to the emission physics of short duration gamma-ray bursts will likely remain.  Indeed, the detection of GW170817 has raised the question about the nature of the EM emission---an off-axis structured relativistic jet or a trans-relativistic shock breakout---and there remains a considerable amount of uncertainty about the emission mechanism, jet physics, and range of properties of such mergers.  As shown in Figure~\ref{fig:grb_detections},  \strobex{} will have the capability to detect -- with positions immediately accurate to $<$2' --  $\sim 10$ short duration gamma-ray bursts (compact binary mergers) per year, contributing to a larger statistical sample that can answer these questions.  

Another outstanding question is whether binary neutron star mergers coalesce directly into black holes or have a short-lived magnetar phase  before such a collapse.  For nearby mergers, with its XRCA, \strobex{} will be able to detect the few-hour X-ray plateau of such a short-lived object, or, alternatively, place constraining upper limits on such radiation.  This magnetar might also emit bar-mode gravitational radiation, which would be detectable with Advanced LIGO within 27 Mpc \cite{corsi09}.  In addition to detecting binary neutron star mergers, gravitational-wave detectors are searching for merging neutron star-black hole systems, which might also be progenitors to some gamma-ray bursts.  The expected detection rate is $\lesssim O(1)$/year with current technology, though there are some expected sensitivity improvements to high-mass-ratio systems in third-generation detectors, so this will be a promising discovery space for  \strobex.
%What is crucial from high energy observatories is to identify the position fast enough to aid optical searches, and then to help characterize the X-ray emission.  The lessons from GW170817 are that X-ray emission after the prompt phase are unlikely to contribute to the localization of the source.

%For continuous wave sources (i.e., neutron stars with deformations leading to quadrupole moments -- e.g., Ref.~\citenum{Owen_etal_1998}), the LIGO positional precision for detections should be $\sim$10~arcsec. The strongest candidates for continuous gravitational wave sources should be fast-spinning young pulsars which tend to have the highest X-ray luminosities, or active accretors; the  \strobex{}/XRCA is uniquely suited to detecting X-ray periodicities from these objects.  Furthermore, it has become clear in recent years that the high energy pulsations are emitted with a wider opening angle than the radio pulsations, so that there can be some radio-quiet high energy pulsars.  \fixme{this needs a refernece, and it's also sort of discussed in the pulsar section}  LEFT IN PULSAR SECTION, REMOVED HERE

Looking forward, we can anticipate some overlap in time between {\em LISA} and  \strobex.  For stellar mass binaries, this will enable precise neutron star mass measurement from white dwarf-neutron star binaries with mass transfer \citep{tauris18}.  Furthermore, for detached double compact object binaries, LISA will discover a large number of neutron stars in binaries solely through gravitation, allowing electromagnetic follow-up in a variety of wavelengths to determine whether pulsations can be seen, and hence giving a method for determining the pulsar beam opening angles from a gravitationally-selected set of pulsars.

Under conditions that will be met a small fraction of the time\cite{2008ApJ...677.1184L}, {\em LISA} will obtain positions accurate to 10 square degrees and with redshifts accurate to a few percent $\sim$1 month before the mergers of supermassive black holes.  This will, in turn, allow searching for the counterparts with the WFM, and then after detection, with quick follow-up with XRCA.  

The  \strobex/WFM will be sensitive to merging supermassive black holes in a higher mass range than the one over which {\em LISA} is sensitive.  The detected sources are expected to be highly obscured, meaning that hard X-rays and radio are the best bands in which to study them\cite{blecha18}.  Strong quasi-periodic oscillations, along with near-Eddington accretion, are also expected from double black holes as they merge \cite{2018MNRAS.476.2249T,2018ApJ...853L..17B}, which would be detectable out to redshift $z\sim$0.6 (where the mergers may be detectable in gravitational waves after the X-ray emission via their ringdowns\cite{2005MNRAS.361.1145R}) with the WFM, and, if one knew where to look, to redshift $z\sim 10$ with the XRCA.

\paragraph{Neutrino Sources: }
%Tom M, Eric Burns to do -- Maria Petropolou did much better job
%Astrophysical neutrinos, the ``ghost particles of the Cosmos'', are unique probes of the physical conditions in their sources, as they can reach Earth almost unimpeded due to their extremely weak interactions with matter and radiation. 
High-energy ($\gtrsim 20$~TeV) astrophysical neutrinos are produced by interactions of relativistic nuclei with radiation or matter, and as a result their detection can unveil the locations and mechanisms of cosmic ray acceleration.%potential sources of high-energy cosmic rays and the underlying particle acceleration mechanisms. 
AGN jets can accelerate and confine multi-PeV cosmic rays \citep{1984ARA&A..22..425H}, which can then interact with various radiation fields in the AGN environment (from e.g., the accretion disk, the broad line region, and the jet) to produce high-energy neutrinos 
\cite{1991PhRvL..66.2697S, 1995APh.....3..295M,2001PhRvL..87v1102A} (for recent reviews, see \cite{2017NatPh..13..232H}). $\gamma$-ray flares from blazars are ideal targets for neutrino searches, since it is expected that electromagnetic and neutrino flaring should be related.
%\footnote{Blazars, a subclass of AGN with relativistic jets pointing towards the observer, are the most numerous extragalactic $\gamma$-ray sources.}  
%have been suggested as ideal periods for the detection of high-energy neutrinos due to the expected lower atmospheric neutrino background and the higher neutrino production efficiency \citep[e.g.][]{2016APh....80..115P, 2017A&A...603A..76G, 2018ApJ...865..124M}. 

The first compelling neutrino source association was based on the detection of a high-energy neutrino (IC170922A) coincident with the flaring blazar TXS 0506+056 in 2017 \citep{2018Sci...361.1378I}.  The detection triggered a series of follow-up observations across the electromagnetic spectrum \citep[][]{2018Sci...361.1378I, 2018ApJ...864...84K}, which resulted in a rich multiwavelength data set. The X-ray observations by Swift/XRT and NuSTAR  ($\sim 2-80$ keV), in particular,  were crucial for constraining theoretical models and explaining the first multi-messenger flare 
\citep[e.g.][]{2018ApJ...864...84K, 2018ApJ...865..124M, 2019NatAs...3...88G}.  STROBE-X with its responsive scheduling and large sky coverage will provide rapid-response detections of flaring blazars with the XRCA at 10$\sigma$ in 2500 seconds for sources 100 times fainter than TXS~0506+056.

An analysis of archival IceCube neutrino data that revealed a ``neutrino flare'' in 2014 from the direction of TXS~0506+056, with duration $\sim 6$ months \citep{2018Sci...361..147I}, but without flaring in hard X-rays \citep[e.g.,][]{2018MNRAS.480..192P}.  STROBE-X's WFM, a factor of 3 times more sensitive than Swift BAT for Crab-like spectra, will help provide better limits on X-ray flaring associated with future neutrino flares.

\paragraph{Gamma-ray Bursts:}
%{\it Adam Goldstein} responsible
In addition to gravitational wave counterpart searches, \strobex{} will also dramatically enhance more traditional studies of gamma-ray bursts (GRBs) and supernovae, matching and exceeding the capabilities of the {\em Swift}/BAT.  The \strobex/WFM is similar to the {\em BeppoSAX} Wide Field Camera but with a much larger field of view.  The population of relatively faint and soft GRBs called X-ray flashes (XRFs) has not been studied since the {\em BeppoSAX} mission ended in 2002, due to the lack of wide-field, medium-energy X-ray instruments.  XRFs are still not well-understood, and may be normal GRBs viewed off axis or at high redshift, or they could be baryon-loaded ``dirty fireballs''\cite{1999ApJ...513..656D}.  By collecting measured spectra of XRFs from HETE-2\cite{Barraud03, Pelangeon08}, Swift\cite{Sakamoto08}, and candidates in Fermi GBM\cite{Jenke16}, folding those spectra through the WFM response and accounting for the WFM FoV, an estimate of the rate of XRF detections by \strobex can be made. Similarly, by collecting measured spectra of canonical GRBs by Fermi GBM\cite{Gruber14}, the short and long GRB rate estimate can be inferred. As shown in Figure~\ref{fig:grb_detections}, in addition to about 100 canonical long GRBs per year, the WFM is expected to detect more than 10 XRFs per year, most of them bright enough to perform detailed spectroscopy, which may help illuminate their relationship to canonical GRBs.  Additionally, the arcminute localization of the WFM will allow positions to be sent immediately to the ground which will be precise enough for IFUs to do immediate spectroscopic follow-up. The automated repointing capability of \strobex{} will allow XRCA spectra for the late prompt emission for fortuitously located GRBs, and the early afterglows for most of them. 

Given its soft energy band, in principle the WFM detection efficiency for short/hard GRBs is expected to be modest, with an onboard detection rate of about 7 events/year (Figure~\ref{fig:grb_detections}). However, based on HETE-II and Swift/BAT measurements, there is evidence that at least a fraction of short GRBs show a weak soft extended emission following the first hard spike. The unprecedented combination of soft energy band, wide field of view and sensitivity of the WFM compared to previous and present GRB monitors, will allow it to detect tens of short GRBs per year through their soft extended emission, thus providing an additional channel for the detection and localization of events produced by NS-NS and NS-BH mergers.

Furthermore, X-ray spectroscopy of GRBs and early afterglows in the energy range where abundant metals have atomic lines and edges can help in a variety of ways.  Redshifts may be potentially obtained with prompt emission; this was done previously with {\em BeppoSAX}\cite{2000Sci...290..953A} but it could be done more frequently, precisely, and reliably with the higher spectral resolution of the WFM. Furthermore, searches for emission and absorption lines with the XRCA in the early afterglows could yield a host of information about the nature and environment of GRBs\cite{2002ApJ...572L..57L}.  The latter case has so far yielded only statistically marginal results, but could be pursued dramatically better with XRCA than with past instruments.
%or proposed future instruments. - we do not know all missions proposed for future ...

\paragraph{Supernovae:}
For supernovae,  \strobex{} has the potential to be revolutionary.  Supernova shock breakouts should be visible to a distance of about 20 Mpc, and a few of these events are expected per year if the 2008 event in NGC~2770\cite{2008Natur.453..469S} is taken as a template.  The hard X-ray emission is the hottest and earliest signature of shock breakout.  Recently, an X-ray transient, AT2018cow, or ``The Cow'', associated with a type Ic supernova at 60 Mpc would have been bright enough for several days for the WFM to detect\cite{2018ATel11737....1R}; this event's nature is still under debate, but it may represent an orphan X-ray afterglow, an extreme supernova shock, the birth of a black hole, or the tidal disruption of a white dwarf by an intermediate mass black hole \cite{riveracow,perleycow,kuincow,margutticow}.   \strobex{} will discover similar events {\it earlier} relative to other proposed instruments (e.g., optical missions) that can also discover these shock breakouts, and its large instantaneous field of view will allow \strobex{} to discover the {\it nearest} of these events, offering unprecedented diagnostic tools. For example, shock breakout arrival times can be compared with the arrival times of neutrinos and gravitational waves to reveal the speeds of blast waves through the envelopes of massive stars in their final moments. A full understanding of core-collapse supernovae has not yet been reached, and full multi-messenger data sets of these events in the local Universe will generate breakthroughs at this frontier. 

The advances made in the arenas of gravitational wave and neutrino detection have broadened our portfolio from multi-wavelengths to multi-messenger, and  time-domain astronomy is poised to create a new golden era of astrophysics. Probing early cosmic star formation with high-redshift GRBs, and exploring the dynamic physics of supernovae in the local Universe will be the focus of instruments that can respond to transients, probe their emission on the shortest timescales directly and provide precise locations and fast characterization to ground and space based observatories ready for follow-up. As demonstrated with missions such as {\em Swift} and {\em Fermi}, wide field-of-view and fast response are key elements of missions exploiting the era of time domain astronomy.  \strobex{} offers unique capabilities that combine all-sky monitoring, sensitivity, rapid slewing and high time resolution, which by themselves and in combination with synergies with other observatories, will greatly advance our knowledge of a broad class of transient phenomena.% such as GRBs, supernovae, shock breakouts, tidal disruption events, and others --- as we briefly discuss in the next section. 

\subsection{The Broader Portfolio of {\em STROBE-X} Astrophysics}

In addition to the core science goals envisioned for {\em STROBE-X}, the mission has unique capabilities to contribute to a wide variety of other scientific fields.  These consist of goals related to accretion physics, stellar evolution, cosmology and galaxy evolution, and nuclear and particle physics.

\subsubsection{Accretion Physics: Tidal Disruption Events}
\label{sec:tdes}

%\subsubsection{Tidal Disruption Events}
%{\it TDE section: Erin Kara}

Tidal disruption events (TDEs) occur when a star on a close approach to a supermassive (or intermediate mass) black hole gets ripped apart by the black hole's strong tidal forces, creating a short lived electromagnetic flare. Three out of the roughly two dozen observed TDEs have been detected in hard X-rays ($L_{\mathrm{X,peak}}\sim 10^{48}$~erg/s), suggesting that we are watching the birth of a beamed relativistic outflow. The Wide Field Monitor will discover tens of new jetted TDEs per year, which can be followed up quickly with the LAD and XRCA. The fast response of {\em STROBE-X} will also allow us to follow-up TDEs discovered by ground-based optical and radio facilities. In this section, we will focus on the scientific impact of these three instruments for TDE science.

\begin{figure}[th]
\includegraphics[width=\textwidth]{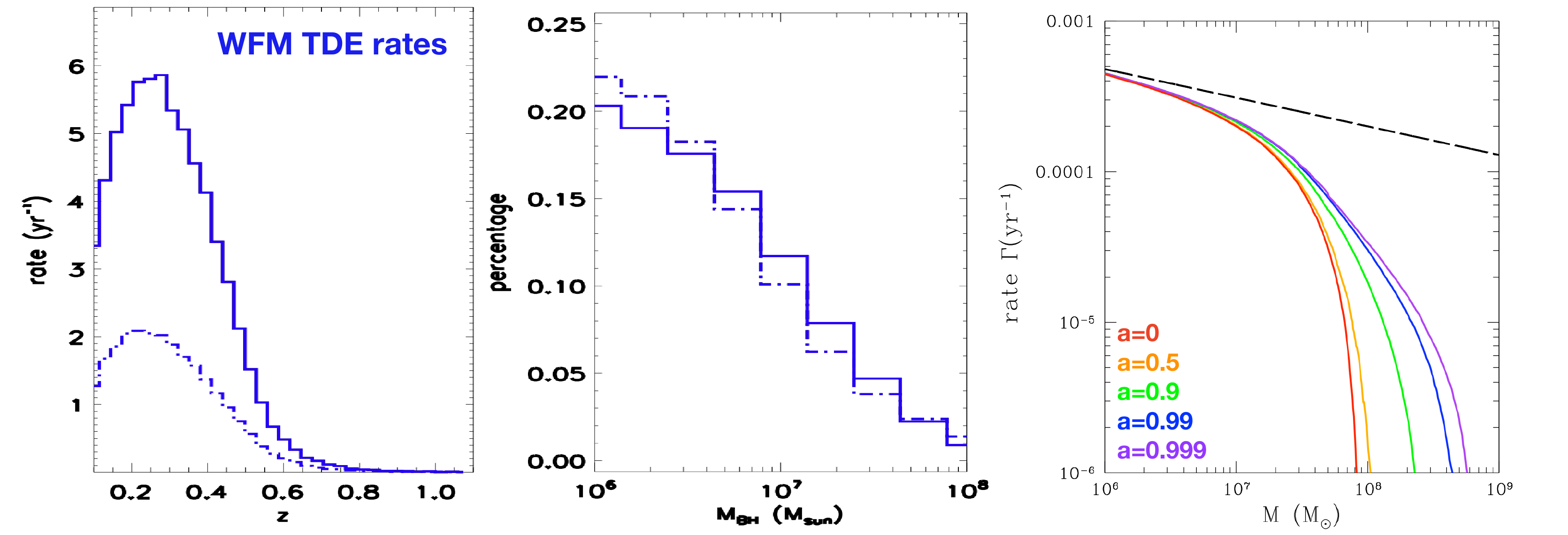}\\
\caption{\footnotesize{{\em Left}: The predicted rate jetted TDEs discovered by {\em STROBE-X} WFM as a function of redshift from \citet{rossi15}. {\em STROBE-X} WFM will discover 24--67 TDEs per year out to a redshift of z=0.6. {\em Middle}: The predicted rate of jetted TDEs discovered by {\em STROBE-X} WFM as a function of mass. In both panels, two different black hole mass functions from \citet{shankar13} are shown. {\em Right}: The rates at which stars are tidally disrupted by SMBHs in powerlaw galaxies obeying the $M-\sigma$ relation (from \citep{kesden12}). The dashed black line is the prediction of \citet{wang04}. The colored curves show the relativistic corrections to this prediction for different black hole spins.}}
\label{fig:TDEfig}
\end{figure}

\paragraph{WFM: Providing an unbiased measure of the spin of quiescent black holes: }
The Wide Field Monitor (sensitive from 2--50 keV) will produce an unbiased, volume-limited sample of jetted TDEs. Thus far, only three such events have been definitively detected in X-rays, and none have been discovered in optical surveys. {\em STROBE-X} WFM will discover 24--67 per year out to a redshift of $z=0.6$  \citep{rossi15} (Fig.~\ref{fig:TDEfig}-left, middle). Below, we describe how this unbiased sample can be used to study spin demographics in quiescent black holes.

Tidal disruption events occur preferentially in lower mass AGN because the tidal disruption radius scales as $r_{\mathrm{td}} \propto M_{\mathrm{BH}}^{1/3}$, while the innermost stable circular orbit scales as $r_{\mathrm{isco}} \propto M_{\mathrm{BH}}$. In black holes with $M_{\mathrm{BH}}>10^8 M_\odot$, the star will be ``swallowed whole'' by the black hole, leaving no electromagnetic trace of its final demise. However, the innermost stable circular orbit scales with the spin of the black hole, and thus, we will only see TDEs of normal stars in high mass black holes ($M_{\mathrm{bh}}>10^8 M_\odot$) when the black hole is spinning rapidly \citep{kesden12} (Fig.~\ref{fig:TDEfig}-right). The unbiased jetted TDE mass function will therefore constrain what fraction of normal black holes are rapidly spinning.  Optical time domain surveys will measure the TDE mass function in thermal TDEs \citep{vanvelzen18}, but the WFM is the only instrument that can make this measurement for jetted TDEs, which potentially occur preferentially in high-spin black holes if jets are powered by some mechanism (e.g., the Blandford-Znajek mechanism) in which jet power scales with spin.\\

% Guillochon double peaked flare of 5e44 erg/s out to 200 Mpc 

\paragraph{LAD for jetted TDE identification and follow-up: }
After a detection in the WFM, the LAD and XRCA will follow-up rapidly, allowing for spectral classification as a jetted TDE. Most TDEs do not evolve extremely rapidly, and so even responding within 1-day of the WFM trigger is sufficient. In principle, the featureless power law spectra will look similar to those of flaring blazars, but by the time {\em STROBE-X} launches, the {\em eROSITA} mission will have identified all blazars with log$(L_X)>44$ out to $z=0.6$. The LAD and XRCA will monitor the source as it rises to peak luminosity 1 month after disruption, and decays as roughly $t^{-5/3}$ over the next several months. 

Only one jetted TDE (Swift~J1644+57) has been followed up in real time, and thus the discovery space for this science is huge \citep{burrows11}. Swift~J1644+57 was highly variable, and follow-up observations with \textit{Suzaku} and \textit{XMM-Newton} in the 0.3--10~keV band allowed for the discovery of quasi-periodic oscillations \citep{reis12}, and reverberation of a highly blueshifted iron~K emission line (suggestive of a ultrafast outflow of $\sim 0.3-0.5$~c; \citep{Kara_etal_2016}). These spectral-timing results can constrain the geometry of the accretion flow, and also can be used to estimate the mass of the central black hole, which can be compared to independent estimates from fitting the long-term lightcurve \citep{mockler18} and from optical-based methods like the $M-\sigma$ relation. Unfortunately, even in a source as bright as Swift~J1644+57, the evolution of the QPO and reverberation lag could not be measured, and therefore, we cannot constrain how the accretion flow evolves in this highly super-Eddington event. With {\em STROBE-X}, we will follow-up on tens of jetted TDEs per year, and obtain precision broadband measurements with high signal-to-noise and no pile-up effects.

\paragraph{XRCA: Rapid soft X-ray follow-up of optically-selected TDEs: }
For the foreseeable future, most TDEs will be discovered in the optical band through large optical time domain surveys.  These ``thermal TDEs'' show a $\sim10^4$~K blackbody spectrum, the origin of which remains strongly debated. It is unknown whether the optical is due to shocks from colliding stellar debris streams (e.g. \cite{piran15}), or reprocessing of X-ray disk emission in a large obscuring torus or wind (e.g. \citep{roth16}; \citep{dai18}). The X-ray emission, however, is much hotter ($\sim10^{5}$~K), which is strongly suggestive of thermal emission from a newly formed accretion disk at tens of gravitational radii from the black hole. Therefore, understanding when X-rays appear relative to optical emission (e.g. \citep{auchettl17}) is essential for understanding the origin of thermal TDEs. 

{\em STROBE-X} will have a response time to optical triggers of $<3$~hours, which will provide our best yet measurements of when the X-rays respond relative to the optical. Moreover, the XRCA has a similar spectral resolution to \swift/XRT, but has an effective area that is $>200$ times \swift, allowing rapid spectral and timing measurements that are impossible with \swift. Finally, unlike \swift, {\em STROBE-X} will be able to observe high count rate sources without the spectral distortions that come from pile-up. 

The large effective area of the XRCA will also allow for the detection of quasi-periodic oscillations in thermal TDEs. Very recently, \textit{XMM-Newton} observations of the brightest and most nearby soft X-ray TDE ASASSN-14li showed evidence for a high-frequency QPO at 7.65 mHz \citep{pasham18}. If this QPO is associated with a particle orbit at the ISCO, it suggests that the black hole in ASASSN-14li is rapidly spinning, with $a>0.7$. With the XRCA's large effective area, we will perform phase-resolved spectroscopy of the QPO in order probe which spectral components are responsible for the oscillations (similar to Ingram et al., 2015 for type-C QPOs in Galactic black hole binaries). Should high-frequency QPOs be present in all thermal TDEs, the XRCA will measure this behavior out to $z=1.5$, providing an unbiased probe of the evolution of black hole spin over cosmic time.

\subsubsection{Accretion Physics: Active Galactic Nuclei}
\label{subsub:agn}
%Alessandra De Rosa & David Ballantyne
\strobex\ will probe the nature of accretion and how relativistic jets are ejected from a variety of actively accreting systems, including black holes across the mass scale and neutron stars. Both the timing and spectroscopic capabilities of \strobex\ will allow unique measurements that will yield invaluable insights into the nature of accretion physics.

%{\em STROBE-X} will probe the nature of accretion and how relativistic jets are ejected from actively accreting systems.  We will achieve a greater level of detail in detected quasi-periodic oscillations, and our studies of the ``noise'' components of power spectra of XRBs can help us understand both power spectra and QPOs and use them to probe accretion geometry.  Supermassive black hole mass estimates made from AGN power spectra\cite{2006Natur.444..730M} will be made at a variety of inclination angles, and can hence be used to tie together the masses from masers and those from reverberation mapping.
% I'm not sure what this last sentence means.  Is this speaking towards an interpolation between two populations of sources or reconciling two different mass determination techniques?

In the case of AGN, a clear example of the promise of \strobex\ is determining the origin of the ``soft excess'', the excess above the main power-law at energies $<$~1 keV.  The origin of this radiation remains uncertain three decades after its discovery. Two compelling models for the excess in unobscured AGN are a ``warm'' corona \cite{2013A&A...549A..73P} that may be between the accretion disk and the hot corona producing the X-ray power-law, and highly blurred relativistic reflection from a high density disk that naturally produces a strong soft excess \cite{2004MNRAS.351...57B,2016MNRAS.462..751G}. The spectra predicted by these two models are almost identical between 1 and 10~keV \cite{garcia18}, but the broadband spectra and high sensitivity of \strobex\ will easily be able to distinguish the two models in $\approx 20$~ks for a bright AGN. The short exposure time needed to make these measurements means that \strobex\ can quickly measure soft excesses and, even more importantly, their variability in multiple AGN.  Moreover, the $0.2$--$30$~keV bandpass produced by the combination of data from the XRCA and LAD will yield measurements of the high-energy cutoff and subsequent hot coronal parameters (such as optical depth and electron temperature) for bright AGN (e.g., $\sim 10^{-10}$~erg~s$^{-1}$~cm$^{2}$ in the \textit{Swift}/BAT band) in roughly half the exposure time currently needed by \nustar\ (assuming a cutoff of $200$~keV).  This opens up the unique possibility of examining relationships between, e.g.,  the ``hot'' and ``warm'' coronae in many AGN.  Such an investigation could only be performed by \strobex\ and will be crucial to understanding the flow of accretion energy through the disk and the putative two-phase corona into the observed X-ray spectrum.

\subsubsection{Accretion Physics: The Absorbing Environment of AGN}
\label{subsub:otheragn}
%Alessandra De Rosa

As described in \S\S \ref{subsub:spins} and \ref{subsub:agn}, \strobex\ will provide crucial insights into AGN physics, shedding light on the details of the inner accretion disk and corona, as well as measuring the BH spins in numerous AGN. However, the combination of monitoring capabilities, large effective area and good spectral energy resolution will allow \strobex\ to make important discoveries in other areas of AGN physics. In this section, we highlight how \strobex\ monitoring of AGN will yield novel measurements of the distant obscuring gas that surrounds most AGN.

The broadband AGN spectra efficiently produced by \strobex\ will be able to quickly measure parameters of the obscuring torus models \cite[e.g.,][]{2009MNRAS.397.1549M, 2011MNRAS.413.1206B,balokovic18} envisaged in the Unified model for AGN emission \cite{1993ARA&A..31..473A, 1995PASP..107..803U}. These measurements require good sensitivity around the Fe K$\alpha$ line and a high-energy response to measure the the shape of the reprocessed continuum, which typically peaks above 10 keV. However, spectroscopy alone gives degenerate constraints on the torus parameters (such as column density $N_{H}$ and covering fraction $f$).
The \strobex\ WFM will provide multi-epoch observations of the torus parameters in different states of absorption. Follow-up LAD observations will allow us to measure the $N_{\rm H}$ variability due to passing clouds along the line of sight.
Simultaneous fitting of the Fe line and the bump above 10 keV available with the XRCA and LAD  data will allow measurement of  the continuum at different levels of absorption.
Indeed, simulations show that 10 \strobex\ observations of $\sim$10~ks are required to measure the mean torus column density and covering factor in a bright, nearby AGN and only $\sim$30 observations are required for AGN that are a factor of a few fainter. The high throughput of \strobex\ means that these measurements can be done for large numbers of obscured AGN, building up a statistical sample of torus properties for determining the correct physical model of the obscuring region.

AGN monitoring observations will also probe the location of the obscuring gas, by reverberation of the \emph{narrow} Fe K$\alpha$ line. The origin of the narrow core of the Fe K$\alpha$ line complex is unknown, but must be emitted by dense gas far from the AGN \cite[e.g.,][]{2006MNRAS.368L..62N, 2011ApJ...738..147S} with distances extending past the broad-line region (BLR) \cite{2013A&A...549A..72P,2015ApJ...812..113G,2015ApJ...812..116B}. The reverberation of the narrow Fe K$\alpha$ line on days-to-months long timescales will be a new probe of the structure of the outer accretion disk and BLR. Previous reverberation experiments with \xmm\ \cite[e.g.,][]{2013A&A...549A..72P} had sparse monitoring and limited Fe K$\alpha$ sensitivity during each observation. The \strobex\ WFM will be able to provide 3$\sigma$ daily detections of Fe K$\alpha$ for bright AGN ($>10^{-10}$~erg~s$^{-1}$~cm$^{2}$) or weekly for weaker sources ($>5\times 10^{-11}$~erg~s$^{-1}$~cm$^{2}$). With pointed observations, the LAD and XRCA instruments will provide well-sampled monitoring of the narrow Fe K$\alpha$ line and its associated reflection spectrum: a 1~ks LAD observation will provide the normalization of a 100 eV equivalent width Fe K$\alpha$ line with $\approx$3\% uncertainty for bright AGN, with the XRCA providing similar measurements for weaker sources due to its lower background. The combination of monitoring cadence, large area and good energy resolution make \strobex\ the only future X-ray mission able to perform such experiments for a large number of bright AGN. 

\subsubsection{Accretion Physics: X-ray Timing Approaches to Black Hole Mass Estimates}
%Tom M, based on stuff Alessandra De Rosa did for LOFT

Characteristic break timescales in X-ray power spectra can provide an excellent mass indicator for AGN if one also knows the mass accretion rate \citep{mchardy06}.  Calibration of the X-ray timing methods against other approaches for measuring black hole masses would be an essential step forward.  The key benefits of X-ray timing measurements are two-fold: first, unlike reprocessing lags, X-ray power spectral break frequencies should be inclination angle independent, meaning that X-ray timing of a large number of reverberation-mapped AGN should be able to sort out and understand the inclination angle effects on optical reverberation mapping.  Second, X-ray timing is one of the most promising means to estimate the masses of high redshift AGN, using an approach that may be calibrated most effectively with a mission like {\em STROBE-X}, and then applied to high redshift objects with other higher angular resolution large collecting area missions.

Both the WFM and pointed observations will greatly expand the number of AGN with good power spectra.  The WFM will monitor more than 100 bright AGN with daily 5$\sigma$ detections, producing good power spectra for them \cite{derosaloft}, increasing the number of AGN with good power spectra by about an order of magnitude ``for free''.   These predominantly nearby, bright, relatively low black hole mass systems will overlap strongly with the ones for which optical reverberation mapping works best, helping to tie these two methods together.

The pointed observations will allow the study of fainter AGN.  In 100 seconds, the {\it XRCA} will detect AGN at 10$\sigma$ at about the ROSAT all-sky survey limit.  The space density of these AGN is about 1 per square degree, and these objects have properties ranging from nearby low luminosity AGN to $z\sim2$ bright quasars. The former allow timing tests of the disk truncation model \citep{scaringi15}, while the latter open up new means for testing the $M-\sigma$ relation at higher redshift than it can current be done \citep{shen15}.

Additionally, simultaneity of \strobex{} with survey projects like LSST will provide the capability of measuring correlations and lags to the optical, and for blazars, the $\approx$10 keV band is ideal for comparison with time series from Cerenkov telescopes.  With pointed instruments, the capability of the XRCA, combined with the flexible scheduling, will allow monitoring campaigns on AGN that are much fainter than those monitored by \rxte{} and Swift.

\subsubsection{Accretion Physics: X-ray binary disks' responses to bursts}
%David Ballantyne
 
 Type I X-ray bursts are nuclear explosions on the surfaces of accreting neutron stars that last less than a minute, but are expected to strongly interact with the surrounding accretion disk \cite{2005ApJ...626..364B,2018SSRv..214...15D,2018ApJ...867L..28F}. The details of this interaction are unknown, but radiative driven outflows, enhanced accretion (due to Poynting-Robertson drag), and significant heating and cooling effects may all be relevant depending on the details of the burst and accretion disk. Determining the response of an accretion disk to such an impulse will provide unique insights into the physical processes at work in the accretion flow. \strobex\ time-resolved spectroscopy of a Type I X-ray burst will be able to use X-ray reflection signatures to map out the response of the accretion disk to the burst in real time \cite{2016ApJ...826...79K}. Modeling the time-resolved reflection signatures will provide a picture of any changes in the accretion disk's geometry during the burst. As these bursts occur several times a day all over the sky, \strobex\ will be able to provide a never-before-seen view of the dynamics of accretion disks. Finally, since the accretion disk itself produces X-rays, understanding the behavior of the accretion disk during an X-ray burst is necessary to ensure the accuracy of using bursts to measure EOS parameters (\S\ref{subsub:eos}).

\subsubsection{Accretion Physics: Jets and Disk Winds in X-ray Binaries}
%Joey Neilsen
%Because of their short variability timescales compared to supermassive black holes (see \ref{subsub:spins}), X-ray binaries represent a critical laboratory for understanding the physics of accretion and ejection processes around compact objects. \strobex{} will enable significant advances in our understanding of these systems, particularly in the areas of accretion disk winds, the source of QPOs, and the origin and evolution of accretion states in both black hole and neutron star systems.

%\textit{Accretion Disk Winds:} 
Accretion disk winds offer powerful diagnostics of the connection between black holes and their environments across the mass scale. We now know that these winds---once thought to be simple ionized disk atmospheres---may carry away the vast majority of the infalling matter \citep[e.g.,][]{Neilsen11,Ponti12}, effectively regulating the accretion rate at the event horizon. Evidence suggests that similar processes play out in both AGN and stellar-mass black holes \citep{King13}. The consequences of extracting so much matter and angular momentum \citep{Miller15} are unknown, but it is clear that without deep insight into the physics of ionized winds, we cannot fully predict the evolution of accreting black holes. How, where, when, and why do winds appear? Are they governed by magnetic processes \citep[e.g.,][]{Miller06}, radiative/thermal processes \citep[e.g.,][]{Neilsen11}, or some combination of the two \cite{Neilsen12}, and how are these processes coupled to the physics of accretion at the event horizon?

\begin{figure}    \centerline{\includegraphics[width=3in]{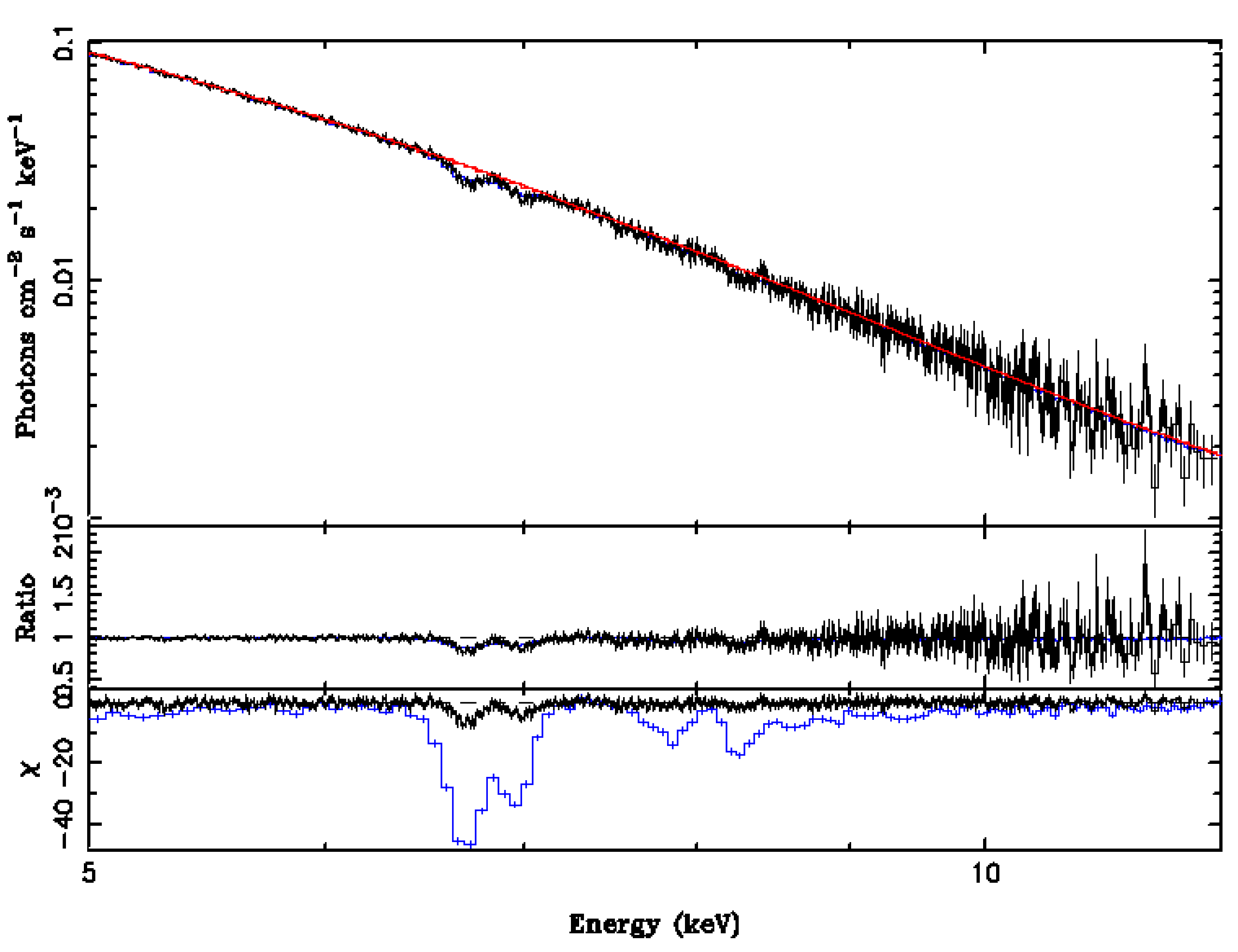}\includegraphics[width=3.6 in]{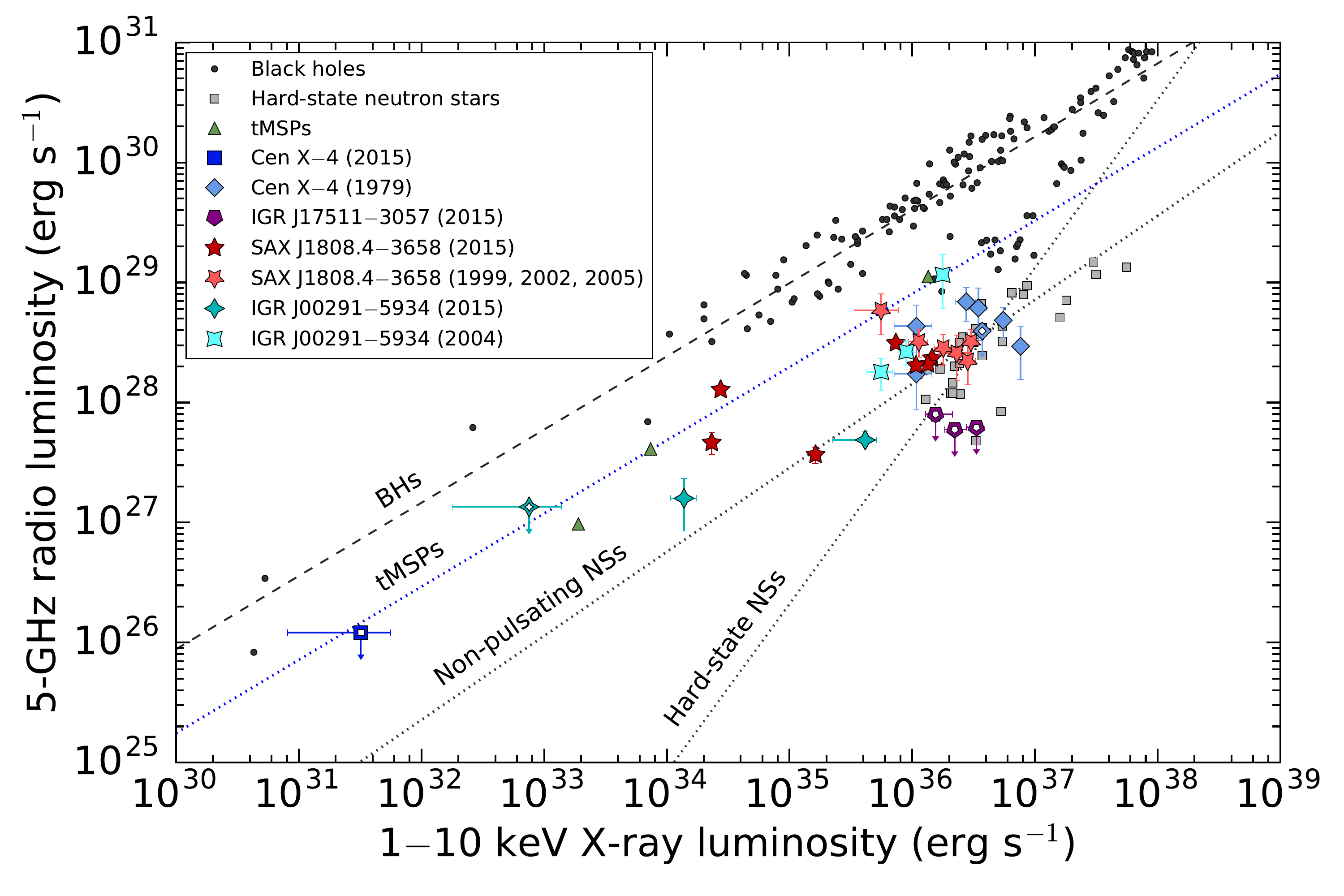}}
\caption{Left:\strobex{} simulation of a 1 ks observation of an accretion disk wind. The incredible sensitivity of the XRCA and LAD enables a highly significant detection of X-ray absorption, facilitating detailed variability studies of ionized outflows from accreting compact objects. Right: The radio/X-ray correlation for black hole X-ray binaries (black dots) and neutron star X-ray binaries (colored symbols) \citep{tudor2017}, indicating that black holes mostly follow a standard power law relation, with some significant deviations, while the neutron star systems are less well-characterized, mostly due to a combination of faintness, faster variability placing stronger constraints on simultaneity.}
    \label{fig:winds}
\end{figure}

Because photoionization couples the observable properties of winds to the local radiation field, precision spectroscopy is required to address these questions. With exquisite sensitivity to ionized absorption from oxygen to iron on time scales as short as seconds, \textit{STROBE-X} is thus uniquely suited to mapping the geometry and plasma properties of these outflows. At an X-ray flux of $4\times10^{-9}$ erg s$^{-1}$ cm$^{-2}$, the XRCA and LAD will detect typical hot disk winds ($\log\xi=10^4$ erg cm s$^{-1}$, $N_{\rm H}=10^{23}$ cm$^{-2}$) in under a minute (see Figure \ref{fig:winds}), enabling the integration of photoionization studies with state-of-the-art timing to reveal how winds respond to rapid variability (through direct methods on $\sim500$ second timescales and Fourier-domain correlations down to sub-second timescales). Via the recombination timescale, the lags and magnitudes of the responses directly determine the density, location, and clumpiness of these winds.  \strobex{} is thus uniquely suited to mapping the geometry and plasma properties of outflows. Ultimately, this will enable a direct comparison between observations and magnetohydrodynamic simulations of winds, revolutionizing our understanding of accretion and ejection processes around black holes.

\strobex{} will also make major contributions to our understanding of the disk-jet connection.  With \rxte, clear relations, albeit with significant scatter, were found between radio and hard X-ray luminosities of sources, and tentative evidence was found for most accreting neutron stars following a different behavior than most accreting black holes \citep{tudor2017,gallo2018}, but with the sample sizes still insufficient for developing a full understanding of the problem. The increased sensitivity of the Wide Field Monitor will allow quasi-simultaneous X-ray data for free with any radio campaign for outbursting X-ray binaries, and pointed XRCA observations enabling monitoring to near quiescence for most objects, while better sensitivity and cadence in the radio will come from the new generation of facilities already built and planned.

Furthermore, the pointed instruments will allow estimation of the time lags from the X-ray emission from the disk to the jet emission at different wavelengths from optical through radio where the jet may often be the dominant source of emission.   These time lags can then be used to infer the jet speeds in different regions and hence the jet's acceleration or deceleration \citep{casella2010,gandhi2017,tetarenko2019}. The WFM will enable the same to be done for a large set of AGN, for which the availability of good sampling has meant that only a tiny set of objects has been studied \citep{bell2011}.  At the present time, these techniques have shown strong potential, but have not been fully exploited because of the relatively small time window of overlap in which both {\it RXTE} and the relevant multi-wavelength facilities were available and the difficulty in scheduling similar observational campaigns with now-existent X-ray facilities (and the fact that for X-ray/infrared cross-correlations, even with RXTE, the X-ray signal to noise was the limiting factor).  The SKA pathfinders can already easily provide good monitoring of the AGN for which the WFM will provide high cadence X-ray light curves, and for the mean fluxes of bright X-ray binaries, while the time lag measurements for bright X-ray binaries, and flux measurements for the faint X-ray binaries will be straightforward in few hour runs with facilities like the ngVLA and SKA.

\subsubsection{Accretion Physics: Low Frequency Oscillations}
%Mostly by Abbie Stevens
Quasi-periodic oscillations (QPOs) have been seen in the X-ray light curves of accreting BHs and NSs for more than 30 years \citep{vanderklis85}, and yet the physical processes that produce them (and even the precise location of those processes) are still a mystery. 
Theories of low-frequency QPOs ($\sim$0.01--30 Hz in BHs, $\sim$1--200 Hz in NSs) invoke intrinsic brightness variations, such as trapped disk oscillations \citep{nowak91} or spiral accretion-ejection instabilities \citep{tagger99,varniere02}, or geometric variations in the observer’s line of sight, such as general relativistic Lense-Thirring precession \citep{stella98, ingram09,liska18}. 
Low-freq.~QPOs are predominantly \textit{hard} features, and spectral-timing analysis such as lag-energy spectra, covariance spectra, and phase-resolved spectroscopy have shown that QPOs have phase-dependent energy variations \citep[for a review, see][]{Uttley_etal_2014}, spanning a soft disk-like blackbody to the Compton hump \citep{Miller_Homan_2005, ingram16, stevens16}.  
These variations for one of the particular subclasses of QPOs are consistent with the expectations of the Lense-Thirring precession model which would require both substantial spins for the black holes and misalignments of the spin and orbital axes.  
Furthermore, Cygnus~X--1 notably {\it does not} show this oscillation, despite enormous amounts of observing time in the appropriate spectral states; 
Cygnus~X--1 also has a high mass relative to other stellar mass black holes and a space velocity consistent with not having formed with a substantial natal kick, providing suggestive evidence that the QPOs in other sources are Lense-Thirring, with the natal kick as the origin of the spin misalignment.  
{\em STROBE-X} has the potential to test this understanding with better measurements on a much broader class of sources.

{\em STROBE-X} will provide the QPO trifecta of sub-millisecond time resolution, CCD-quality energy resolution, and simultaneous soft and hard X-ray coverage, to enable transformational studies of these still-enigmatic signals. 
Observing QPOs in bright ``Z-source'' NSs will probe the dynamics and configuration of near-Eddington accretion flows, and catching rapid transitions between types of QPOs in both BHs and NSs is necessary for understanding the inflow-outflow connection between accretion disks and relativistic jets. 
{\em STROBE-X}'s low integration time, due to its large collecting area and high throughput, will make it the most sensitive observatory for detecting and tracking low-freq.~QPOs. 
Finally, far more low-freq.~QPOs are known in BHs than in NSs \citep{motta17}. 
Population studies will be crucial to identify key properties of QPO classes, as opposed to peculiarities of single occurrences, and draw comparisons between the strong-gravity effects on accretion around NSs and BHs.  
If it can be established that the QPOs expected to be Lense-Thirring precession are absent in a subset of neutron star systems despite high quality searches, this could potentially be evidence that these neutron stars have spins well aligned with their orbits, which would likely result only if those neutron stars form from accretion-induced collapse events.  
This would then help identify the best targets for follow-up to determine if the masses of these neutron stars are systematically low.

\subsubsection{Accretion Physics: Magnetized Accretors}
%From silas Laycock

High-mass X-ray binaries (HMXBs -- the X-ray binaries with massive donors, which tend to have higher magnetic field neutron stars because the NS magnetic fields have not had time to decay) are excellent laboratories for studying the physics of accretion onto highly magnetized neutron stars. Single pulses from HMXB pulsars will be detectable throughout the Galaxy and Magellanic Clouds. Virtually all past and contemporary work uses pulse-profiles and spin frequencies derived by averaging many tens to hundreds of spin-cycles of the neutron star.  Changes in the accretion stream and magnetosphere occurring on the natural dynamical timescales are lost in this process. The ability to perform advanced analysis techniques such as acceleration searches to find pulsars with large frequency derivatives (whether of orbital or torque origin) hinges on being able to reduce the averaging-window to a minimum, ideally to the single-pulse level to attain true phase connectivity. Long duration ``deep" pulsar surveys \citep{Hong2017} have shown that large changes in both absorption and accretion rate can occur on timescales of minutes to hours, masking the physics occurring inside the magnetosphere.   The magnetic fields of NS are in principle open to determination via several approaches, these fall broadly into torque-based\citep{klus14}, cyclotron-line \citep{staubert18} and indirect methods \citep{campana18}, all of which \strobex{} can revolutionize. {\em STROBE-X} will provide frequency derivatives, and pulse profiles {\it in the critical low-luminosity regime} at which the propeller effect is thought to halt accretion, and reverse the torque balance. 

Wind-fed supergiant X-ray pulsars are also interesting candidates for {\em STROBE-X}. Thanks to the combination of fast timing resolution and high sensitivity, both the LAD and XRCA on board {\em STROBE-X} can  accurately measure spectral variations with high SNR on very short time scales. Crucially, this will provide the capability to measure dynamical timescale phenomena within their coherence time, probing the binary environment at unprecedentedly small spatial scales. It has been demonstrated that studies of such systems can furnish a new independent way to measure NS masses \citep{Manousakis2012}, as well as to probe the large-scale structures at work in these systems (such as the accretion wake in Vela X-1), to understand the origin of the well-known yet poorly understood “off-states”, and to help probing the ambient wind structure through spectral analysis \citep{Malacaria2016}. The latter objective is particularly advantageous for {\em STROBE-X} (with respect to, e.g., \nicer) because of its broader spectral range (which would help constraining the spectral shape variability along the binary orbit).

\subsubsection{Accretion Physics: Ultraluminous X-ray Sources}
%Not sure where this came from
It has now been established that a substantial fraction of the ultraluminous X-ray sources (sources emitting above the Eddington luminosity for a 10$M_\odot$ black hole) show pulsations. 
The XRCA, with its modest background and high throughput, is ideal for detecting new pulsations 
from known ultraluminous sources.  In the \nustar{} soft band from 
3--10 keV, the RMS amplitude of the pulsations from M82 X-2 is about 5--10\% \citep{bachetti14}.  Such amplitudes of 
pulsation can be detected with the XRCA in 10 kiloseconds at distances of 20 Mpc for sources with $L_X>10^{40}$ erg/sec.  Thus, even in a relatively pessimistic case, we should expect {\em STROBE-X} to
be sensitive to all pulsating ULXs within 10 Mpc in a reasonable exposure time.  By executing an efficient survey, \strobex{}
is likely to identify $\sim$30 pulsating ULXs, an order of magnitude larger than the currently known sample. A follow-up program can then 
monitor them, measuring spin-up or spin-down rates, and measure the spectra of the pulsations to search for cyclotron lines that may indicate proton cyclotron emission due to ultra-high magnetic fields \citep{brightman18}.

Moreover, there are suggestions that the majority of the non-pulsating ULXs may host
magnetized NSs \citep{2017A&A...608A..47K}, where pulsations can be lost due to optically-thick envelopes that engulf the NSs \citep{2017MNRAS.467.1202M} and outflows
\citep{2017MNRAS.468L..59K}. The interaction of the pulsed signal with these optically-thick
structures can imprint distinct signatures in the power-density spectra that should be detectable with \strobex{}  \citep{2019MNRAS.484..687M}.

\subsubsection{Stellar Evolution: Extreme Coronal Activity}
%Written by Adam Kowalski
Stellar flares are transient bursts of radiation across the electromagnetic spectrum, from the $\gamma$-rays to the radio.  Flares are thought to result from catastrophic reconfiguration of magnetic fields in the corona, producing accelerated beams of nonthermal electrons and ions at mildly relativistic energies that propagate down to the stellar chromosphere where they deposit all of their kinetic energy via collisions with the dense ambient plasma.  This heating of the chromosphere causes the explosive evaporation of material into the reconfigured fields, forming flare loops that emit at temperatures of tens of million K.   

One of the important outstanding questions of stellar flare studies is the extent to which the solar paradigm
holds under very different conditions of temperature, density, magnetic field strength, and length scales. 
 Many stellar flares, primarily from RS CVn binaries or dMe stars, have been observed by \chandra,  \xmm{}, and \swift{} at low X-ray energies ($0.2 - 12$ keV), but much remains unconstrained about the properties of individual flares and the physics that produce the hot, thermal plasma.  Typically, several dMe flares must be co-added to obtain a temperature estimate using the bremsstrahlung continuum shape \citep{Osten2005}.  
 %This makes it very difficult to test radiative-hydrodynamic models of the coronal density and temperature response to the various heating mechanisms.  While electron beams are thought to be the primary driver of the bulk of thermal flare radiation, additional heating mechanisms could also be important: return current electric fields \citep{Holman2012}, direct heating by slow shocks during the reconnection process \citep{Longcope2011, Longcope2016}, low-energy proton beam energy deposition \citep{Allred2015}, and Alfv\'en wave heating \citep{Reep2016} have all been suggested to alleviate some of the observational challenges \citep{Fletcher2007, Warmuth2016} presented if only electron beam heating is considered.  Furthermore, the thermal energy content of the hot ($0.1 - 100$ MK) plasma in solar flares has recently been found to be more than 10x larger than previously thought \citep{Aschwanden2015}, which makes it far more likely that multiple mechanisms contribute to the energetics of solar flares.
%One way to critically test the standard electron beam models of solar flares is to extend our predictions to a much more energetic regime of flares than ever is observed on the Sun: the regime of the so-called superflares.

Superflares from rapidly rotating G dwarfs recently have been detected in white light with \emph{Kepler};  simple scaling relations suggest that strong magnetic fields (and larger flaring areas in the low atmosphere) may be able to explain the three-to-four orders of magnitude larger energy release compared to even the largest solar flares \citep{Maehara2012, Namekata2017}.  However, no coronal (i.e., X-ray) observations of these superflares have yet been obtained.   The rate of superflares from these 
rapidly rotating G dwarfs is only one per $\sim10$ d for the most active G dwarfs and one per several hundred days for the least active G stars \citep{Shibayama2013}.  Thus,  triggered observations offer a reasonable approach to characterize the flaring coronae of such extreme, rare events.  With the ability to detect the most extreme flares with the WFM, and then to make automated slews to follow-up with XRCA, \strobex{} is ideally suited to make X-ray observations of these events.

{The high sensitivity of {\em STROBE-X} from $0.2 - 30$ keV will allow regular, detailed measurements of the hottest bremsstrahlung continua during individual flares from M dwarf stars, RS CVn systems, and G dwarf superflare stars.}  The spectral resolution of the XRCA will allow measurements of emission line strengths from elements like iron and nickel, allowing crucial diagnostics of the likely multi-temperature emission regions.  The ability of the wide field monitor to detect the most extreme stellar flares and trigger slews onto them will allow the evolution of these flares to be studied.

The X-rays from flares have several very important impacts on exoplanetary systems.  The XUV radiation from flares can ionize protoplanetary dust, thus providing a mechanism for removal and grain growth inhibition \citep{Osten2013}, and the XUV can heat planetary atmospheres, thus causing atmospheric mass loss from potentially habitable planets \citep{Owen2016}.   However, the $E\gtrsim1.2$ keV X-rays are generally assumed to be a minor contribution to planetary atmospheric chemistry, and the irradiance from X-rays during superflares is generally ignored in photochemistry modeling due to a lack of observational constraints \citep{Segura2010}. The importance of these energetic X-rays even on Earth's atmosphere is only now beginning to be realized \citep{Sojka2013, Sojka2014}; X-ray flaring has also recently been seen to affect protostellar disk chemistry, as measured by ALMA,  dramatically \citep{cleeves17}.  Multi-wavelength observations of many stellar flares triggered by \strobex{} will provide valuable and comprehensive constraints on the evolution of planetary and disk irradiation within the first Gyr of formation when stars are rapidly rotating and very magnetically active, while young planets (and their atmospheres) establish their compositions and surface conditions.  

In addition to flare radiation, coronal mass ejections (CMEs) and associated energetic particles (SEPs) are thought to influence the habitability for close-in planets around magnetically active stars \citep{Segura2010}.  Though CMEs and SEPs comprise a large fraction of the energy released in solar eruptive events \citep[SEEs;][]{Emslie2012}, they are the least observationally constrained for other stars. Searches for radio emission produced directly by CMEs so far have turned up null results \citep{Crosley2016,CrosleyOsten2018a,CrosleyOsten2018b}, calling into question the applicability of extrapolating solar flare scaling relations into the stellar regime of more energetic and more frequent flaring \citep{OstenWolk2015}. The idea that transient ejections of matter during the early stages of flare brightening can introduce temporary increases in the hydrogen column density, known as absorption dimming \citep{Mason2014}, has been claimed in at least one large stellar flare \citep{Moschou2017}. The broad bandpass and large collecting area of \strobex{} will enable the routine search for absorption dimming signatures to determine whether magnetic reconnection events on stars are accompanied by significant eruptions of matter as well. The mass of the event can be determined by integrating the differential $N_{H}$ signature over time, and velocities inferred from time-domain analyses. 

Exoplanetary aurorae are also a novel technique to search for exoplanets \citep{AndersonHallinan}.  These searches are currently aided by use of Swift-BAT flares.  The increased sensitivty and the more optimal bandpass of \strobex{} mean it should detect about 100 times as many stellar flares as Swift-BAT \citep{drake15}, making it a far more useful source of triggering information for this type of work.

With the WFM, \strobex{} will detect more than 100 flares per year, and it will be straightforward to slew immediately onto the brightest of these to obtain detailed XRCA, and. in some cases, LAD, light curves and spectra of them to observe how parameters, even including chemical abundances, change \citep{drake15}.  Furthermore, the optimized capabilities mean that rare types of events, such as superflares, have a reasonable chance of showing their first detections.

\subsubsection{Stellar Evolution: Evolution of X-ray Binaries}
%LMXBs written by Tom M, HMXBs mostly by Silas Laycock

The evolution of binary stars is one of the most challenging and
fundamentally important problems in stellar evolution.  {\em STROBE-X} will
aid in our understanding of binary evolution through discovery of a
large number of new X-ray binaries, and can also do excellent
characterization of both the spin period derivative and orbital period
derivative of well-known systems.

Collecting large samples of X-ray binaries and estimating the masses
of the compact objects in them is of vital importance for
understanding supernovae.  A gap has been seen between the heaviest
neutron stars and the lightest stellar mass black holes \citep{Ozel10,Farr11}, but the statistical evidence for this gap is marginal.  If established to be real, it would indicate that the instabilities that
produce the outward shocks that lead to the actual explosions of
supernovae must have growth timescales of $\sim$10-20 milliseconds
after the collapse of the core \citep{Belcz12}, something considered quite unlikely prior to the tentative discoveries of the mass gap.

Because the Wide Field Monitor of {\em STROBE-X} is about an order of
magnitude more sensitive in 2-10 keV X-rays (i.e. hard enough to be
seen through the Galactic Plane) than past instruments and gives
$\sim$ arcminute positions, it should detect the outbursts of many
more objects than past wide field monitors \citep{Arur18}.  Furthermore, with the new-found discovery of a method to infer the radial velocity amplitudes of quiescent binaries from emission lines\citep{casares16}, it is possible to make good mass estimates for objects which are much more heavily reddened than in the past.  

\begin{figure}[t]
\includegraphics[width=3.25in]{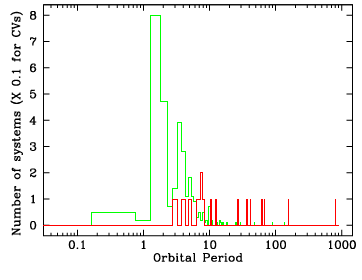}
\includegraphics[width=2.75in]{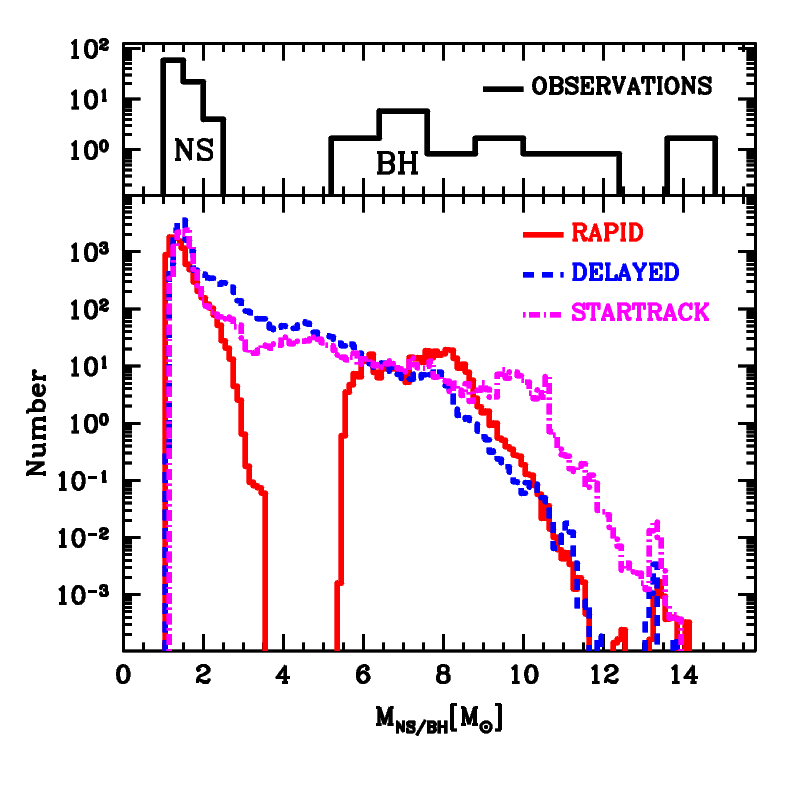}
\caption{\footnotesize{Left: The distribution of orbital periods of low mass X-ray binaries (red) and cataclysmic variables (green), illustrating the selection bias against short period X-ray binaries, from \citep{loftbinev}. Right: The distribution of masses of compact objects plotted along with several models, indicating that the tentative discovery of a gap between the highest mass neutron stars and lowest mass black holes would indicated a rapid supernova mechanism is necessary.  Taken from \citep{Belcz12}.}}
\label{perioddist}
\end{figure}

It is well-established that the orbital period distribution of {\it
  known} X-ray binaries is strongly skewed toward much longer periods
than the orbital period distribution of cataclysmic variable stars (see Figure \ref{perioddist}), despite strong similarities in their expected
evolutionary processes \citep{Patterson84}.  Thus either the
orbital period distribution is giving fundamentally new information
about binary evolution, or, more likely, there are several selection
biases against the shortest period X-ray binaries.

The selection bias is well-explained in terms of the fact that almost
all known X-ray binaries have been discovered in outburst, and their
outburst peak luminosities are strongly correlated with their orbital
periods, so that the X-ray selection favors discovery of longer period
systems.  Typical outbursts of long period systems are already
detectable throughout the Milky Way, even with relatively insensitive
wide field monitors, but the outbursts of much shorter period, and
hence intrinsically faint black hole X-ray binary outbursts will be
detectable with {\em STROBE-X}.  {\em STROBE-X} should thus discover 3-4 of these short period transients per year and about 1 long period transient per
year, meaning it should dramatically expand the sample of black hole
X-ray binaries \citep{loftbinev}.

\paragraph{The spin period distribution and spin period evolution: }

The observed spin-period distribution for low-mass X-ray
binaries is different from that for millisecond radio pulsars \citep{Tauris12}, but it is not yet clear if this is due to spin-down during the pulsar phase or observational selection effects.  {\em STROBE-X} would dramatically enhance the sample of neutron star spin periods in X-ray binaries. Development of a period distribution with enough objects to make clean comparisons between the two sets of objects is one major goal of
period distribution studies.  The other key goal is to determine if a
large sample retains a sharp cutoff at frequencies lower than the
neutron star breakup spin, which would provide support for the idea
that gravitational radiation limits the spin-up of neutron stars
through either r-modes or the presence of small ``mountains'' \citep{Ho2011,JohnsonMcDaniel13}.

{\em STROBE-X} will expand the spin period distribution of X-ray binaries
both by discovering sources too faint to have been discovered with
past wide-field instruments and by making spin measurements of weaker
pulsations than have past missions.  The improved sensitivity of the
wide-field monitor relative to past missions is crucial because about
half the known accreting millisecond pulsars have orbital periods
shorter than 2.5 hours, despite the strong selection effects against
short period transients being detected \citep{Patruno_Watts_2012}.  Given
that these systems are predominantly within a few kpc, we can expect
an increase in the discovery rate by about a factor of 10 by having a wide-field monitor 15 times more sensitive than RXTE's.  

Further, {\em STROBE-X} will have a much greater ability to detect
weak pulsations from accreting neutron stars.  Several X-ray binaries showed intermittent pulsations with \rxte, and theory predicts that intermittent episodes of channeled accretion should take place even onto neutron stars with high accretion rates and low magnetic fields \citep{Romanova_etal_2008}.  The example which shows the most promise for a higher collecting area
instrument is that of Aql~X-1 \citep{Casella_etal_2008} where a pulse
train lasting about 100 seconds showed the spin period already known
from burst oscillations.  The intermittent pulsation showed a
$\approx$ 5\% rms amplitude at 10 keV and an extremely hard spectrum
while Aql X-1 was at a brightness level of about 500 mCrab.  With
{\em STROBE-X}, 100 second pulse trains would be detectable at about 150
mCrab with 0.4\% rms amplitudes, dramatically opening up the parameter
space for their discovery in exposure times short enough that orbital
smearing will be negligible.  

Spin period evolution is also something which has been challenging
with \rxte, and could be done dramatically better with {\em STROBE-X}.  Spin period measurement precision will scale linearly with count rate, so period derivative precision will be about 9 times as good as from \rxte.  Period derivatives from accreting
millisecond pulsars with RXTE were typically near the limits of detection, with different harmonics showing apparently different behavior \citep{Riggio11}, suggesting that an improvement of a factor
of 9 in measurement quality will dramatically increase our ability to
make sense of the data.  Combined with precise luminosity measurements
that will come from a combination of radio and {\it Gaia} parallaxes, measurement of a sample of accretion-powered pulsar
period derivatives will yield a real understanding of the torques on
accreting neutron stars.

\paragraph{High-Mass X-ray Binaries: }

{\em STROBE-X} will open up the parameter space for studies of HMXBs as probes of stellar evolution to inform two important (and independent) questions. Firstly,  HMXBs count among their ranks the leading progenitor-channels for double degenerate (DD) binaries (NS+NS, NS+BH, BH+BH) and hence GW sources from inspiral events. Secondly, the source of energy in the early universe that ended the cosmic dark ages, ushering in in the era of re-ionization, is currently under hot debate. In star-forming galaxies today, HMXBs dominate the production of hard photons, their powerful winds (including associated supernovae and pulsar winds) also supply kinetic energy to the ISM, sweeping out cavities on parsec scales, enabling that ionizing radiation to penetrate throughout their host galaxies. If as the EDGES result implies, the era of re-ionization began too early for AGN, and primeval stars were cocooned in dense material, HMXBs could emerge as the leading ionizer.

Both of these questions hinge on better knowledge of (A) the production rates of HMXBs, (B) the mass distribution of their compact components, and (C) the fate of HMXBs, i.e. whether the further mass transfer evolution is dynamically stable or unstable (the latter leading to a 
common envelope solution). {\em STROBE-X} XRCA %will push our discovery-space for HMXB pulsars out into the Local Group, providing an order of magnitude increase in the number known,
will be able to detect 10\% rms pulsations out to 2 Mpc for $10^{37}$ erg/sec sources in 2000 seconds, opening up the nearby star-forming galaxies to searches and enabling a statistically robust association of HMXB production rate with other tracers of age and environment.  Landmark results on the evolution of HMXB populations \cite{2019arXiv190101237A,2010ApJ...716L.140A,2016MNRAS.459..528A} has stemmed from this type of investigation, and its extension to the youngest star-forming environments \citep[e.g.]{2018ApJS..239...13W,Laycock17} needed to discover how the production rates of the BH-containing DDs develop in the early stages (first few million years) of a starburst. This information will inform binary population synthesis models, and confront the ongoing GW event-rate determinations.  The high sensitivity and subsecond time resolution of {\em STROBE-X} combine to make this possible. 

{\em STROBE-X} will also be able to measure orbital parameters for a wide variety of HMXBs.  Both timing residuals from the accreting pulsars and the orbital modulations from stellar winds will be useful in measuring not just orbital periods, but, crucially, orbital period derivatives.  With the likelihood of new infrared and/or radio facilities that will help provide precise measurements of the stellar wind mass loss rates, the measurements of the orbital period derivatives can then be used to estimate the total system masses.  These can give estimates of accretor masses even in systems where the mass function cannot be measured directly.  This provides strong constraints on the models for the evolution of gravitational wave sources.

\subsubsection{Stellar Evolution: Chemical Composition of Supernova Remnants}
%Written by Katie Auchettl, heavily edited/reduced by Tom M.
Supernova remnants (SNRs) are the long lived structures that result from the explosive end of a massive star (core-collapse), or from the thermonuclear explosion of a white dwarf found in a binary system (Type Ia) in a supernova explosion (SN). The expanding shock-front produced by the SN heats and mixes the metal-rich supernova ejecta and swept-up ISM to X-ray emitting temperatures, while also accelerating electrons and ions to energies approaching $10^{15}$ eV (see reviews e.g., \cite{2008ARA&A..46...89R, 2012A&ARv..20...49V, 2015SSRv..188..187S}). With the advent of high spatial and spectral resolution X-ray instruments such as \textit{ROSAT}, \textit{ASCA}, \chandra, \textit{Suzaku} and \textit{XMM-Newton}, it has been possible to gain insight into various processes associated with these sources. This includes the nucleosynthesis yields of the original star and explosion, ejecta asymmetries, properties of the shock and the X-ray emitting plasma, as well as evidence of significant particle acceleration in the form of X-ray synchrotron radiation. Current observations require $\sim$ day-long exposures, even for Milky Way sources, to be able to obtain sufficient statistics to be able to probe the properties and nature of these sources. As such, we are currently limited to the very brightest and nearby sources found in our Milky Way and Magellanic Clouds, with a number of remnants only marginally detected in X-rays. 

%In particular, Strobe-X will allow us to study a much larger and more complete sample of SNRs in both the Milky Way, and Magellanic Clouds, and even in local group galaxies such as M31 and M33. With very short exposures, one can systematically map these galaxies to search for X-ray emission from known radio SNRs (and other sources) to flux limits less than that currently probed (e.g., current X-ray survey of the LMC has a flux coverage limit of $\sim 10^{-14}$ erg cm$^{-2}$ s$^{-1}$ arcmin$^{-1}$ \citealt{2002astro.ph..3233H, 2016A&A...585A.162M}). This will provide us with a more complete understanding of the general population of SNRs in these Galaxies, which will allow us to probe the properties of these sources as whole which is currently difficult to do. (Removing because I'm not sure this is feasible, and anyway opens up only a small amount of parameter space).

{\em STROBE-X} will obtain the first, detailed CCD quality spectra for a large number of remnants in reasonable exposure times that will allow us to detect emission lines that we can use to constrain ejecta abundances. As both Type Ia and core-collapse remnants naturally produce different yields of both intermediate mass (O, Ne, Mg, Si) and heavy (e.g., Fe, Mn) elements (e.g., \cite{1999ApJ...513..861U, 2002RvMP...74.1015W, 2003ApJ...598.1163M, 2016ApJ...821...38S}), we can begin to characterize the possible progenitor and thus explosion mechanism of these events. In particular, the measurements of K-shell emission from stable Fe-peak elements is an important discriminator between Type Ia and core-collapse remnants \citep[e.g.,][]{2014ApJ...785L..27Y, 2015ApJ...801L..31Y,2015ApJ...803..101P}. With current X-ray instruments, an example of the state of the art with current instrumentation is the spectrum of 3C~397, a 1 mCrab source for which a 70~kilosecond \textit{Suzaku} observation could establish its Type Ia nature (see e.g., \cite{2015ApJ...801L..31Y}). Given the exposure time needed for this work, covering a large sample of Galactic objects is prohibitive in satellite time, and covering any but the brightest Magellanic Cloud remnants is impossible.   The effective area of {\em STROBE-X} (about 100 times that of Suzaku) and excellent coverage of the energy range associated with these Fe-peak elements ($\sim6-10$ keV) will mean we can easily detect these lines and independently characterize the explosive origin of these objects. As such, by being able to observe about 100 of these remnants in the Milky Way and Magellanic Clouds, we can begin to observationally constrain nucleosynthesis models and the influence of e.g., metallicity on these yields while also determining whether the number of remnants detected is consistent with the expected rate of core-collapse and Type Ia's detected from optical SN surveys, and current star formation rate maps of these galaxies.

\subsubsection{Isolated Neutron Stars: Nearby, Thermally Emitting Neutron Stars}
\begin{figure}
\centering
\includegraphics[clip, trim=1.3cm 7.0cm 1.7cm 3.7cm, width=0.4\textwidth]{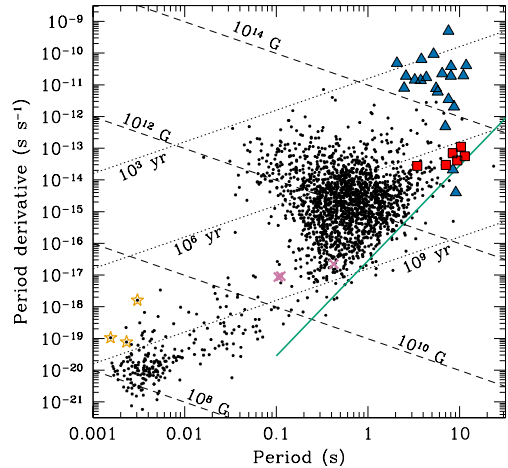}
\caption{The period -- period derivative ($P-\dot{P}$) diagram for pulsars based on data from the ATNF pulsar catalog \citep[v1.59][]{2005AJ....129.1993M}. The purple crosses mark the only three compact central objects (CCOs) with detected periodicities; the red squares show the six of the Magnificent 7 with measured spin properties; the blue triangles show the current sample of magnetars (anomalous X-ray pulsars and soft gamma repeaters). The orange stars mark the three millisecond pulsars discussed in \S\ref{sec:ptas}.  The dotted and dashed lines show constant characteristic age ($\tau\equiv P/2\dot{P}$) and surface dipole magnetic field strength ($B_{\rm surf} \propto \sqrt{P\dot{P}}$), respectively. The green line shows the theoretical pulsar death line from \citet{1992A&A...254..198B} below which pulsars are expected to cease producing radio emission.   \label{fig:ppdotf}}
\end{figure}

 The \rosat\ All-Sky
Survey (RASS; \cite{rbs}) showed that our census of cooling, nearby neutron stars
was incomplete: it contained not just the known cooling (i.e., young) radio pulsars
such as PSR~B0656+14, but also seven isolated neutron stars \citep[the Magnificent Seven or M7;][red squares in Figure~\ref{fig:ppdotf}]{2007Ap&SS.308..181H,2009ASSL..357..141T,2008AIPC..968..129K}.  The M7 are
nearby ($<1\,$kpc), young ($<1\,$Myr) cooling neutron stars with
very soft ($kT<100\,$eV) largely thermal X-ray spectra, long periods ($>3\,$s), faint optical
counterparts, and no  radio emission. These are
interesting both because of their abundance (although not nearly as many as predicted initially; \cite{1993ApJ...403..690B}) and because of the promise
of inferring neutron-star parameters from modeling their thermal
emission \citep[e.g.,][]{2002ApJ...564..981P}.    Large
investments of time with \chandra\ and \xmm\ have confirmed that the
emission is thermal, but we currently understand neither the chemical composition nor state (gaseous, condensed) of the
surface \citep{2007MNRAS.375..821H,kvk09,kvk09b,2014PhyU...57..735P,2017A&A...601A.108H}.

X-ray timing \citep{kvk09,kvk09b,2011ApJ...740L..30K,2008ApJ...673L.163V,2005ApJ...635L..65K,2005ApJ...628L..45K,2017A&A...601A.108H} has shown that the spin-down implies a magnetic field of $\sim 10^{13}$\,G, between those of normal rotation-powered pulsars and magnetars \citep{2017ARA&A..55..261K}.  The M7 have remarkably similar magnetic field
strengths, 1.0 to $3.5\times10^{13}\,$G, and  they
all have characteristic ages of several Myr, substantially in excess
of the $\sim\!0.5\,$Myr inferred from cooling and
kinematics \citep{msh+05,vkk07,kvka07,mph+09,tnhm10}. Both
properties follow naturally if the isolated neutron stars (INSs) initially had much stronger
fields, which decayed.  This was predicted theoretically by
\citet{pmg09}: while for initially weak magnetic fields, field decay leads to
only a factor $\sim\!2$ change that is essentially
unnoticeable \citep{ppm+10}, 
field decay becomes increasingly important for fields above $\sim\!10^{13}\,$G, with predicted final
fields that are always a few$\times 10^{13}\,$G, independent of initial
values.  This is just like we observe for the M7.  Including more
rapid spin-down for an initially stronger magnetic field, one also
recovers the current long spin periods and characteristic ages.  Furthermore, field-decay induced heating
helps explain the observed preponderance \citep{kvk09b,2013MNRAS.434..123V,2015SSRv..191..239P} of sources like the M7 compared
to ``normal'' middle-aged pulsars.

Despite the initial success from \rosat, we have been slow to add to the population of similar sources. Only a single neutron star candidate among the sample of \rosat\ sources has been confirmed -- the nearby 59\,ms X-ray only pulsar PSR J141$2+$7922 (aka Calvera; \cite{2008ApJ...672.1137R,2011MNRAS.410.2428Z}), which is unrelated to the M7.  The rotation-powered pulsar PSR~J0726$-$2612 may represent a source that will evolve into something like the M7 \citep{2011ApJ...743..183S}, and considerable effort has identified several \xmm\ sources that also appear similar, albeit fainter \citep{2012A&A...544A..17P,2015A&A...583A.117P}. 
 The forthcoming \eROSITA\ mission wwill do a significantly deeper soft X-ray survey of the sky and is expected to discover $\sim 100$ thermally emitting neutron stars like the M7 \citep{2018IAUS..337..112P}.  This will expand the need for precision X-ray timing dramatically.   Pulsations will be identifiable for fainter sources in short observations of  $<10$\,ks.  Fully-coherent timing solutions will be more demanding, requiring $\sim 100\,$ks spread over 1--2\,years.  This may not be possible for the full population, but we will take advantage of the larger population to make inferences with limited detailed information about individual sources.

Understanding the distribution of spin-period and magnetic field across the population will allow tests of coupled magneto-thermal evolution \citep{2013MNRAS.434..123V,pmg09,ppm+10} and searches for correlation among intrinsic parameters (such as X-ray temperature) that are hinted at by the current data sets on related populations \citep{2009ApJ...704.1321Z,2011ApJ...740L..30K,2016ApJ...833...59S}, as well as doing detailed studies of individual objects.  We will also be able to search for variability in their timing and spectral properties, which so far has only been seen for one of the M7 \citep{dvvmv04,vdvmv04} and whose origin is still not understood \citep{htdv+06,vkkpm07,hhv+09,2017A&A...601A.108H} but may hint at directly-observed reconfiguration of the magnetic field or even free precession.

\subsubsection{Isolated Neutron Stars: Precision X-ray timing of Pulsars}

\paragraph{Timing in Support of Pulsar Timing Arrays:}\label{sec:ptas}
Pulsar timing arrays (PTAs; \cite{2015arXiv151107869B,2015RPPh...78l4901L}) employ long-term precision pulse timing of an ensemble of millisecond pulsars (MSPs) at radio wavelengths in an effort to detect gravitational waves with frequencies $\sim 10^{-9}$\,Hz that are expected to be produced by supermassive black hole binaries starting long before the system mergers \citep{2001astro.ph..8028P,2003ApJ...583..616J,2013MNRAS.433L...1S}, among other more exotic sources \citep{2016APS..APR.K5003B}. A major hindrance towards improved sensitivity of PTAs arises from propagation effects of the radio emission through the interstellar medium (ISM), especially non-deterministic variations in dispersion on long time scales and scattering \citep[e.g.,][]{2017MNRAS.468.1474L,2016ApJ...821...66L,2016ApJ...818..166L}. These deleterious effects cannot be easily disentangled from intrinsic timing noise of the pulsars under consideration or from timing perturbations caused by gravitational waves. In contrast, X-ray observations do not suffer from these ISM propagation effects.

The feasibility of precision X-ray timing for determining the long-term stability of three rotation-powered MSPs exhibiting sharp non-thermal pulsations desirable for precision timing (PSRs J0218+\linebreak[0]{}4232, B1821$-$24, and B1937$+$21; orange stars in Figure~\ref{fig:ppdotf}) was recently demonstrated using monitoring observations with \nicer{}  \cite{Deneva2019}. Based on the $\sigma_z$ measure of timing stability (which uses the average of the cubic coefficients of polynomial fits to subsets of timing residuals; see \cite{1997A&A...326..924M} for details), \nicer\ is so far achieving timing stabilities of $\sigma_z \approx 3 \times 10^{-14}$ for PSR B1937+21 and $\sim 10^{-12}$ for PSRs B1821$-$24 and J0218+4232. Within the span of the \nicer\ X-ray timing data (1 year for PSRs J0218$+$4232 and B1937$+$21, and 9 months for PSR B1821$-$24), there is no break point in the slope of $\sigma_z$; such a break would indicate that further improvement in the cumulative root-mean-square (RMS) timing residual is limited by timing noise. Such a  break point is seen in the comparison radio timing data for PSR~B1821$-$24 and PSR~B1937+21 on time scales of $> 2$~years. 

 At present, generating X-ray pulse time of arrival measurements (TOAs) with $\le 1$\,$\mu$s uncertainty  needed for use in PTAs requires exposure times of $\sim$150~ks for PSR~B1821$-$24, $\sim$50~ks for PSR~B1937$+$21, and $\sim$1 Ms for PSR~J0218+4232 with \nicer. The cadence and uncertainty of X-ray TOAs can be substantially improved with the large increase of effective area for \strobex\ compared to \nicer\ with a modest exposure requirement (few to tens of ks). Therefore, long-term precision timing with \strobex\ would complement radio PTAs by providing the means to evaluate and mitigate the effects of red noise on radio PTA data sets, as well as by producing X-ray TOAs on a subset of the PTA MSPs that may be used in conjunction with radio TOAs. 

\paragraph{Timing in Support of Targeted Gravitational Wave Searches:}

%\begin{itemize}
%\item getting and maintaining precise X-ray pulsar ephemerides for NSs that are promising %sources of continuous gravitational waves for LIGO CW  searches (include a Table of targets?):
%\end{itemize}

  GWs from isolated neutron stars are generated when there is a quadrupolar mass deformation or fluid oscillation, and stars with significant such asymmetry and faster spin frequencies are a stronger source of GWs \citep{Owen_etal_1998,2017arXiv170907049G,2017MPLA...3230035R}.  %The most sensitive GW searches are targeted searches of known pulsars because their accurately measured position and timing properties greatly reduce the parameter space that needs to be examined and allow phase-coherent integrations over long periods of time.  
  Searches of known pulsars can be more sensitive than blind searches because the knowlege of the position and spin-period greatly reduces the parameter space over the search must be conducted.
  Thus far, searches of known pulsars rely on timing information from primarily radio telescopes and the \textit{Fermi Gamma-ray Space Telescope}, and these searches obtained constraints on GW emission for about 200 radio and $\gamma$-ray pulsars \citep{2014ApJ...785..119A,2017ApJ...839...12A}.  However there are a number of pulsars whose timing properties can only be determined from X-ray measurements, due to these neutron stars being intrinsically radio and gamma-ray quiet, the pulsar beams of radiation not being visible from Earth, or strong interstellar dispersion at radio wavelengths.  Two examples of pulsars that are only seen in X-rays are PSR~J0537$-$6910 (also known as the Big Glitcher) and PSR~J1412+7922 (Calvera).  PSR~J0537$-$6910 is an excellent target for a monitoring program with \textit{STROBE-X}, as its frequent glitches ($\sim 3.5$ per year) demand regular observations to maintain timing precision and its (almost) glitch predictability offers the possibility of observing a glitch as it occurs; it is also a particularly interesting pulsar because its timing properties suggest that it could be a strong GW emitter \citep{2018ApJ...864..137A}.  Meanwhile PSR~J1412+7922 could be relatively young and is nearby \citep{2013ApJ...778..120H}, making it a good target for LIGO.  Conversely, if all-sky searches by LIGO detect a previously unknown neutron star, \strobex{} is also ideal for follow-up, given that the XRCA will be, by far, the most sensitive X-ray instrument for detecting pulsations and because the gravitational wave searches will be nearly independent of the pulsar beam orientation, so the energy band with the widest opening angle is optimal for searches.\footnote{The positional uncertainty of a continuous wave search detection from LIGO should be significantly smaller than the collimator opening angle for XRCA because the continuous wave detections take advantage of the motion of LIGO due to the Earth's orbit and have a baseline of 2 AU.}

\subsubsection{Cosmology and Galaxy Evolution: Spectroscopy of Clusters and Groups}

The XRCA is particularly well-suited to the study of high redshift ($z>1$) galaxy clusters, which are now routinely being discovered by SZ surveys such as the South Pole Telescope (SPT) and the Atacama Cosmology Telescope (ACT). Due to their immense mass, galaxy clusters provide a fairly accurate accounting of the universal allocation of baryons. Studies at low and intermediate redshift have found that the metallicity of the diffuse intracluster medium, which has a temperature of $>$10$^7$\,K and glows in X-rays, is roughly a third solar, with no appreciable dependence on cluster mass or redshift, and no measurable cluster-to-cluster scatter \cite{2016ApJ...826..124M,2017MNRAS.472.2877M,2018SSRv..214..129M}. 
This universal metallicity of the hot, diffuse universe suggests a rapid enrichment at early times ($z\sim2$), perhaps during the peak of star formation. 
%With existing X-ray telescopes, measurement of the X-ray metallicity in clusters at $z>1.5$ is exceptionally challenging -- as an example, we required $\sim$600\,ks with \emph{XMM-Newton} to constrain the metallicity of the most distant SZ-selected cluster ($z=1.75$) to better than 30\% accuracy. 
Based on the proposed sensitivity and expected backgrounds of the XRCA, we expect that we will be able to constrain the metallicity of the intracluster medium for the most massive (M$_{500} \gtrsim 1.5\times10^{14}$ M$_{\odot}$) clusters at $z>1.5$ to better than 30\% accuracy in $\sim$100\,ks with\strobex. Such a measurement for a sample of $\sim$10 massive clusters at $z\sim2$ would provide our first look at the metallicity of the diffuse intracluster medium during the initial formation and collapse of these systems.

\begin{figure*}[ht]
\begin{tabular}{c c}
\centering
\includegraphics[width=2.6in, trim=0cm -3.5cm 0cm 0cm]{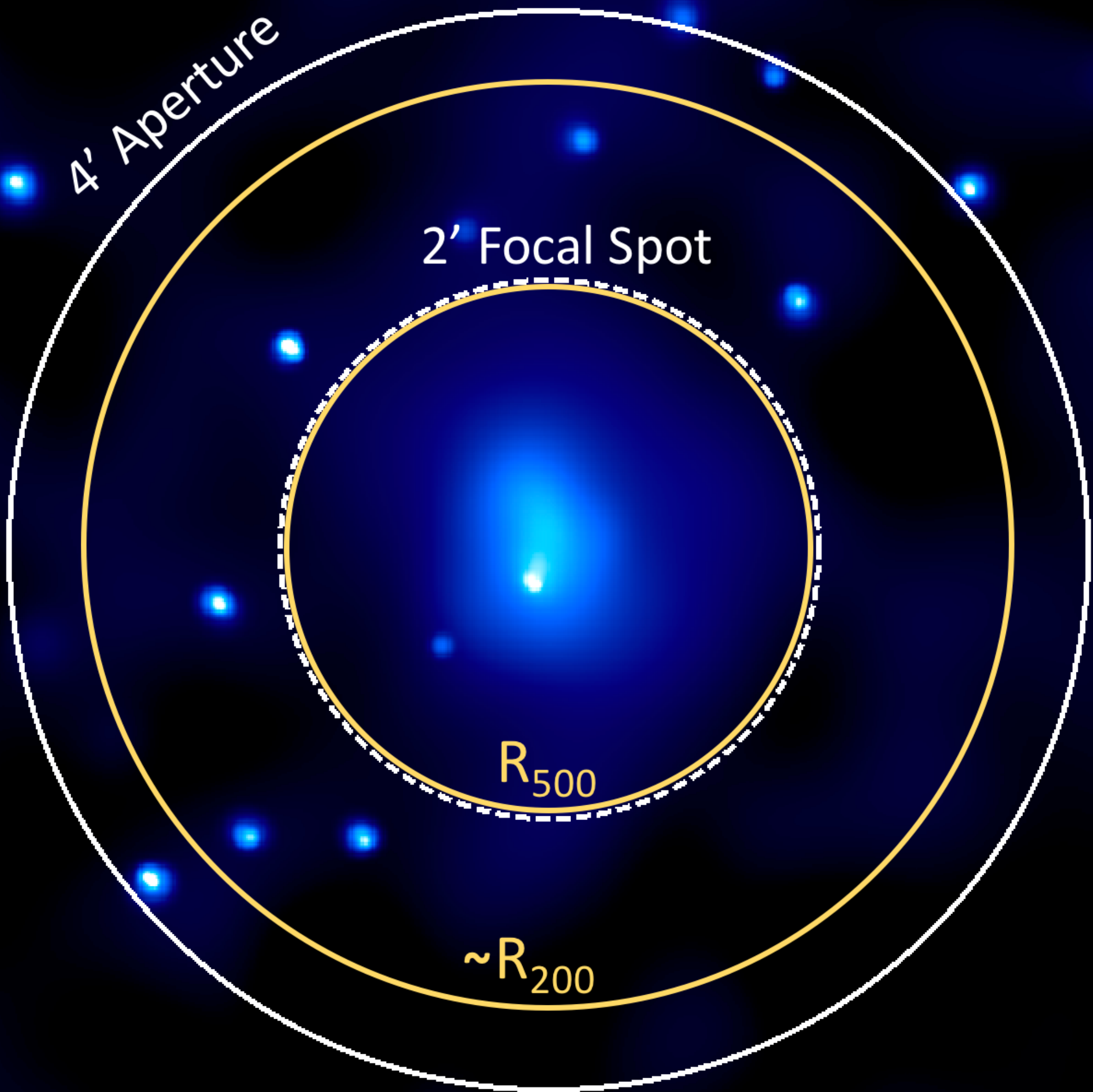} &
\includegraphics[width=3.8in]{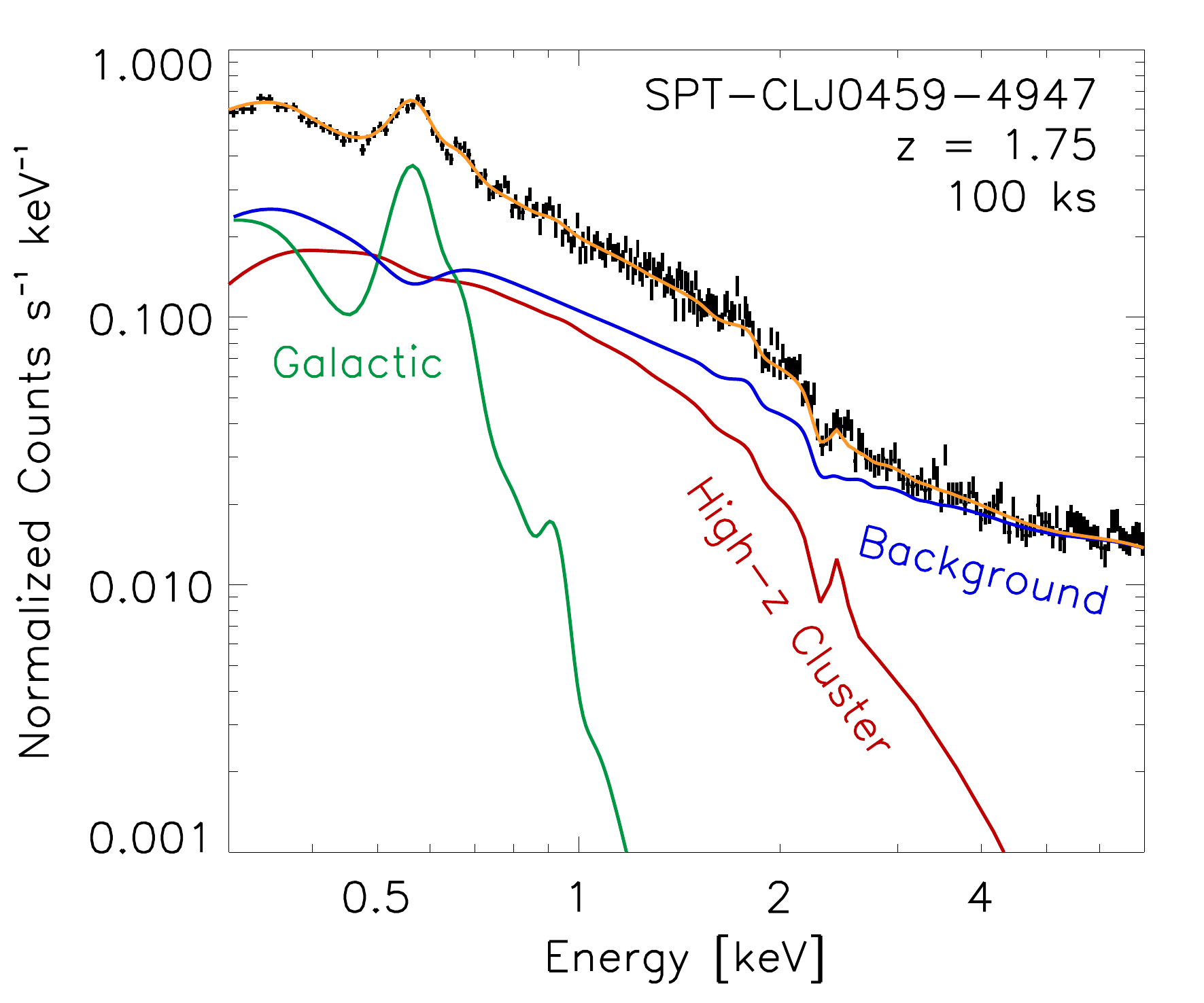}
\end{tabular}
\caption{\emph{Left:} Chandra ACIS-I image of SPT-CLJ0459-4947, a massive galaxy cluster at $z=1.75$. This is the most distant galaxy cluster for which the intracluster medium has been robustly mapped in the X-ray. This 100\,ks exposure yields an accurate surface brightness (electron density) profile, but is of insufficient depth to constrain the temperature or metallicity. In yellow, we show the estimates of R$_{500}$ and R$_{200}$, along with the size of the XRCA focal spot and aperture, which are well-matched to the size of a galaxy cluster at $z>1$. \emph{Right:} Simulated X-ray spectrum of SPT-CLJ0459-4947 with the XRCA. The emission from the high-$z$ cluster is at a similar surface brightness to the emission from our own galaxy in the soft X-rays, and is sub-dominant to the cosmic X-ray background at hard energies. Despite this, with exposures of $\sim$100\,ks, we expect to be able to constrain the metallicity to better than 30\% in clusters at $z>1.5$, providing some of the best constraints on the enrichment history of the diffuse universe.
\label{fig:highz-cluster}}
\end{figure*}

\subsubsection{Cosmology and Galaxy Evolution: An All-Sky Medium Energy X-ray Survey}

%Written by Tom M.
 \strobex{} would also complete an all-sky survey in the medium-energy X-rays that would surpass the current standard, the {\em RXTE} Slew Survey.  The \strobex{} survey would also fill in the gaps in the {\em RXTE} survey's coverage and, importantly, would improve the spectral resolution of that survey by a factor of $\sim5$.  In addition to providing sensitivity to highly obscured Galactic sources of moderate brightness, this survey would exceed the sensitivity of {\em eROSITA} to Compton-thick AGN by taking advantage of  \strobex's superior collecting area at 6-7 keV, identifying the Compton-thick AGN using their very high equivalent width iron emission lines \citep{Matt1996}, and taking advantage of the expected 50 Msec exposure time.  Furthermore, it would allow the discovery of the highest power, highest redshift blazars, as these objects would be detected by the {\em STROBE-X}/WFM but not by {\em Swift}/BAT in the same way that $z\sim4$ blazars are detectable by {\em Swift}/BAT but not by {\em Fermi}/LAT \cite{2011MNRAS.411..901G}.

%{\em STROBE-X} data collected in conjunction with data from other facilities can be used for multi-wavelength timing which can help illuminate the details of the disk-jet connection\cite{2015arXiv150102770D} \cite{2015arXiv150102766C}.  Jetted tidal disruption events (TDEs) can be discovered with {\em STROBE-X}\cite{2015arXiv150102774R}, and the details of the accretion flows in TDEs can be accurately characterized.  

%\paragraph{Abundance Measurements from High-Throughput Spectroscopy}
%{\it Clusters \& Groups -- McDonald}
%{\it Coordinate other stuff -- Maccarone -- Magellanic Clouds could go under stellar evolution, solar system not really in our decadal, so mention, but just a paragraph}

%
%[expand this, also look into whether we can do cluster work after all]

%\subsubsection{Nuclear and Particle Physics: X-ray Bursts}
%{\it Deepto to lead this one, David Ballantyne to help}

%Its observations of long-duration nuclear bursts, especially given its capability to obtain XRCA and LAD data on the tail ends of these bursts, will provide empirical constraints on the properties of nuclear reactions of proton-rich isotopes, which is one of the central unanswered questions in nuclear physics \cite{2017ApJ...844..139S}. 

\subsubsection{Nuclear and Particle Physics: Axion Searches}
%Written mostly by Chanda Prescod-Weinstein

\begin{figure}
\centering
\includegraphics[width = .75\linewidth]{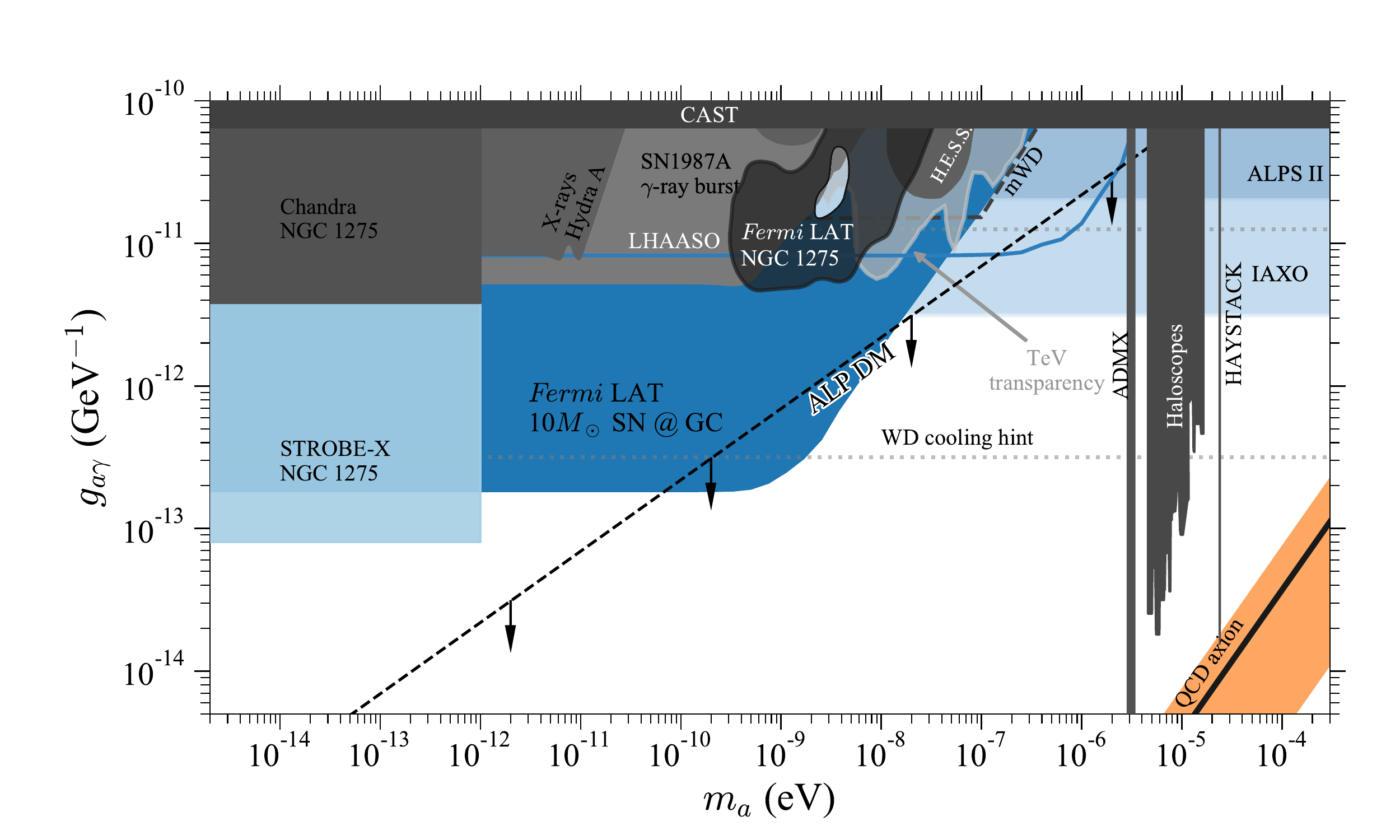}
\caption{\label{fig:photon-ALP}Limits, future experimental sensitivities, and theoretically preferred regions of the axion and ALP coupling to photons as a function of mass~\cite[see, e.g.,][]{Irastorza2018}. 
Sensitivity projections for future experiments, including {\em STROBE-X}, are shown in blue, current limits are shown in gray. The QCD axion band is shown in orange.}
\end{figure}

The quantum chromodynamics (QCD) axion is a hypothetical particle produced in the Peccei-Quinn mechanism, which resolves an open problem in the Standard Model of particle physics~\citep{PhysRevLett.38.1440}. The Peccei-Quinn mechanism allows for production of axions non-thermally in the early universe~\citep{DINE1981199}. Astrophysically, the QCD axion is of interest as a potential dark matter candidate because the non-thermally produced axion is potentially abundant and cold~\citep{PRESKILL1983127}, and it may have distinct phenomenology from WIMP candidates, see, e.g.~\cite{PhysRevD.92.103513}. QCD axions are part of a larger class of spin-0 dark matter candidates that are often referred to as ``axion-like particles'' (ALPs). ALPs arise naturally in beyond Standard Model particle physics, such as string theory~\citep{MARSH20161}. 

X-ray observations with {\em STROBE-X} will be able to probe axions and axion-like particles (ALPs) in two possible ways. The first is via photon-ALP conversion in galaxy cluster magnetic fields -- due to a mild axion-electromagnetic coupling -- which will leave distinctive oscillatory features in spectra of active galactic nuclei located at the cluster's center. This interaction generates quasi-sinusoidal fluctuations in the spectra of dark matter dominated objects \citep{MARSH20161,marsh2017}. This interaction is particularly efficient on megaparsec scales, and in the presence of strong magnetic fields making clusters prime targets for ALP searches from X-ray to $\gamma$-ray energies \citep{Abramowski:2013oea,TheFermi-LAT:2016zue,conlon2016, conlon2017, conlon2018}. 

{\em STROBE-X} observations of bright active galactic nuclei in the central regions of nearby clusters can be used to search for ALP oscillatory modulations and constrain the mass and coupling of ALPs. 250\,ks {\em STROBE-X} observations of NGC1275 in the center of the Perseus cluster will improve the current limits provided by the \chandra{} observations \citep{berg2017} over a factor of 5 as shown in Figure \ref{fig:photon-ALP}, assuming a similar energy resolution of {\em STROBE-X} and \textit{Chandra} and that the limits scale as the signal-to-noise ratio of the spectrum. 

On the other hand, axions (or ALPs) produced in the interior of stars get trapped in neutron stars after the supernova explosions and provide an additional cooling though a neutrino-like axion pair-breaking process, offering a potential resolution to the discrepancy between theory and observations of neutron star temperatures~\citep{Nakagawa1988,Iwamoto2001,Kolomeitsev2008,Keller2013,PhysRevD.93.065044}. The axion cooling would change the observed surface temperature of neutron stars measured from X-ray spectral observations  \citep{PhysRevD.93.065044,Blaschke2012,2018arXiv180607151H}. X-ray monitoring of isolated neutron stars over their evolution tracks provides a unique avenue to test and constrain axion cooling models. {\em STROBE-X} monitoring of isolated neutron stars observed with \nicer{} will probe a time scale $>14$ years in neutron star surface temperature evolution.  Furthermore, thermal axion production in strongly magnetized environments is possible, for example via conversion of thermal photons on the surface of a neutron star in the magnetic field of the star \citep{2012ApJ...748..116P}.

\subsection{Science Traceability Matrix }

\begin{figure}
    \centering
    \includegraphics[width=5.5in]{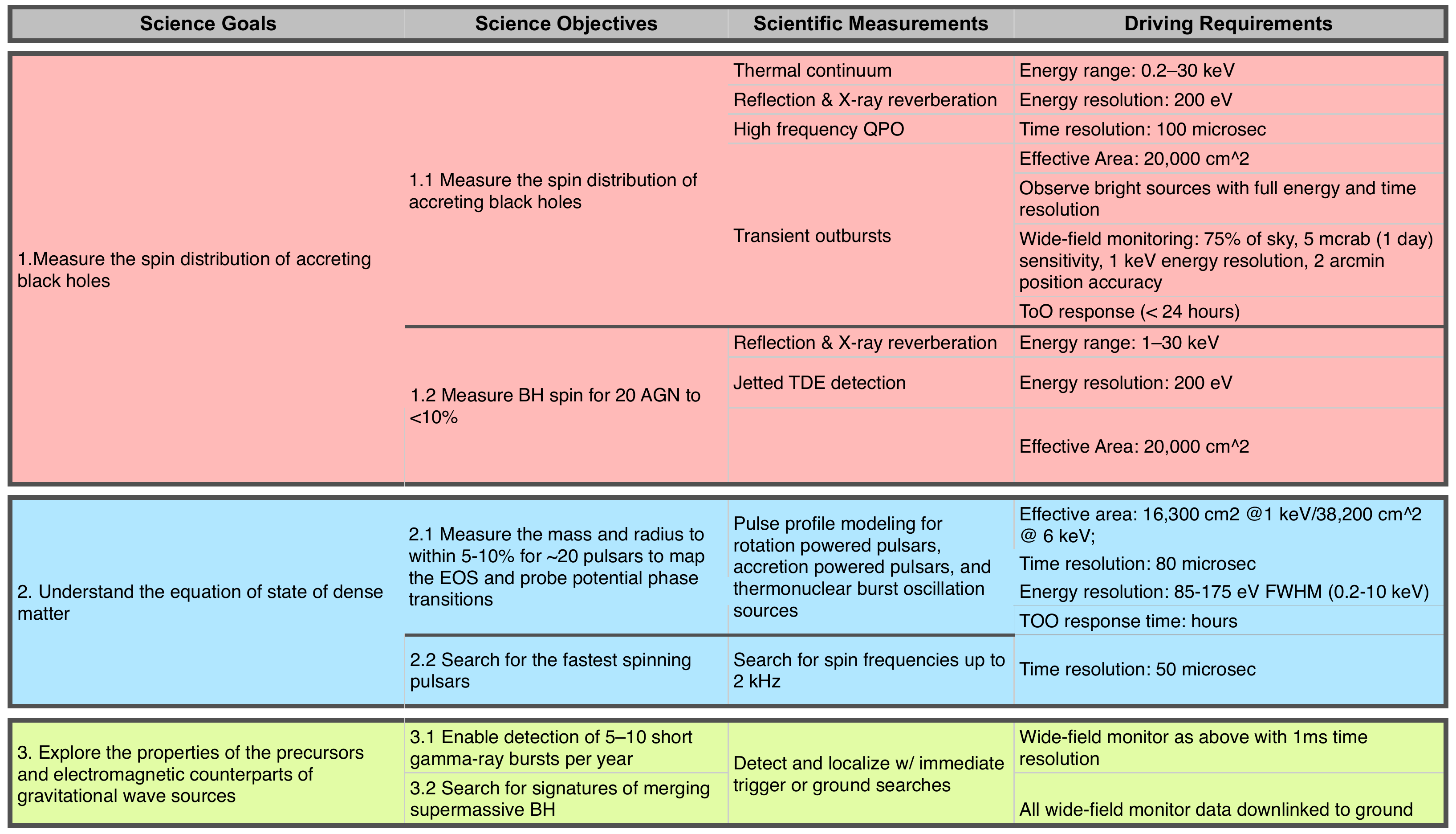}
    \caption{Science traceability matrix}
    \label{fig:traceability}
\end{figure}

While \strobex{} will enable a broad science program that will 
have many impacts across the astrophysics community, the design
of the mission and instruments is fundamentally driven by 
a flowdown of requirements from the key science projects.
The traceability from measurement objectives to performance requirements 
is displayed in Figure~\ref{fig:traceability}.

\section{Instruments}

The {\em STROBE-X} instrument suite has a strong heritage from the\nicer{} mission in the U.S. and the {\em LOFT} mission concept that 
has been under study for many years in Europe. The team has created detailed designs and prepared thorough cost estimates
during a study at the NASA/GSFC Instrument Design Lab (IDL) in 2017 November and December. A major result of this study was the
division of the primary instrument into four identical ``quadrants,'' each with a composite optical bench for the XRCA and a 
deployable panel for the LAD. This design has several advantages: Firstly, integration and test flow is simplified, can incorporate 
parallelism, is reduced in cost, and 
requires smaller facilities than if the instruments were a monolithic unit.  Second, system reliability is improved because of the 
modularity that allows any one quadrant to fail without bringing the observatory capabilities below the science requirements.  Finally, 
the composite optical bench has reduced mass, increased stiffness and a reduced coefficient of thermal expansion relative to 
earlier aluminum structural designs.  We describe the individual instruments in the sections below.

\begin{figure}
\includegraphics[width=3.25in]{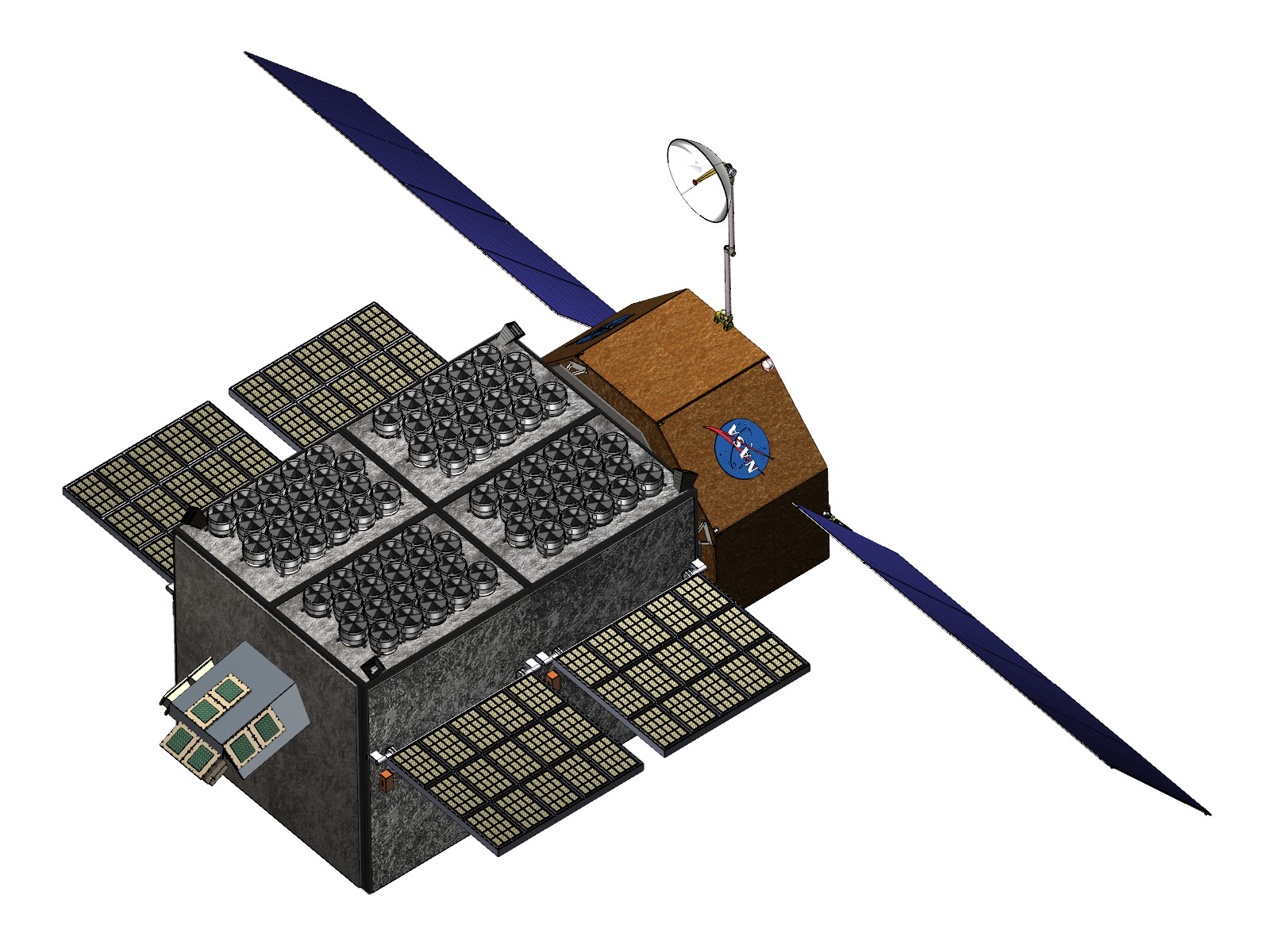}
\includegraphics[width=3.25in]{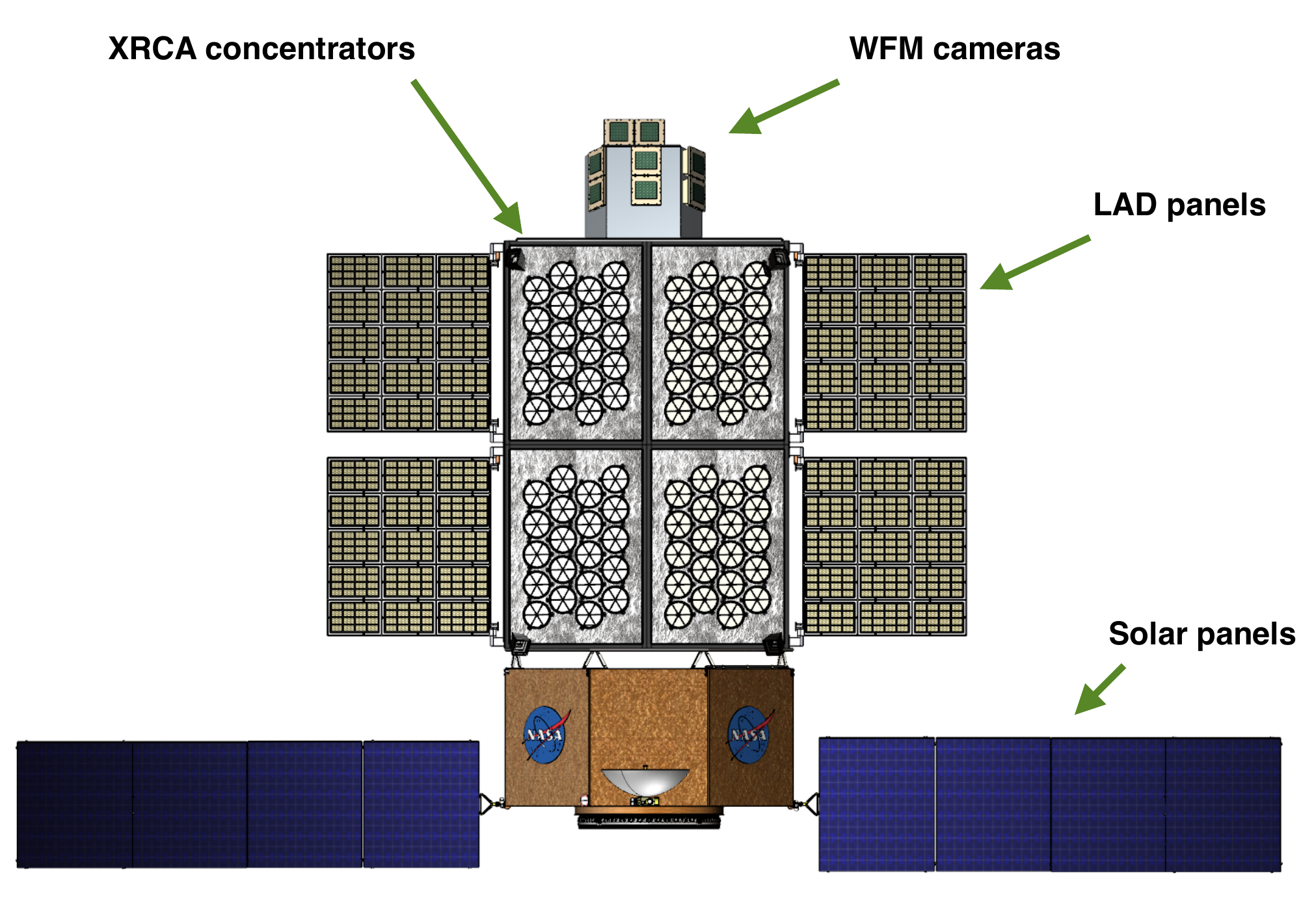}
\caption{Detailed design renderings of the {\em STROBE-X} mission from the NASA/GSFC Instrument Design Laboratory (IDL) and Mission Design Laboratory (MDL).\label{fig:strobex}}
\vspace*{0.1in}
\end{figure}

\subsection{X-ray Concentrator Array (XRCA)}

\begin{figure}
\includegraphics[width=3.0in]{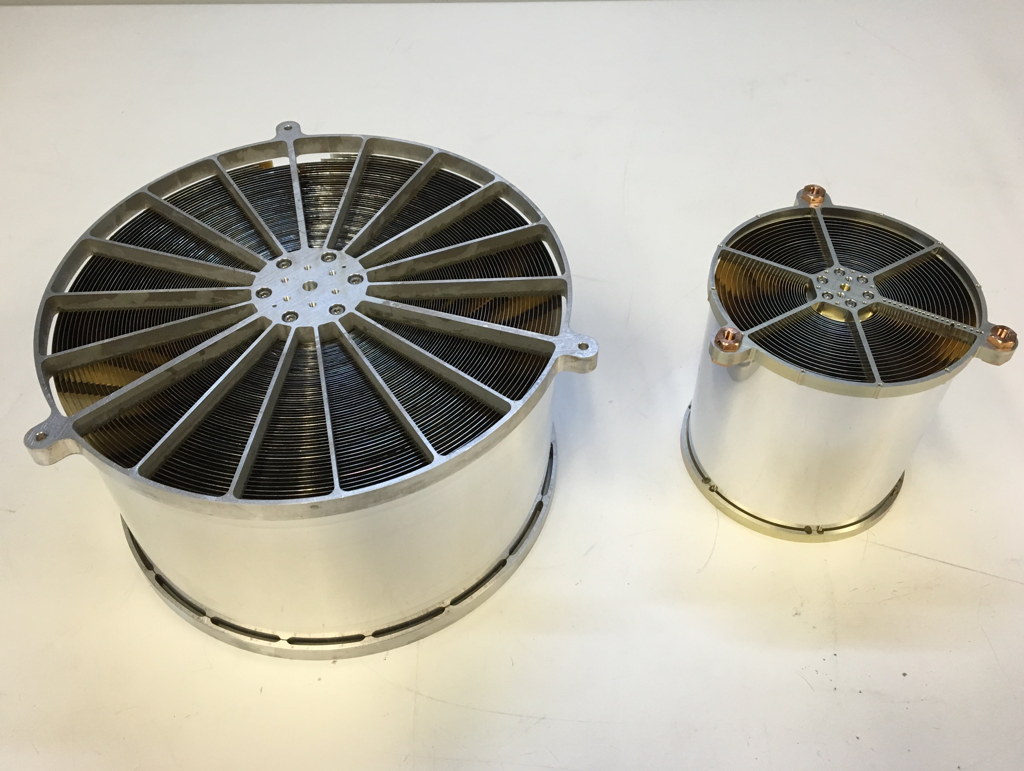}
\hfill\includegraphics[width=3.5in]{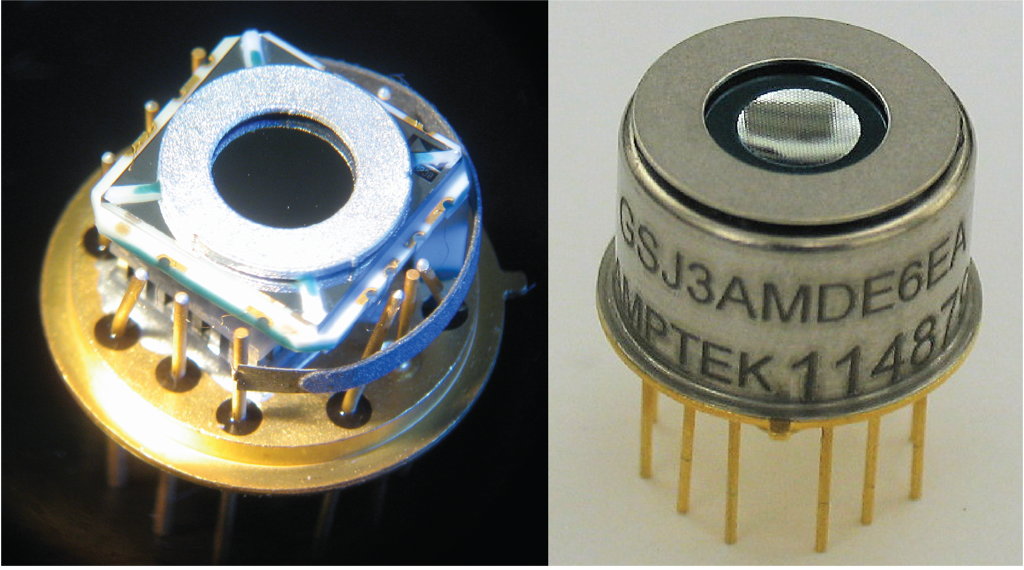}
\caption{Optics and detector technology for {\em STROBE-X}/XRCA. Left: X-ray concentrators from \textit{XACT} (larger) and \textit{NICER} (smaller). \strobex{} would use optics very similar to \textit{XACT}, so no new technology development is required. Right: Si drift detectors from \textit{NICER}, which meet all the requirements for \strobex{}.\label{fig:XRCAopt}}
\end{figure}

\strobex\ covers the soft X-ray (0.2--12 keV) band with the XRCA instrument, a modular
collection of identical X-ray ``concentrator'' (XRC) units that
leverage the successful design and development efforts associated
with GSFC's X-ray Advanced Concepts Testbed (\textit{XACT}) sounding-rocket
payload \cite{doi:10.1117/12.926152} and the \nicer{} mission of opportunity
\cite{doi:10.1117/12.2231304,doi:10.1117/12.2234436}. Flight-like
builds of the \textit{XACT} and \nicer{} XRCs are pictured in
Fig.~\ref{fig:XRCAopt} (left). 

A concentrator is a high-throughput optic that is optimized for
collecting photons from a point-like (less than $\sim 2$ arcmin in
extent) source over a large geometric area and delivering them onto
small detectors. With reduced detector size, particle interactions
that mimic cosmic X-ray detections are minimized, reducing
background by orders of magnitude. 

The design of the XRCA's concentrators benefits from technological 
heritage and manufacturing experience spanning decades of X-ray
astronomy missions, including \textit{BBXRT}, the \textit{SXS}
sounding-rocket payload, \textit{ASCA}, \textit{Suzaku},
\textit{InFocus}, \textit{Hitomi}, and \textit{NICER}. These optics focus X-rays
using grazing-incidence reflections. The individual optical elements
are nested aluminum-foil shells that are inexpensive to fabricate
and provide high reflectivity in soft X-rays by virtue of a smooth
($\sim 5$\,\AA\ roughness) replicated gold surface. The diameters
and paraboloidal figures of the shells are chosen so that graze
angles do not exceed approximately $2^\circ$.

As implemented on \textit{XACT} and \textit{NICER}, XRCs depart from the
GSFC imaging optics of earlier missions in two ways that enhance
their throughput and point-spread function (PSF) performance, with
no added risk. First, they are formed to have paraboloidal figure
instead of conical approximations, an enhancement that significantly
improves their PSF and vignetting characteristics, with the
attendant benefit of reduced background as detector apertures can be
made smaller. Second, because imaging is not a requirement,
concentrators eliminate the secondary mirrors that are needed for
true imaging optics. 
%Concentrators, optimized for sub-arcminute sources, use only a single reflection to deliver the photons from sky to detector. 
Thus, 1) they only suffer reflection inefficiencies once, resulting
in enhanced effective area; 2) the number of optical elements
required is half that of an imaging configuration, resulting in
substantial cost and schedule savings; and 3) with no need to align
primary and secondary optics, integration is significantly
simplified.

The \strobex\ XRCA is four quadrants of 20 identical concentrator units each (i.e., 80 units total), which are scaled-up versions of the 56 \textit{NICER} X-ray Timing Instrument
concentrators. Each XRC has a focal length of 3.0 m, with a set of
107 nested foil shells spanning a range of diameters between 3 cm
and 28 cm. The foils are held in place by a spoked-wheel (or
``spider'') structure with mount points that are used to adjust the
XRC alignment in tip and tilt. 

The \strobex\ XRC design retains approximately the same
focal ratio as \textit{NICER}'s optics, so that $\sim 2$ arcmin on-axis focal
spots are again achieved, while the longer \strobex\ focal length
enhances throughput at energies above 2.5 keV.  Detectors will be
masked with apertures corresponding to a 4 arcmin diameter FOV, to
fully capture the PSF while minimizing diffuse sky background
focused into the aperture.

In its baseline configuration, the \strobex\ XRCA adopts \textit{NICER}'s 
silicon-drift detectors (SDDs) and readout 
electronics \cite{doi:10.1117/12.2231718}.  Within these SDDs, a 
radial electric field guides ionization charge clouds from a large (25 
mm$^2$) area to a central low-capacitance readout anode; the resulting 
charge pulse is amplified and shaped to enable measurement of its 
height and the time at which it triggers digitization and further 
processing. Built-in thermoelectric coolers (TECs) maintain each 
detector at $-55^\circ$ C to minimize thermal electron 
noise. The \strobex\ SDDs offer CCD-like energy resolution, 85 to 175 
eV FWHM over the 0.2--10 keV range. They also enable
%, with appropriate readout design, 
very precise photon detection time-stamping, well 
under 100 ns RMS. The detectors are thick enough to 
provide $\sim 50$\% quantum efficiency at 15 keV, and are packaged 
with an aluminized thin-film window that offers good transparency to 
photon energies as low as 0.2 keV while maintaining a hermetic 
seal.
%---the overall XRCA effective area shown in Fig.~3 assumes 
%NICER-like window thicknesses of thermal and optical-blocking filters 
%for both the detector and the XRC units.

The XRCA detector architecture consists of analog electronics 
including a charge-sensitive preamplifier 
%integrated into the detector's mechanical supporting structure (the Focal Plane Module, FPM),
integrated within the SDD assemblies,
and power, TEC control, and digital electronics  
%support multiple detectors---e.g., groups of 8---simultaneously (the Measurement/Power Unit, MPU). The MPU
that communicate X-ray event data (arrival time, photon energy, 
and event quality) to the observatory's command and data-handling 
system for multiple detectors---e.g., groups of eight---simultaneously. 
The XRCA is capable of 
supporting high photon count rates by virtue of both its modularity and 
the fast readout characteristics of the individual SDD channels: even 
with the seven-fold increase in collecting area of a single \strobex\ XRC 
relative to \textit{NICER}, an SDD module built to \textit{NICER} specifications would 
not be affected by pile-up for incident fluxes below 2 crab, and this 
performance can be improved further with modest changes to the
readout electronics.

\begin{table}
\caption{Parameters for the LAD and XRCA instruments\label{fig:ladxrca}}
\includegraphics[width=6.5in]{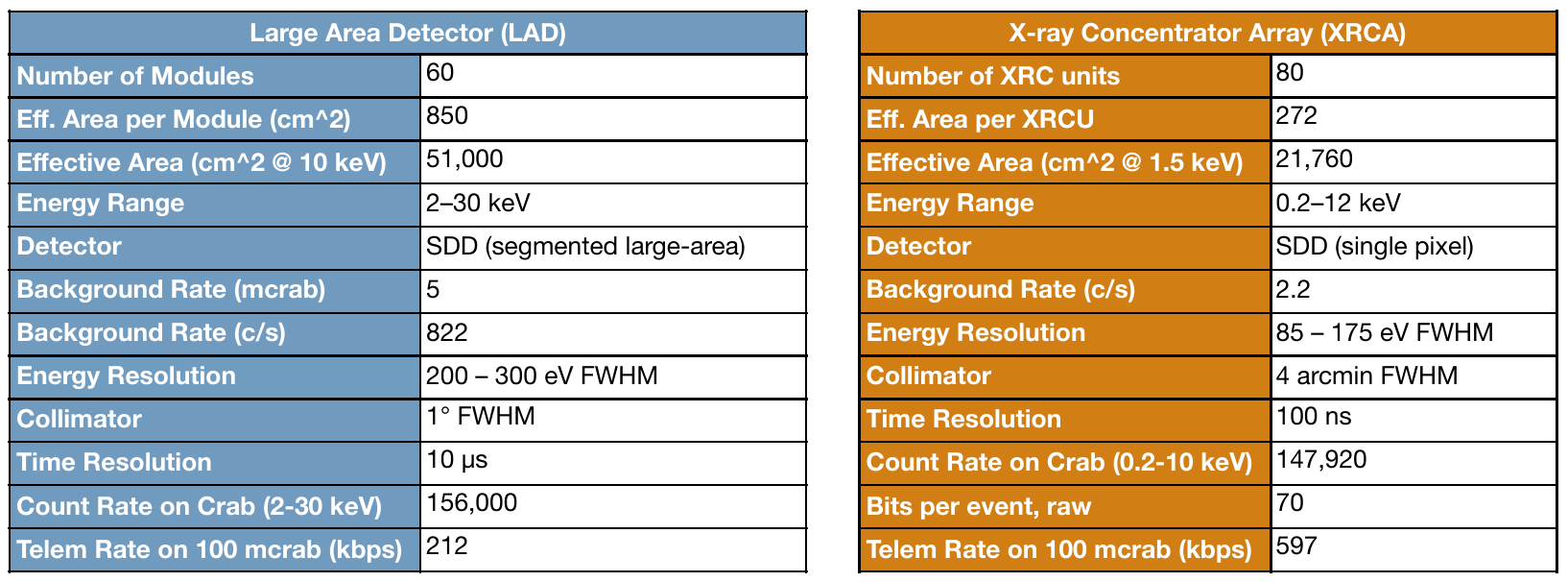}
\end{table}

%\fixme{Is this number right? I thought the current
%electronics were worse than this. Should we say that switching to an FPGA would remove this issue?}

\subsection{Large Area Detector (LAD)}

The Large Area Detector (LAD) is a large-area, collimated instrument, operating in the 2--30 keV
nominal energy range. The instrument is based on the technologies of the large-area Silicon Drift 
Detectors (SDD) and capillary plate collimators, enabling several square meters to be deployed in space
within reasonable mass, volume and power budgets. The concept and design of the LAD instrument is 
based on the same instrument proposed as part of the scientific payload of the {\em LOFT} mission concept 
\cite{LOFTExpA,Zane_etal_2014}. 

\begin{figure}
\centering
\includegraphics[width=4.0in]{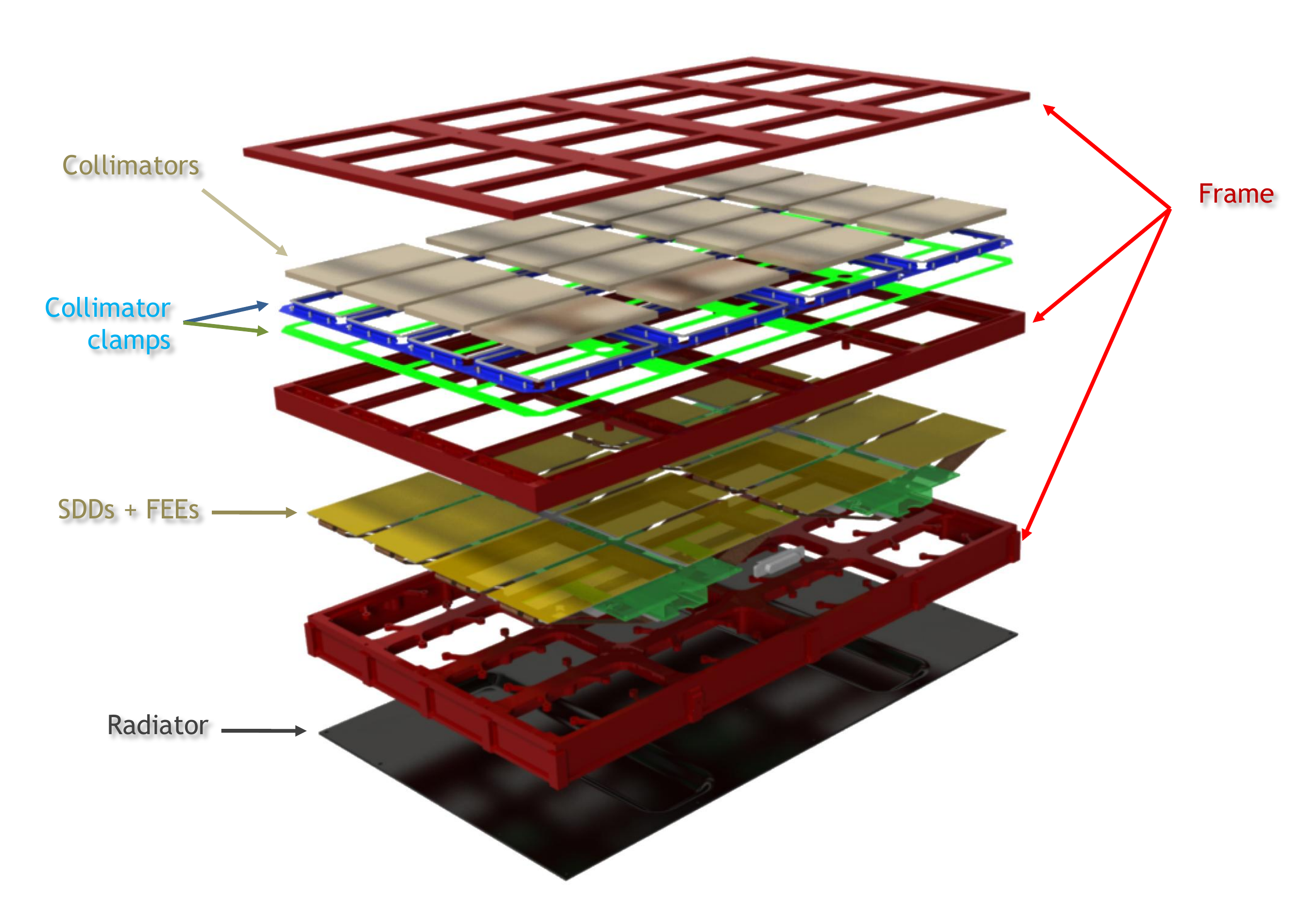}
\caption{An exploded view of the {\em STROBE-X}/LAD module frame design, based on the design for the ESA {\em LOFT} mission concept.
\label{fig:module_with_legend}}
\end{figure}

The LAD's unprecedented collecting area is achieved through a modular and intrinsically highly redundant design. Each LAD Module hosts a set of 4~$\times$~4 detectors with their front-end electronics  and 4~$\times$~4 collimators, supported by two grid-like frames. The back-side of the Module
hosts the Module Back-End Electronics (MBEE). It controls SDDs, FEEs and Power Supply Unit (PSU), 
reads-out the FEE digitized events, generates housekeeping and ratemeters, formats and time-stamps 
each event and transmits it to the Panel Back End Electronics (PBEE). A 300 $\mu$m Pb back-shield 
and a 2 mm Al radiator complete the Module structure, with the tasks of reducing the background events and dissipating heat from Module box, respectively. An exploded view of the LAD Module and its components is shown in Figure \ref{fig:module_with_legend}.

The LAD Modules are organized in 4 large Panels, deployable from each of the {\em STROBE-X} 
quadrants (see {Fig.~\ref{fig:strobex}}). The total effective area is about 5m$^{2}$ at 8 keV.
Each of the LAD Panels hosts 15 (5~$\times$~3) Modules, for a total of 60 Modules or 960 detectors, 
and a PBEE, for a total of 4 PBEEs, in charge of interfacing the 15 Modules to the central 
Instrument Control Unit (ICU). The main parameters of the LAD are listed in Table~\ref{fig:ladxrca}.

\begin{figure}[t!]
%%%%%% Insert Figure %%%%%%%%%%%
%\vspace*{0.3in}
\centering\includegraphics[width=4.0in]{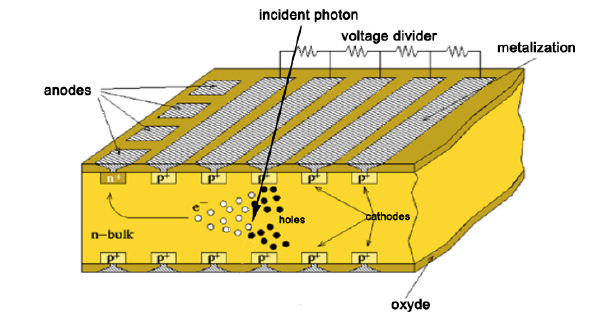}
\caption{Working principle of the Si drift detector used for the {\em STROBE-X}/LAD (see text for details).}

\label{fig:sdd_working_principle}
%%%%%% Insert Figure %%%%%%%%%%%
\end{figure}

The design of such a large instrument is feasible thanks to the detector technology  
of the large-area Silicon Drift Detectors (SDDs, \cite{gatti84}), developed for the ALICE/LHC experiment 
at CERN \cite{Vacchi_etal_1991} and later optimized for the detection of photons to be used on 
{\em LOFT} \cite{2014JInst...9P7014R}, with typical size of 11~$\times$~7 cm$^{2}$ and 450 $\mu$m thickness. 
The key properties of the Si drift detectors  are their capability to read-out a large photon collecting 
area with a small set of low-capacitance (thus low-noise) anodes and their very low mass 
($\sim$1 kg m$^{-2}$). The working principle is shown in Figure \ref{fig:sdd_working_principle}: 
the cloud of electrons generated by the interaction of an X-ray photon is drifted towards the read-out 
anodes, driven by a constant electric field sustained by a progressively decreasing negative voltage 
applied to a series of cathodes, down to the anodes at $\sim$0 V. The diffusion in Si causes the 
electron cloud to expand during the drift. The charge distribution over the collecting anodes depends on the absorption point in the detector. The maximum drift time of $\sim$ 7 $\mu$s, which is the highest detector contribution to the uncertainty in the determination of the arrival time of the photon. Each LAD detector is segmented in two halves, with 2 series of 112 read-out anodes (with 970 $\mu$m pitch) at two edges and the highest voltage along its symmetry axis. 

The high-density anodes of the detector require a read-out system based on ASICs. The requirement on the energy
resolution implies that low-noise and low-power ASICs are needed (17 e$^{-}$ rms noise with 650 $\mu$W\/channel). The read-out of each LAD SDD detector is performed by 8 full-custom 32-channel ASICs inherited from the IDeF-X HD development\cite{2012NIMPA.695..415G}, with A/D conversion carried out by one 16-channel OWB-1 ASIC \cite{2017ITNS...64.1071B}. The dynamic range of the read-out electronics is required to record events with energy up to 80 keV. The events in the nominal energy range (2--30 keV) are transmitted with 60 eV energy binning.
Despite detecting as many as 156,000 counts per second from the Crab, the segmentation into 960 detectors 
and 215,000 electronics channels means that the rate on the individual channel is very low even for very bright sources, removing any pile-up or dead-time issues.  To maintain good energy resolution throughout the mission lifetime, the detectors need to be cooled to reduce the leakage current.  To keep the energy resolution below 300 eV at end-of-life, the detector temperature must be kept below $-30^{\circ}$C.  Operating at higher temperatures is allowable, but the energy resolution will be degraded. Passive cooling is to be used, given the large size of the instrument.

Taking full advantage of the compact detector design requires a similarly compact collimator design. This is provided by the capillary plate technology.
In the LAD geometry, the capillary plate is a 5~mm thick sheet of lead-glass ($>$40$\%$ Pb mass fraction) with same size as the SDD detector, with round micro-pores 83 $\mu$m in diameter, limiting
the field of view (FoV) to 0.95$^{\circ}$ (full width at half maximum). The open area ratio of the device is 75$\%$. 
The thermal and optical design is then completed by an additional optical filter, composed by a thin (1 $\mu$m thickness) 
Kapton foil coated with $>$100 nm aluminium. This guarantees 10$^{-6}$ filtering on IR$/$Visible$/$UV light, while transmitting $>$90$\%$ at 2 keV and above.

\subsection{Wide Field Monitor (WFM)}

\begin{figure}[t!]
\includegraphics[width=2.5in]{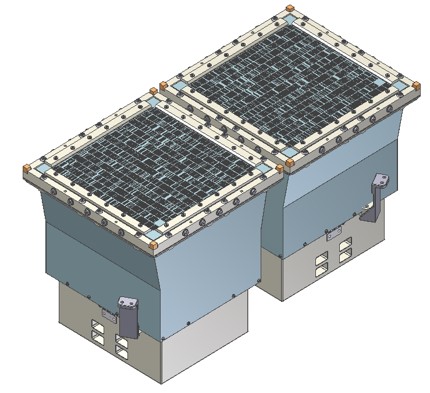}\hfill
\includegraphics[width=3.0in]{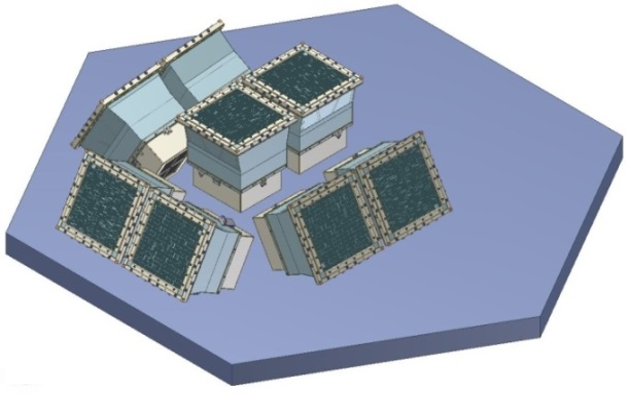}
\caption{The {\em STROBE-X}/WFM instrument. Left: A WFM camera pair. Right: The WFM assembly with four camera pairs.}
\label{fig:WFM_cameras}
\end{figure}

The Wide Field Monitor (WFM) is a coded mask instrument consisting of four pairs of identical cameras, with position sensitive detectors in the (2--50) keV energy range. The same Silicon Drift Detectors (SDDs) of the LAD are used, with a modified geometry to get better spatial resolution. These detectors provide accurate positions in one direction but only coarse positional information in the other one (1.5D). Pairs of two orthogonal cameras are used to obtain precise two-dimensional (2D) source positions (see Fig. \ref{fig:WFM_cameras}, left). The design of the WFM is modular, so that there is no need to put the two cameras of each camera pair together. The concept and design of the WFM is inherited from the {\em LOFT} WFM instrument \cite{LOFTExpA,WFM2014}. 

The effective field of view (FoV) of each camera pair is about 70$^{\circ} \times 70^{\circ}$ ($30^{\circ} \times 30^{\circ}$ fully illuminated, $90^{\circ} \times 90^{\circ}$ at zero response). A set of four camera pairs is foreseen, with three pairs forming an arc covering 180$^{\circ}$  along the the sky band accessible to the LAD and XRCA, and the fourth pair aimed to monitor the anti-Sun direction (see Fig. \ref{fig:WFM_cameras}, right). 

\begin{figure}[t!]
\centering\includegraphics[width=4.0in]{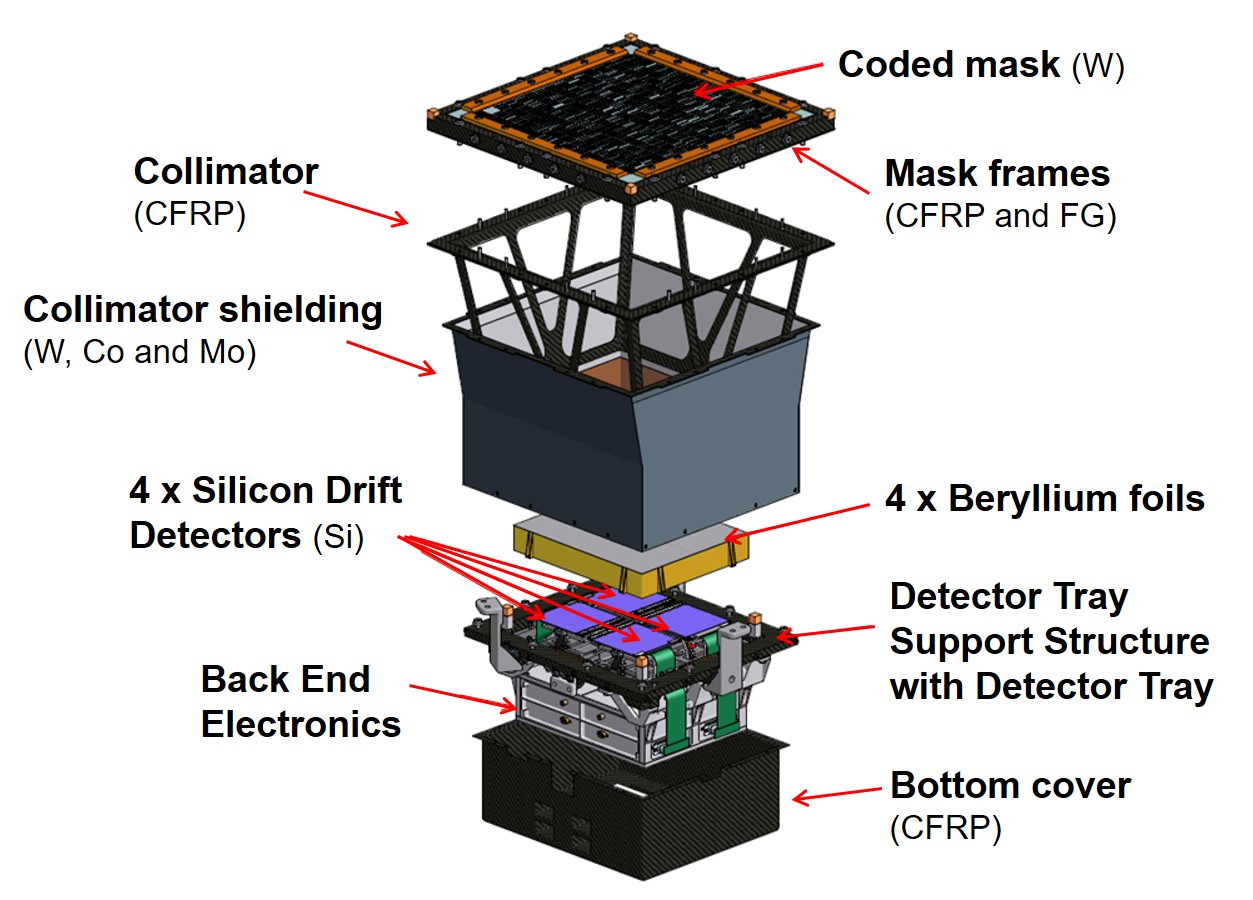}
\caption{An exploded view of a {\em STROBE-X}/WFM camera, indicating all its components.}
\label{fig:WFM_exploded_view}
\end{figure}

Each camera includes a detector tray with four Silicon Drift Detectors, four Front-End Electronics, four Be windows, one Back End Electronics assembly, a Collimator and a Coded Mask with a Thermal Blanket, as shown in Fig. \ref{fig:WFM_exploded_view}. In addition, two Instrument Control Units (ICUs), in cold redundancy, are required. 

The WFM SDDs, ASICs and Front End Electronics are similar to those of the LAD, except that the SDD anode pitch is reduced (145 $\mu$m versus 970 $\mu$m) to improve spatial resolution. There is a higher number of ASICs per SDD: 28$\times$ IDeF-X HD ASICs, with smaller pitch, and 2$\times$ OWB-1 ASICs. Also the PSU (Power Supply Unit) and the Back End Electronics are similar to those for the LAD, but with the BEE providing additional capability to determine photon positions. The ICU controls the eight cameras independently, interfaces with the Power Distribution Unit, and performs on board location of bright transient events in real time.

A beryllium window above the SDDs, 25 $\mu$m thick, is needed to prevent impacts of micro-meteorites and small orbital debris particles (see Fig. \ref{fig:WFM_exploded_view}).

The coded mask of each WFM camera is made of tungsten, with an area of $260\times 260$ mm$^2$ and a thickness of 150 $\mu$m. The mask pattern consists of $1040 \times 16$ open/closed elements, with a mask pitch of 250 $\mu$m $\times$ 14 mm. The dimensions of the open elements are 250 $\mu$m $\times$ 16.4 mm, with 2.4 mm spacing between the elements in the coarse resolution direction for mechanical reasons. The nominal open fraction of the mask is 25$\%$. The detector-mask distance is 202.9 mm. The corresponding angular resolution (FWHM) for the on-axis viewing direction is equal to the ratio of the mask pitch to the detector to mask distance: 4.24 arcmin in the high-resolution direction and 4.6 degrees in the coarse resolution direction. 
The mask must be flat, or at least maintain its shape, to $\pm 50$ $\mu$m over its entire surface across its full operational temperature range. In the \strobex{} design, the spacecraft is pointed so that the XRCA optical bench acts as a sun shield for the WFM, preventing sunlight-induced temperature gradients on the mask. %The design of the mask frame also helps to fulfill this requirement.

The collimator supports the mask frame and protects the detector zone from  X-rays coming from outside. It is made of a 3 mm thick open Carbon Fiber Reinforced Plastic (CFRP) structure, covered by a shield with the following structure: a 1mm thick CFRP layer outside covered by a 150 $\mu$m Tungsten foil, and inside covered by Cu and Mo 50 $\mu$m thick foils. Cu and Mo are included for in-flight calibration purposes (see Fig. \ref{fig:WFM_exploded_view}). 

The BEE box is located at the bottom of the camera. It includes the BEE board and the power supply unit. It is protected by a CFRP cover. The ICU includes the WFM Data Handling Unit, with mass memory and the Power Distribution Unit; there is a main and a redundant unit, in separate boxes.  

\begin{table}
\caption{Parameters for the WFM instrument and the overall {\em STROBE-X} Mission \label{fig:wfmstrobex}}
\includegraphics[width=6.5in]{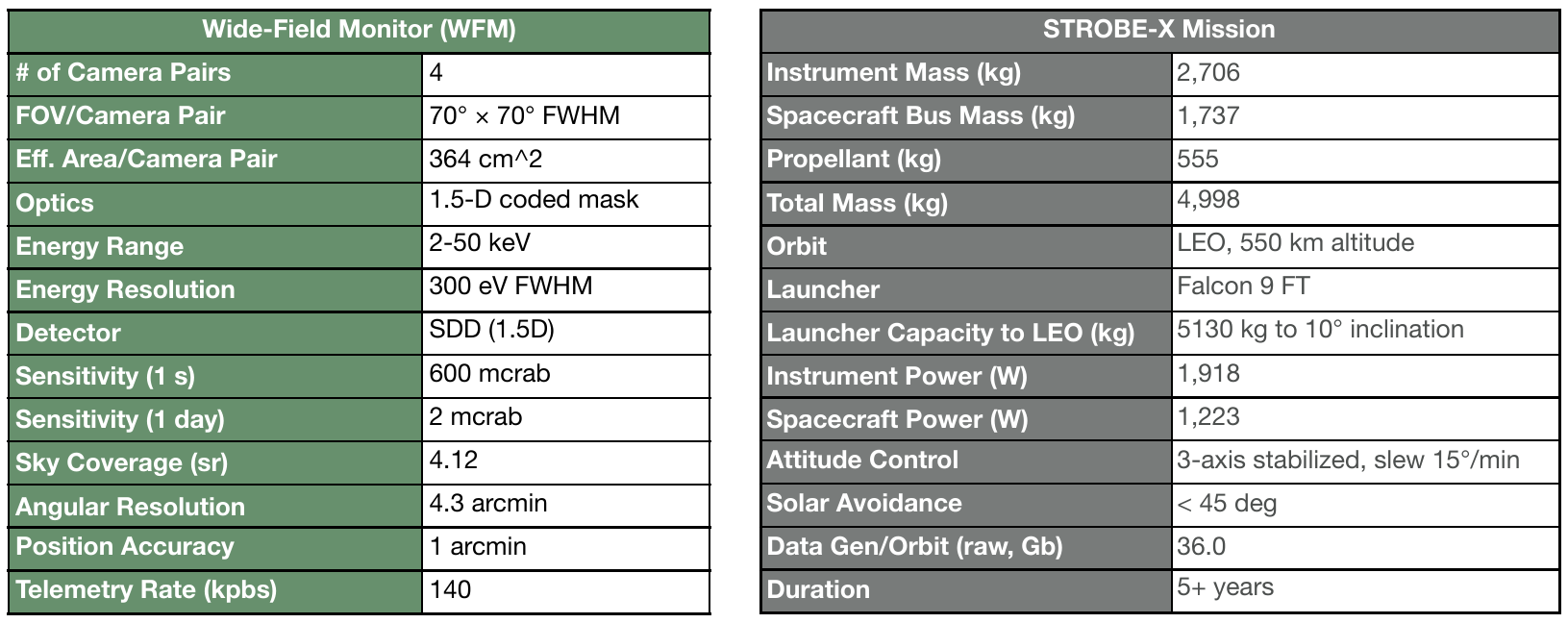}
\end{table}

\section{Design Reference Mission}

The overall mission concept is an agile X-ray observatory in low-Earth orbit, similar to previous missions like {\em RXTE} and 
{\em Swift}. A study in 2018 April at the NASA/GSFC Mission Design Lab (MDL) developed the spacecraft bus design and other aspects of the mission, as described below.

\begin{figure}
    \centering
    \includegraphics[width=5.0in]{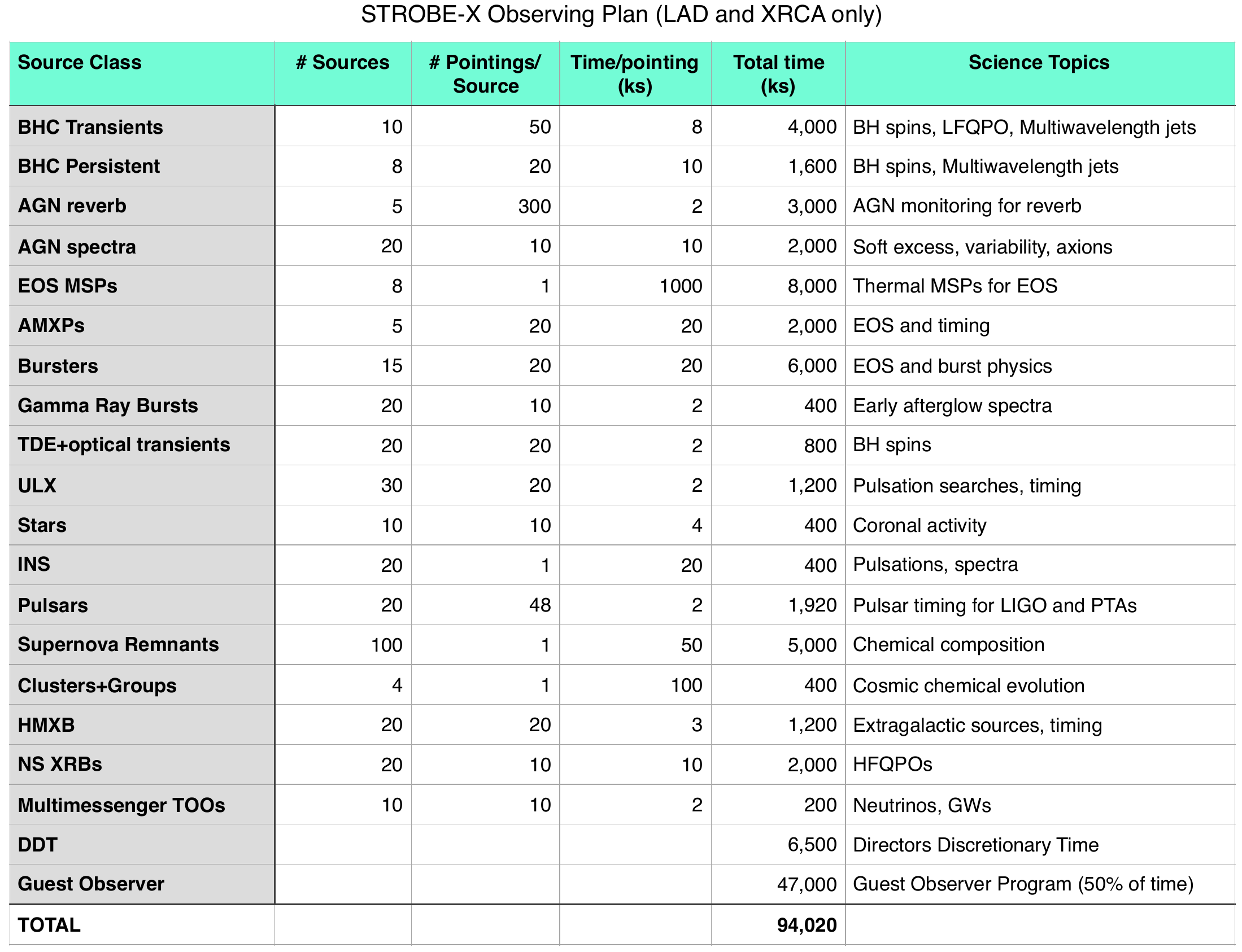}
    \caption{Candidate observing plan for the pointed instruments to achieve the core science 
    described in \S\ref{sec:sci}.  A total of 47 Ms (50\%) is allocated for the core
    science, with 47 Ms for a Guest Observer program.
    %\fixme{Please look at the plan for your favorite source class and comment!}
    \label{fig:obsplan}}
\end{figure}

The scheduling will be highly dynamic, with planning being done frequently ($\sim$ daily) 
based on the currently
active sources, TOO requests, and coordinated observations as well as the long-term plan for 
observations of steady sources and monitoring campaigns. To demonstrate that we can accomplish the
priority science during the prime mission, we have developed an example observing plan (Figure 
\ref{fig:obsplan}). This is based on 5 years of observations, with 50\% observing efficiency, holding 50\%
of the available time for an openly-competed guest observer program. This results in a total
of 47 Ms of observations for the core science program. This includes a 
substantial allocation of discretionary time, which has proved highly 
valuable for \nicer{} to respond to numerous unexpected observing opportunities
and to apply additional time to aspects of the core program as needed.

\subsection{Launch and Orbit}
The LAD detectors are sensitive to non-ionizing energy losses from radiation exposure (which causes increased leakage current), so minimizing the time spent in the 
South Atlantic Anomaly (SAA) by going to as low an orbital inclination as possible is desirable. 
We evaluated {\em STROBE-X} using the performance parameters provided by SpaceX via NASA Launch Services for a Falcon 9 launch vehicle, 
assuming an expendable first stage. The mass that can be put into a 600 km circular orbit is a very strong function of the
desired inclination, with a capacity of 5130 kg to 10$^\circ$ inclination and 7730 kg to 15$^\circ$ inclination. We thus 
plan for a 550 km altitude circular orbit at an inclination of 10$^\circ$.  If additional mass margin is needed, a small 
inclination increase can easily allow the launcher to accommodate that. And, if an equatorial launch site becomes available
from SpaceX or as a European contribution, then {\em STROBE-X} could avoid the SAA entirely, which would increase efficiency and
reduce the cooling requirements on the LAD.

\subsection{Propulsion}

Avoiding a propulsion system altogether would be desirable from a cost and complexity perspective. 
However, there are three reasons that one could be necessary: (1) if the casualty probability from an 
uncontrolled reentry exceeds $10^{-4}$, (2) if reboosting is required to achieve the goal of 10-year orbital lifetime, 
or (3) if a maneuver capability is needed to avoid orbital debris collisions. While none of these are formally required,
we have taken the conservative approach and included a propulsion system capable of all three.

\subsection{Pointing and Attitude Control}

{\em STROBE-X} must be able to slew rapidly over the full sky outside of the 45$^\circ$ Sun avoidance region in order to
follow transients, make monitoring observations, and respond rapidly to targets of opportunity. The minimum required
slew rate is 5$^\circ$/minute, with a goal of 15$^\circ$/minute. While the minimum rate could be achieved with conventional
reaction wheels, this would be inefficient for short observations and delay getting to fast transients. We thus chose
to use Honeywell M50 control moment 
gyroscopes\footnote{\url{https://aerocontent.honeywell.com/aero/common/documents/myaerospacecatalog-documents/M50_Control_Moment_Gyroscope.pdf}} 
(CMGs) for attitude control. These are somewhat more expensive than 
reaction wheels but allow us to reach our 15$^\circ$/minute goal, and are at high TRL. 
CMGs have somewhat larger jitter than reaction wheels, which can be an issue for high-resolution imaging instruments, but {\em STROBE-X}'s instruments only require arcminute-scale stability.

Attitude knowledge is provided by star trackers and coarse Sun sensors and momentum unloading is accomplished with magnetic torquers.

Most maneuvers are planned on the ground and uploaded as time tagged commands, but the spacecraft also has the ability
to perform an autonomous repoint based on a message from the WFM processor, which can be programmed to trigger on specific
transients, like GRBs and superbursts.

%\fixme{SOME STATEMENTS HERE ABOUT THE POINTING REQUIREMENTS, JITTER REQUIREMENTS AND CAPABILITY TO MEET THEM USING CMGS?}

\subsection{Telemetry and Data Rates}

We use TDRSS Ka band downlink via a high gain antenna to achieve 300 Mbps, enabling downlink of 540 Gb/day average.  A key driver for telemetry capacity is the ability to observe bright sources with full energy and time resolution. Table 1 shows telemetry rates for the LAD and XRCA, assuming they are observing a typical 100 mcrab source, resulting in an easily accommodated telemetry rate of of 5.5 Gb/orbit. To estimate the telemetry capacity needed to accommodate full resolution observations of bright sources and to allow more extended-time observations of bright sources to be downlinked over several orbits, we assumed that the two {\em STROBE-X} pointed instruments are observing a 1 mCrab source 25\% of the time, a 500 mCrab source 50\% of the time, a 5 Crab source 5\% of the time, and background 20\% of the time, with 15 minutes per orbit in SAA for an entire day, resulting in 540 Gb/day (36 Gb/orbit average). This high telemetry capacity enables {\em STROBE-X} to downlink all events all the time, including for very bright sources, and all events for the WFM.  This capability is an important enhancement relative to the {\em LOFT} concept.

TDRSS S-band Multiple Access through a pair of omnidirectional antennas allows broadcasting of burst and transient 
alerts to the ground in less than 10 seconds as well as rapid commanding from the ground 
in response to a TOO request. The burst and transient alerts will be rapidly followed by localizations and quick look light curves, similar to {\em Swift} and {\em Fermi}/GBM. These S-band antennas can also communicate with ground stations for contingencies and 
launch and early orbit operations.

\subsection{Ground Segment and Mission Ops}

The ground system and mission operations will be similar to previous missions like \rxte{} and \fermi.
The baseline plan utilizes the GSFC Multi-Mission Operations Center where spacecraft commanding, mission planning/scheduling, telemetry monitoring and Level 0 data processing take place. A science operations center will handle higher level data processing, experiment planning, and instrument health monitoring.
The final data products will be archived and made available via the HEASARC. Most functions are provided 
by existing COTS or GOTS software, with some development of mission-unique capabilities.

\subsection{Data Analysis}
\label{sec:calibration}
The data analysis for spectral-timing X-ray data is well established. The foundation is the FITS format data files archived
at the HEASARC, and the HEASoft suite of FTOOLS for manipulating them. On top of that many groups have developed software
for specific tasks such as ISIS and XSPEC for spectral fitting, Stingray for spectral-timing analysis, 
PINT for pulsar timing analysis, and many more. These tools have been developed for missions such as \rxte{} and \nicer{}
and will continue to evolve over the coming years with more high-level language interfaces and capabilities, as well as 
optimizations for handling very large datasets, but the analysis capabilities to achieve \strobex's science goals
are already in place.

The large areas of the XRCA and LAD arrays on \strobex{} mean that systematic effects can dominate over statistical errors unless methods for controlling and quantifying uncertainties in instrument responses, calibrations, and backgrounds are employed.  The \strobex{} team, based upon experience with past and current missions, is dealing with these issues by considering the roles of instrument design, characterization and operations -- to minimize systematic uncertainties -- and modern data analysis techniques -- to address residual systematic effects within the \strobex{} instruments.  X-ray astronomers have a rich legacy of tackling such issues within our broader community, e.g., the International Astronomical Consortium for High Energy Calibration (IACHEC), and these approaches (see \cite{IACHEC2019}) are informing the \strobex{} team's work.

The LAD shares heritage from \textit{LOFT} development, while the XRCA benefits from \nicer{} flight experience.  \rxte{} experience is also relevant.  Key areas of focus have been divided in topic areas that include effective area (and absolute calibration), response matrix development and maintenance, dead time effects, the chain for absolute timing of photon arrivals, and instrumental and astrophysical backgrounds.  The approach is to assign a small team to assess each concern individually, review experience from prior missions and design work, explore recourses, and work relevant trades at a level appropriate to this stage of development. Particular attention is being devoted to mitigation strategies, e.g., for background reduction or gain monitoring, that might be best implemented in hardware or that require critical pre-launch calibrations.

Experience with both \rxte{} and \nicer{} has shown that the implications for spacecraft operations also must be carefully considered. This includes determining the cadence and duration of observational time devoted to blank fields for background model development.  (Upcoming all sky X-ray surveys, e.g., with \textit{eROSITA} also will be leveraged in these studies.) Dead time implications will be studied in practical environments using \nicer, which will inform non-paralyzable designs for event-processing electronics and bookkeeping of corrections for exposure resulting from merging of detectors into modules.

Newly-developed techniques for fitting bright sources with simultaneous parametrization of both the source spectrum itself and adjustments to response matrices have been developed for other missions and will be carried over to \strobex{} analyses.  The success of post-launch fitting of detector response functions has been demonstrated for \rxte{} observations of bright sources.

Trades between different available strategies (hardware, operations, and data analysis) will weigh the costs of dealing with specific issues in design, with calibration during I\&T, and with on orbit calibration and data analysis mitigation.  The latter strategies primarily impact observing times and efficiencies and flight support labor. These trades, however, will be undertaken only after initial consideration of the full range of options for each important issue. The goal is to reduce the impact of systematic effects on scientific results to the point that photon statistics remains the leading term in the error budget for assessing feasibility of scientific goals.

\section{Cost, Risk Assessment}

To be considered as a candidate probe class mission, NASA requires that the total lifecycle mission cost estimate (Phases A--F)
be less than \$1B in FY 2018 dollars. NASA has provided guidance that: (1) \$150M should be held for launch costs, 
(2) unencumbered cost reserves should be 25\% of Phases A/B/C/D costs, 
(3) cost assumptions should be for unmodified Class B missions, (4) assume a Phase A start date of 2023 October 1.

\begin{wrapfigure}{r}{0.5\textwidth}
%\begin{figure}
    \centering
    \includegraphics[width=3.0in]{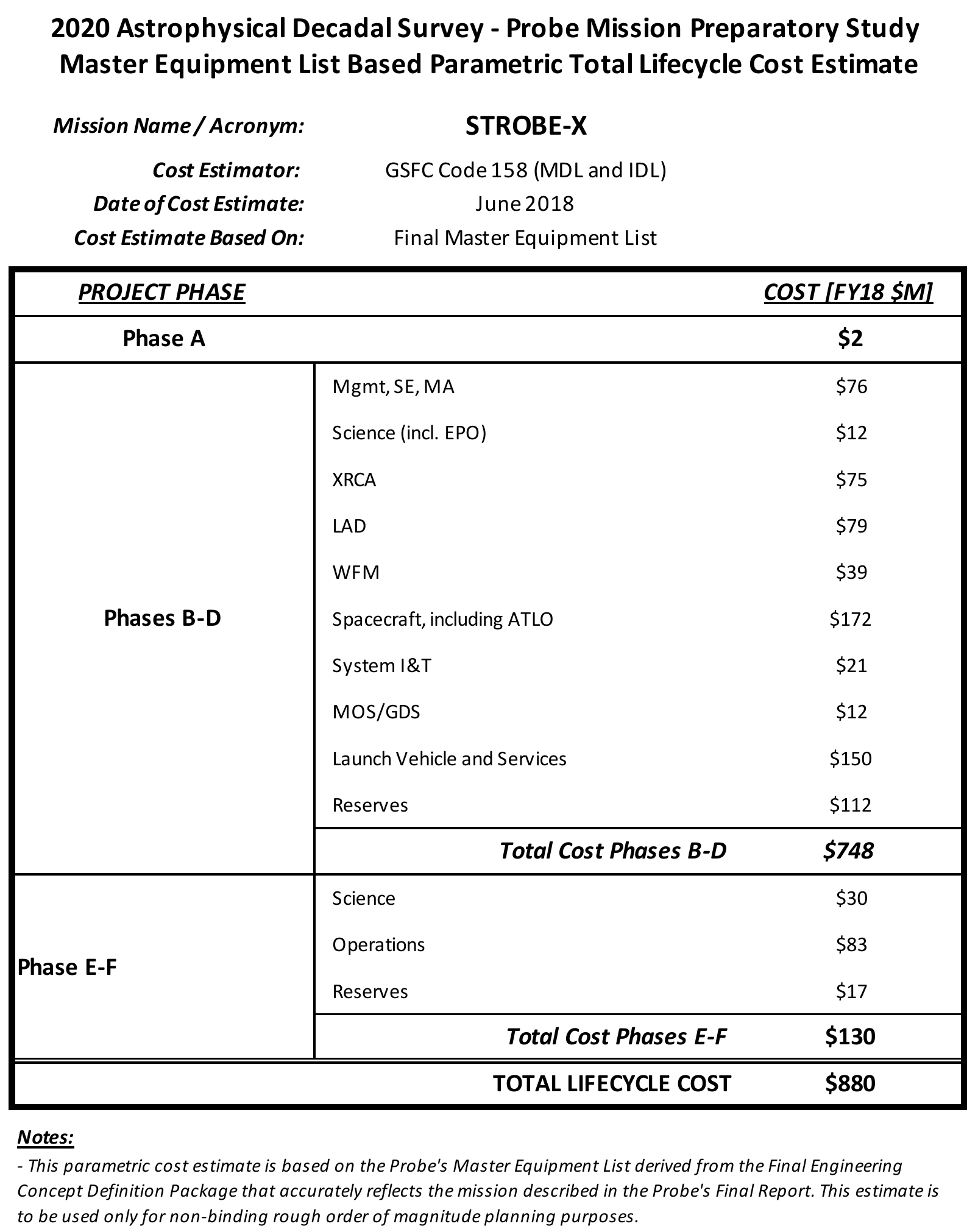}
    \caption{Top level mission cost estimate (format prescribed by NASA) for the design developed in the GSFC IDL and MDL studies. }
    \label{fig:cost}
%\end{figure}
\end{wrapfigure}

The IDL and MDL studies provided detailed cost estimates for the instruments, spacecraft bus, ground systems, integration and 
test, and downlink costs, using a combination of parametric and grass roots costing. To that, we have added standard
percentage ``wraps'' for WBS items like Project Management, Systems Engineering, Safety and Mission Assurance, Science, and 
Education and Public Outreach. For mission operations, we found that the wrap was a significant underestimate compared to NASA
guidance and historical precedent from missions like \textit{Fermi}, so we increased it to \$15M/year for the prime mission. This process
resulted in a total cost estimate comfortably below the probe class cost cap, with over 10\% margin (above and beyond the mandated 25\% reserves), giving us very high confidence that this mission can be executed as a probe.  We note that this cost estimate is very conservative in that it assumes all costs are borne by NASA. In reality, there would surely be a significant contribution from Europe that would reduce the cost to NASA.

For the mission schedule, we followed the NASA guidance for probe class missions being proposed to the 2020 Decadal Survey and
used a Phase A start date of 2023 October 1. Based on this, we constructed a detailed construction, integration and test 
plan, including 8 months of funded schedule reserve, giving an estimated launch date of 2031 January 1.

As another demonstration of feasibility, we note that the last NASA Astrophysics mission in Probe class was \fermi,
which has been operating highly successfully on orbit for a decade. \fermi{} had a total wet mass of 4400 kg
and a total cost of \$800M in FY17 dollars (with reserves). The \fermi{}/LAT instrumented and read out 80 m$^2$ of
silicon strip detectors, and the mission included an all-sky instrument, the GBM, which triggers autonomous
repointing for transient events. So, in mass, electronic complexity and onboard processing, \fermi{} gives confidence
that \strobex{} can be executed as a probe class mission.

\subsection{Risk Assessment}

Designing for high system reliability and robustness was an important consideration from the beginning.  
The philosophy of a highly modular design where many individual component failures can be tolerated while still fully meeting the 
science requirements was very successful for \nicer.  This concept allows substantial savings on parts acquisition and integration and test, compared to missions that rely on single apertures or focal planes where a single failure
can be mission ending.

As part of our studies at the GSFC IDC, 
reliability analyses were performed using Fault Tree Analysis, 
Failure Modes Effects and Criticality Analysis, Parts Stress Analysis, and 
Probabilistic Risk Assessment.
With the redundancy built into the design of the instruments we find a probability of $>92$\%
that we will meet our science performance requirements at the end of 5 years.
In addition, the spacecraft analysis yielded a predicted reliability of $>85$\% at 5 years. Both
of these meet the guidelines for Class B missions.

\subsection{Heritage}

The \strobex{} design makes use of many components with flight heritage and the technology readiness level for the parts that have not yet flow is already quite high, as we describe here.

The XRCA instrument is directly based on the currently flying \nicer{} instrument on the ISS. The biggest differences 
are the larger concentrators and the composite optical bench. Larger concentrators have already been built for the XACT
sounding rocket payload and thus are at TRL 6, while the composite optical bench structure is
new, but many similar composite structures are in use. \fixme{Keith and Zaven, please flesh this out a bit.}

The LAD design is directly inherited from the LOFT 3-year assessment study within the ESA M3 context. The experiment is based on two very solid and mature technologies: 1) large-area Silicon Drift Detectors, with strong heritage in the Inner Tracking System of the ALICE/LHC at CERN, in which 1.4 m$^2$ of SDD with approximately the same design successfully operating since 2008; 2) the capillary plate collimators, that is the structure of the microchannel plates, successfully flown on several space missions in the past decades, including \chandra. 

The WFM design is a conventional coded mask experiment but with the enhanced performance and low resources enabled by the same SDD as the LAD. 
A very similar design is operating onboard the AGILE mission since 2007.

The spacecraft design from the MDL uses only TRL 7--9 components.

\subsection{Descopes}

All of the \strobex{} instruments are highly modular, providing a smooth descope path, should cost savings be required.  Among the potential descope options are:
\begin{itemize}
\itemsep 0in
\itemindent 0in
\topsep 0in
\parskip 0in
\partopsep 0in
    \item Reduce number of XRC units per quadrant (or LAD modules per panel).  Benefit: reduces costs to manufacture, integrate and test, and shortens schedule. Risk: Cuts into science yield, but some of this can be made up with additional exposure time.
    \item Switch from control moment gyros to reaction wheels.  Benefit: Saves a few million dollars. Risk: Hurts TOO response time, particularly for autonomous triggers to burst events, and worsens observing efficiency.
    \item Reduce field of regard. Benefit: Simplifies and reduces cost of thermal system. Risk: Many transient and variable targets may be unobservable when science return would be highest, but many of the targets can just be scheduled at the right times with no harm.
\end{itemize}

\section{Management Plan}

NASA has not decided on how probe-class missions will be managed, if they are implemented in the next decade. As a baseline, we assume the mission management will be similar to \fermi, the last astrophysics probe. This was done as a strategic mission, with leadership from a NASA center. The instruments were PI-led and competed.

\begin{figure}
    \centering
    \includegraphics[width=5.5in]{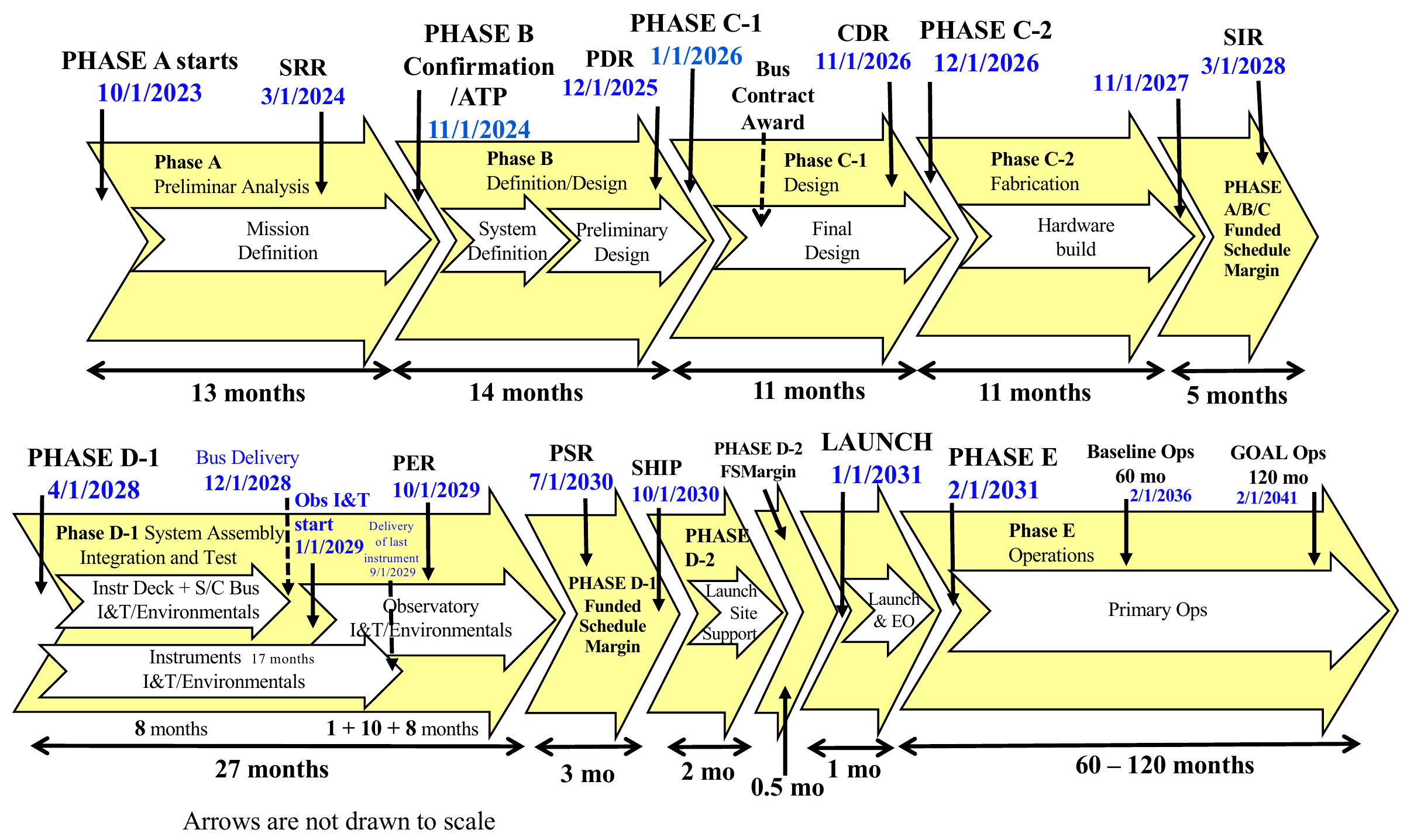}
    \caption{Top-level \strobex{} schedule developed in the MDL study.}
    \label{fig:schedule}
\end{figure}

During our study, we developed a baseline schedule for the mission, which is shown in Figure~\ref{fig:schedule}. 

\section{Technology Maturation Plan and Future Work}

No new technologies are required to execute \strobex; however some investment in raising the technology readiness level of the 
instruments would greatly increase the fidelity of the cost estimates, and save money by shortening the development schedule. 
This could be done with a small number of APRA/SAT-scale programs.  In addition, there are a few places where some investigations 
into alternatives to the baseline implementation could improve performance. Some of the
small investments are:
\begin{itemize}
\itemsep 0in
\itemindent 0in
\topsep 0in
\parskip 0in
\partopsep 0in
    \item Update microcontroller in the XRCA detector readout to an FPGA-based design. This
    will greatly reduce the deadtime and significantly improve high count rate capabilities.
    \item Develop ASIC for reading out the LAD and WFM detectors. Similar ASICS exist that
    demonstrate the functionality and noise performance required, but doing a custom ASIC
    now will shorten the development time and enable prototype modules to be built and tested.
    \item Build a single LAD module with detector, readout and collimator, and test to bring to TRL~6.
    \item Build a single WFM camera and test to reduce risk and understand thermal performance.
\end{itemize}

The initial design presented here, and studied at the IDL and MDL, meets the \strobex{} science goals and probe class 
constraints and many aspects of the design are highly mature. However, in this limited study, we did not have the time or
resources to iterate on the design.  During the study we identified several areas for future work, which will be addressed
either in the next iteration of the design study or by technology development programs that might be incorporated into
the design in a future trade study.  These include:
\paragraph{Thermal Design} The thermal design for the LAD is challenging with only passive cooling, because of the large range of Sun angles required to cover the full field of regard. When the radiators are sized sufficiently to keep the instrument cool in the
    hot case, a large amount of heater power is required to keep them warm in the cold case. In the current design, a large
    amount of power is allocated to these heaters, with effects on solar panel and battery sizing. Preliminary investigations
    indicate that using variable-conductance heat pipes (VCHPs) could be used to maintain the operating temperature with much
    lower heater power.

\paragraph{Alternative LAD Collimators} The baseline glass micropore collimators are lightweight, high-TRL and sufficient for the job. 
    However, improving their performance at high energies can reduce the LAD background and extend its useful energy range.
    Investigations are ongoing at NRL into using atomic layer deposition (ALD) to coat the micropore walls
    with high-Z metals, increasing their stopping power.
    A trade study including cost, mass and manufacturability impacts would be required before new collimators
    are included.
    
\paragraph{Calibration and Background} As described in Section \ref{sec:calibration}, we are considering whether any
    hardware modifications to improve the instrument calibration or background knowledge are worth including in the design.

%\fixme{Should we mention considering a small gamma-ray detector or a fast slewing optical camera to enhance GRB science?}

\section{International Context}

Looking at the European Space Agency (ESA) planning for medium and large class missions, none of the approved M class missions 
cover high energy astrophysics, with only a high-$z$ gamma ray burst mission (\textit{THESEUS}) being one of the three missions currently being
studied for the M5 launch slot\footnote{A precursor mission to \textit{THESEUS} called \textit{SVOM}, a French-Chinese collaboration to study GRBs and compact object mergers, is being developed for a launch in 2021}. One of ESA's planned large missions, \athena{}, will provide
imaging and high-resolution spectroscopy, as well as some timing capabilities, in the 2030s. \athena's high-resolution
spectroscopy and imaging will be highly complementary to \strobex{}. However, with only 2500 cm$^2$ effective area around the 6.4 keV iron line, 
no coverage of the Compton reflection hump around 30~keV, and limited slewing/rapid response capability, \athena{} is not duplicative of \strobex's strengths.

The extensive study of the LOFT mission as part of the ESA M3 process both demonstrated the scientific promise and
feasibility of a high-throughput spectral timing mission and built up a worldwide collaboration of scientists 
advocating for these capabilities.  \strobex{} builds on and extends these capabilities to the soft X-ray band and
has broadened the community's support.

One indication of this excitement is the work on the \textit{eXTP} mission in China, which has several similarities to \strobex, with
Europe contributing LAD and WFM instruments.  There are important differences between \textit{eXTP} and \strobex{} that mean 
they cover a different phase space: \textit{eXTP} has about 1/2 the collecting area, it slews about 1/2 as fast as \strobex, and can
only access 50\% of the sky. 
China does not yet have a long track record for successfully building and operating astrophysics space missions, 
so it is hard to predict how likely \textit{eXTP} is to be completed and meet its schedule and performance 
goals to have a significant science impact. NASA must pursue its own vision in this area, and must not cede leadership to China. 
If \textit{eXTP} is completed, access to data by US scientists is likely to be extremely limited. For example,
few US scientists have been able to get access to data from China's \textit{HXMT} mission or India's \textit{Astrosat} mission. 
NASA is prohibited
from working directly with China and the NSF does not fund work on space mission data. US scientists
may only be able to participate as Co-Is on proposals led by European or Chinese researchers, and the usefulness of the data
is strongly determined by the ease of access and the analysis tools and user support, such as the HEASARC provides for NASA high energy missions.  Interestingly, a similar situation existed 30 years ago, when {\em Ginga} flew before {\em RXTE} and had similar capabilities, but was less powerful and had substantial data access challenges for US investigators.  {\em Ginga}'s discoveries, while quite limited relative to {\em RXTE}'s, helped significantly in laying out the paths that would be the most productive uses of {\em RXTE}.

\section{Conclusions}

The \strobex{} mission concept presented here demonstrates the transformative science that can be done in a probe-class mission.
\strobex{} will do breakthrough science in its core topics of measuring black hole spin across the full range of mass scales, understanding
the properties of dense matter in neutron stars, and time-domain and multimessenger astrophysics. It will also serve a broad community
studying energetic processes throughout the Universe, providing complementary capabilities to the ESA \athena\ mission 
and be a critical facility in the era of time-domain astronomy. We strongly encourage the creation of a probe class line in the 2020s
to allow NASA to execute this mission.

\clearpage
\pagenumbering{alph}

\section*{Acknowledgments}

The {\em STROBE-X} mission concept study is funded by the NASA Astrophysics Probes 
program (16-APROBES16-0008). 
The Italian authors acknowledge support from ASI, under agreement 
ASI-INAF n.2017-14-H.O, INAF and INFN. 
The Spanish authors acknowledge support from MINECO grant ESP2017-82674-R and FEDER funds. 
A.L.S. is supported by an NSF Astronomy and Astrophysics Postdoctoral Fellowship under award AST-1801792.
A.L.W. acknowledges support from ERC Starting Grant 639217.

We gratefully acknowledge the superb engineering teams at the NASA/GSFC Instrument Design Lab and Mission Design Lab,
Steve Kenyon, and Takashi Okajima, for their important contributions to the {\em STROBE-X} technical concept. 
We thank Lorenzo Amati and Giulia Stratta for useful discussions.

\bibliographystyle{apj} 
\bibliography{report} 

\end{document}